\documentclass[twocolumn]{aastex631}

\usepackage{multirow}
\usepackage{array}
\usepackage{makecell}
\usepackage{tablefootnote}
\usepackage{url}
\usepackage{boldline}
\usepackage{colortbl}
\usepackage{subfigure}
\usepackage{soul}

\accepted{}

\shorttitle{Retrievals for Terrestrial Atmospheres with Transmission and Direct Imaging}
\shortauthors{Gilbert-Janizek et al.}

\graphicspath{{./}{figures/}}

\begin{document}

\title{Retrieved Atmospheres and Inferred Surface Properties for Exoplanets Using Transmission and Reflected Light Spectroscopy}

\correspondingauthor{Samantha Gilbert-Janizek}
\email{samroseg@uw.edu}

\author[0009-0004-8402-9608]{Samantha Gilbert-Janizek}
\affiliation{Department of Astronomy and Astrobiology Program, University of Washington, Box 351580, Seattle, Washington 98195}
\affiliation{NASA NExSS Virtual Planetary Laboratory, Box 351580, University of Washington, Seattle, Washington 98195, USA}

\author[0000-0002-1386-1710]{Victoria S. Meadows}
\affiliation{Department of Astronomy and Astrobiology Program, University of Washington, Box 351580, Seattle, Washington 98195}
\affiliation{NASA NExSS Virtual Planetary Laboratory, Box 351580, University of Washington, Seattle, Washington 98195, USA}

\author[0000-0002-0746-1980]{Jacob Lustig-Yaeger}
\affiliation{Johns Hopkins Applied Physics Laboratory, Laurel, Maryland 20723, USA}
\affiliation{NASA NExSS Virtual Planetary Laboratory, Box 351580, University of Washington, Seattle, Washington 98195, USA}

\begin{abstract}
Future astrophysics missions will seek extraterrestrial life via transmission and direct imaging observations. To assess habitability and biosignatures, we need robust retrieval tools to analyze observed spectra, and infer surface and atmospheric properties with their uncertainties. We use a novel retrieval tool to assess accuracy in characterizing near-surface habitability and biosignatures via simulated transmission and direct imaging spectra, based on the Origins Space Telescope (\textit{Origins}) and LUVOIR mission concepts. We assess our ability to discriminate between an Earth-like and a false-positive O$_3$ TRAPPIST-1 e with transmission spectroscopy. In reflected light, we assess the robustness of retrieval results to un-modeled cloud extinction. We find that assessing habitability using transmission spectra may be challenging due to relative insensitivity to surface temperature and near-surface H$_2$O abundances. Nonetheless, our order of magnitude H$_2$O  constraints can discriminate extremely desiccated worlds. Direct imaging is insensitive to surface temperature and subject to the radius/albedo degeneracy, but this method proves highly sensitive to surface water abundance, achieving retrieval precision within 0.1\% even with partial clouds. Concerning biosignatures, \textit{Origins}-like transmission observations ($t=40$ hours) may detect the CO$_2$/CH$_4$ pair on M-dwarf planets and differentiate between biological and false positive O$_3$ using H$_2$O and abundant CO. In contrast, direct imaging observations with LUVOIR-A ($t=10$ hours) are better suited to constraining O$_2$ and  O$_3$, and may be sensitive to wavelength-dependent water cloud features, but will struggle to detect modern Earth-like abundances of methane. For direct imaging, we weakly detect a stratospheric ozone bulge by fitting the near-UV wings of the Hartley band.

\end{abstract}


\section{Introduction} \label{sec:intro}

The discovery of a growing number of rocky habitable zone planets \citep{gillon2016temperate, gillon2017seven, anglada2016terrestrial, dittmann2017temperate} has provided several plausible targets for future telescopic searches for life beyond our solar system.  Upcoming astrophysics missions will be designed to characterize rocky planets and search for life, observing these planets in transmission, emission, or via reflected light spectroscopy (direct imaging) \citep{national2021pathways}. In the short-term, we will have access to transmission spectra of M-dwarf terrestrial atmospheres from the recently-launched JWST \citep{batalha2018strategies, wunderlich2019detectability,lustig2019coronagraph}, and from the next generation of ground-based telescopes which will see first light later this decade \citep{guyon2012elts, snellen2013finding, rodler2014feasibility}. In the coming decades, the Astronomy and Astrophysics 2020 Decadal Survey has prioritized the development of a space-based IR/O/UV direct imaging telescope for characterizing Earth-like planets around Sun-like stars (FGK), as well as a probe-class far-IR telescope, which may be capable of transmission spectroscopy for characterizing Earth-like planets around M-dwarf stars \citep{national2021pathways}. Both the flagship direct imaging mission and the probe-class transmission telescope will seek to uncover life on exoplanets by searching for signs of habitability and biosignatures. A concept for a large space-based MIR interferometric mission to obtain spectral emission from habitable zone planets and search for signs of life is also being developed \citep{quanz2021large}. 

To assess habitability and life on exoplanets, it will be critical to characterize the planets' surface and/or near-surface environment \citep{robinson2017characterizing}, where life is most likely to reside. In preparation for the future missions that will attempt to characterize potentially habitable planets, it is therefore crucial that we understand their respective strengths and weaknesses for probing planetary surface habitability and performing a robust search for biosignatures 
-- especially in the context of the planetary targets most amenable to transmission and direct imaging observation. Transmission is primed to study the atmospheres of habitable zone planets in compact systems orbiting cooler M-dwarf stars, such as TRAPPIST-1 e, f, and g \citep{gillon2016temperate}, due to the high planet-star radius ratio and frequent transits of these planets \citep{suissa2020dim}. Direct imaging will target systems with sufficient planet-star separation to study the atmospheres of Earth-like planets around more Sun-like stars (FGK dwarfs) \citep{luvoir2019luvoir, gaudi2020habitable}. Transmission probes the terminator of the transiting planet, and is most sensitive to the upper troposphere and stratosphere, primarily due to the intrinsic molecular and scattering opacity of the atmosphere \citep{lincowski2018evolved}. In some cases, refraction can also significantly restrict access to the lower atmosphere in transmission observations \citep{munoz2012glancing, betremieux2014impact, misra2014using}. Furthermore, cloud and aerosol extinction is known to have a significant impact on transmission spectroscopy for larger planets \citep{fortney2005effect, kreidberg2014clouds}, and this effect is likely highly wavelength dependent \citep{charnay20153d}. Modeling suggests that similar effects may occur on terrestrial worlds seen in transmission \citep[e.g.,][]{arney2017pale}. In contrast to transmission, and by analogy with solar system observations, observations taken in reflected light could probe the planetary surface if there is broken cloud cover, and over a shorter atmospheric path, potentially making it less susceptible to obscuration by atmospheric gas and aerosol extinction \citep{fortney2005effect, kreidberg2014clouds}.

Upcoming telescopic capabilities will provide our first assessments of the surface environment of potentially habitable exoplanets and raise the question: how well can we infer the conditions on the planetary surface from transmission versus direct imaging, and do the differing capabilities of these techniques impact the assessment of habitability and biosignatures? 
\par 

Future missions will use transmission and direct imaging to probe rocky exoplanet atmospheres and look for direct and indirect signs of habitability, but these techniques will face unique interpretation challenges. Key habitability indicators include the surface temperature, surface pressure, and surface liquid water content \citep{robinson2018characterizing, meadows2018habitability}. Habitability may be inferred by identifying and quantifying potential greenhouse gases, such as CO$_2$ and H$_2$O which may allow us to model the planet’s greenhouse warming and resultant surface temperature \citep{catlingdavid2018exoplanet}. Identifying the presence of an atmospheric ``cold trap'' may also aid in characterizing the habitability of the planetary surface by indicating warmer surface temperatures and volatile (e.g., water) condensation\citep{wordsworth2014abiotic}. 

Transmission may be particularly sensitive to tropospheric temperature and pressure, since the atmospheric scale height ($\frac{RT}{\mu g}$) sets the size of the spectral features \citep{robinson2018characterizing}. However, the continuum pressure need only supply a lower limit on the surface pressure, as sources of refraction and opacity may limit the transparency of the atmosphere to the surface \citep{misra2014effects, lincowski2018evolved, lustig2019detectability}. In the absence of surface reflectivity variations and clouds, direct imaging could use the Rayleigh scattering slope and the width of molecular bands to help probe surface temperature and pressure, but this will likely be challenging due to the truncation of the visible path by even partial cloud cover, as well as degeneracies in pressure, gravity, and mean molecular weight, and poor prior constraints on related planet properties \citep{robinson2018characterizing}. 

While habitability markers allow us to assess whether the surface can support stable liquid water, biosignatures provide a means of directly investigating how life has altered its environment and will be a critical target for future observations of rocky exoplanet atmospheres. In particular, pairs of gases that are not in thermodynamic equilibrium could indicate active fluxes of gases at the surface that are produced by life.  Three prominent pairs to consider for future telescopes are O$_2$/CH$_4$ \citep{hitchcock1967life}, O$_3$/CH$_4$ \citep{des2002remote} and CO$_2$/CH$_4$ \citep{krissansen2016detecting, schwieterman2016identifying}. 

The canonical biosignature pair is O$_2$ and CH$_4$. Without life, these molecules should destroy each other on geological timescales such that their simultaneous presence in an atmosphere strongly implies efficient production that is inconsistent with known abiotic sources \citep{hitchcock1967life}. O$_3$ is a photochemical byproduct of O$_2$ in our atmosphere, and may therefore act as a proxy for O$_2$ when O$_2$ is not observationally accessible \citep{des2002remote}. This substitution gives the O$_3$/CH$_4$  biosignature pair.  However O$_3$ production is non-linear with O$_2$ abundance \citep{kastingozone}, especially for cooler M-dwarf stars \citep{segura2003ozone, kozakis2022ozone}, and inferring the O$_2$ abundance from a measurement of atmospheric O$_3$ may be challenging. 
Photochemistry will also complicate the task of interpreting biosignature pairs since incident stellar energy may enhance or suppress abiotic and biogenic molecules \citep{meadows2018exoplanet}, leading to gradients in the chemical profile \citep{segura2003ozone, segura2005biosignatures}. Detecting these chemical gradients may provide insight into photochemical processes and thus a more robust interpretation of biosignature pairs in the context of their environment. 

Finally, both CH$_4$ and CO$_2$ have been present and potentially detectable \citep{Meadows2023} throughout Earth’s Archean epoch despite drastic changes in the composition of the atmosphere, making it advantageous for applications on exoplanets that may be inhabited but lack a photosynthetic biosphere \citep{krissansen2016detecting}. Additionally, this biosignature pair may be particularly favorable for planets around M-dwarf \citep{segura2005biosignatures} and K-dwarf \citep{arney2019k} stars, which may experience a photochemical enhancement of CH$_4$. 

One advantage of biosignature pairs is that in addition to indicating a disequilibrium, they increase the probability that at least one of the  gases is due to a biological process by helping to rule out false positive mechanisms. For the O$_2$/CH$_4$ pair the CH$_4$ is not only an indication of a methanogenic biosphere, an additional biological process to the photosynthesis generating the O$_2$, but it also acts as a false positive discriminant, making abiotic O$_2$ production a less likely explanation for observed O$_2$ and O$_3$. However, false positive O$_3$ can potentially be generated by the photolysis of CO$_2$ in an environment that does not contain the catalysts needed for rapid recombination, such as H$_2$O \citep{gao2015stability}.  In this context, high abundances of free CO \citep{schwieterman2016identifying}, and vanishingly low abundances of H$_2$O may be useful discriminants to help identify a potential false positive \citep{gao2015stability}. 

Spectroscopic retrieval is a powerful technique to infer planetary and atmospheric properties from remote spectroscopic data \citep{barstow2020comparison} and to forecast observational needs and interpretation challenges prior to future data acquisition. Retrievals have been used extensively to analyze solar system bodies \citep[e.g.,][]{Mahieux2010, Arney2014, Irwin2008, Spurr2001, Kleinb2009, vinatier2007vertical, kim2011retrieval}. More recently, pioneering retrieval work has inferred atmospheric properties of giant exoplanets using data from current observatories \citep[e.g.,][]{kreidberg2015detection, wakeford2017complete, benneke2019water, line2021solar}. However, these applications have also identified critical challenges, such as degeneracies between clouds and mean molecular weight \citep{Line2016} and biases due to non-uniform/multiple thermal profiles \citep{Feng2016nonuniform, feng20202d, Taylor2020biases}, and others \citep[e.g.,][]{caldas2019effects, pluriel2020ares, macdonald2021trident, nixon2022aura}, that motivate more investigations into how model over-simplifications can impact planetary scale interpretations.  

The lessons learned from giant exoplanet work are only just beginning to be applied to rocky exoplanets, and many gaps still remain in identifying sources of potential biases and confounding factors when interpreting habitability and biosignatures from the simulated spectra of terrestrial exoplanet atmospheres. While \cite{feng2018characterizing} assessed the instrument requirements needed to make significant detections of water vapor, ozone, and oxygen on a directly-imaged Earth-twin with patchy water vapor clouds, their investigation was limited to 0.4~\textendash~1 $\mu$m, thereby excluding methane features as well as the effects of clouds in the near-IR. By contrast, \citet{damiano2022reflected} included methane features in their study and fit for the vertical structure of water vapor in the atmosphere to include the effects of clouds. Previous studies have also assessed our ability to spectrally retrieve the characteristics of potentially habitable planets around M-dwarfs in the near and mid-IR \citep{tremblaydetectability2020} and planets around white dwarfs \citep{kalteneggerWDs2020} with transmission observations, but these studies were limited to non-self-consistent modern Earth atmospheres. Furthermore, \citet{feng2018characterizing}, \citet{tremblaydetectability2020}, and \citet{damiano2022reflected} produced their synthetic data from atmospheres generated with unrealistic uniform vertical profiles for temperature and molecular abundances. Though \cite{zifanabiotic2021} investigated our ability to differentiate self-consistent prebiotic and modern Earth-like atmospheres on TRAPPIST-1 e with high resolution JWST transmission observations, they did not assess our ability to discriminate between an abiotic planet and an inhabited one. Finally, existing studies have yet to fully assess the comparative (and potentially complementary) strengths and weaknesses of transmission and direct imaging spectra for assessing habitability and biosignatures, and probing the surface environment, on rocky, habitable-zone planets.

\par Here we present a study to understand the relative effectiveness of transmission observations and direct imaging for characterizing the near-surface environment of exoplanets to assess habitability and biosignatures. We investigate performing spectroscopic retrievals in the presence of clouds for both transmission and direct imaging, as well as a false positive biosignature assessment for transmission observations. Despite the compelling scientific prospects of emission spectroscopy, habitable zone planets may be too cool to achieve comparably high signal-to-noise ratios in thermally emitted light. Therefore, in this study, we exclusively focus on transmission spectroscopy for transiting planets orbiting M-dwarf stars due to its advantageously high signal-to-noise ratio. For both transmission and direct imaging, we assess our ability to probe the photochemical productivity of the atmosphere via the vertical structure of O$_3$. In the following sections, we describe our methods for comparing the capabilities of transmission and direct imaging spectroscopy (Section \ref{sec:methods}) and the self-consistent planet/atmospheric cases we will explore (Section \ref{sec:experiments}). We provide a summary of our results in Section \ref{sec:results}, and discuss the impacts they have for assessing habitability and biosignatures on terrestrial, habitable zone planets in Section \ref{sec:discussion}. Finally, we provide a summary of these takeaways and identify areas for future exploration in Section \ref{sec:conclusion}.

\section{Methods}

\label{sec:methods}
We use a retrieval model with and without vertically-resolved molecular abundance profiles to infer the composition and physical properties of the atmospheres of terrestrial exoplanets from simulated observations, and thereby characterize the surface environment and habitability of these worlds. Part of assessing the composition of the atmosphere is understanding the vertical resolution of some species that have distinctive structure with altitude, such as photochemically-generated or destroyed molecules (e.g. O$_3$) and water vapor, whose abundance is strongly tied to the vertical temperature structure. By probing the vertical structure we can better understand these important gases in the full context of the planetary environment. Our novel retrieval model solves for the distribution of atmospheric states that best fits the simulated spectrum, assuming either evenly-mixed or vertically-resolved molecular profiles. This state-of-the-art terrestrial retrieval model generates spectra and fits them to the simulated data to infer atmospheric parameters and likely abundance profiles. We are capable of running the model for vertically-resolved molecular profiles, where the algorithm is either constrained to pre-selected pressure points or allowed to freely fit for the pressure points. For each observatory, we determine whether the vertically-resolved models appreciably improved the fit compared to the simpler evenly-mixed case. Finally, to assess the interpretability of the results, we compare the retrieved abundance profiles to the true profiles used to generate the simulated data. The models and our inputs are detailed below.

\subsection{Simulated Exoplanet Spectra}
To produce synthetic transmission and reflectance observations of exoplanets, we use a coupled 1-D climate-photochemistry model to produce self-consistent atmospheres, a line-by-line radiative transfer model to generate high-resolution spectra, and an instrument simulator to degrade the resolution of the spectra to that of the relevant instruments and add astrophysical, telescope, and instrumental noise. These synthetic spectra therefore incorporate the full atmospheric complexity rendered by our models, including a non-isothermal temperature structure, heterogeneously mixed gas abundances, and pressure-broadened absorption lines. As a result, our subsequent retrieval experiments using these simulated observations reflect the ability of the model to accurately infer realistic atmospheric states subject to the necessary simplifying retrieval model assumptions.   

\subsubsection{Simulated Atmospheres \& Synthetic Spectra}
\label{sec:firstsmart}

To generate thermally and photochemically self-consistent atmospheres, we couple the VPL Climate model to \textit{atmos}, a photochemistry model. VPL Climate  \citep{lincowski2018evolved} is a 1-D radiative-convective equilibrium model that uses the Spectral Mapping and Radiative Transfer (SMART) code to internally calculate radiative fluxes and heating rates.
SMART is a line-by-line, multi-stream, multi-scattering radiative transfer code, which includes layer-dependent gaseous and aerosol absorption, emission, and scattering \citep[][originally developed by D. Crisp]{meadows1996ground, stamnes1988numerically}. Finally, \textit{atmos} is a 1-D atmospheric chemistry model that has been used to model the photochemistry of various terrestrial exoplanets in a range of studies \citep{segura2005biosignatures, arney2017pale, meadows2018exoplanet}. Here, we use an updated version of the publicly-available \textit{atmos} model as described in \cite{lincowski2018evolved}.

Once we have generated self-consistent atmospheres, we use the Line By Line ABsorption Coefficients (LBLABC) model and SMART to generate high-resolution spectra. LBLABC calculates rotational-vibrational line absorption coefficients for the gases in the atmosphere under consideration \citep{meadows1996ground} using the HITEMP2010 and HITRAN2012 line lists \citep{rothman2010hitemp, rothman2013hitran2012}. LBLABC assumes air (N$_2$+O$_2$+Ar) line broadening with a Lorentzian wing cut-off of 1000 cm$^{-1}$ for CO$_2$ and H$_2$O, as they are known to have broad wings for Earth and Venus. For all other gases, we use a wing cut-off of 50 $cm^{-1}$. For a complete discussion of how LBLABC calculates the line absorption coefficients, please see \citet{lustigyaeger2023earth}. Line absorption coefficients are calculated by LBLABC for each molecular species at high spectral resolution to fully resolve individual lines and only convolved to the lower resolution of the data after the radiative transfer calculation has been computed via SMART. SMART uses the absorption coefficients output by LBLABC to solve the radiative transfer equations and produce a high-resolution spectrum. SMART is capable of generating atmospheric spectra in transmitted, emitted, and reflected light, and has been rigorously validated on observations of the Earth \citep{robinson2011earth} and Venus \citep{meadows1996ground, Arney2014}. SMART has also been used to model simulated spectra for theoretical habitable M dwarf planets \citep{segura2003ozone, segura2005biosignatures, meadows2018exoplanet} and uninhabitable planets \citep{lincowski2018evolved, lincowski2019observing}. SMART calculates cloud opacities independently from rotational-vibrational line coefficient opacities. SMART is also capable of computing layer-dependent flux Jacobians and top-of-atmosphere radiance Jacobians. Radiance Jacobians are matrices consisting of the partial derivatives of the top-of-atmosphere radiance field with respect to each user-specified property of the atmosphere at each atmospheric layer. These matrices thus relate the spectrum to small perturbations to the atmospheric state, allowing us to quickly determine which vertical sections of the atmosphere the spectrum is sensitive to. SMART can compute radiance Jacobians with respect to temperature, gas abundances, and surface albedo. While SMART does not currently support the calculation of transit depth Jacobians, this is being considered for future work.

\subsubsection{Instrument Simulator}

To simulate the \textit{Origins} and LUVOIR instrument data, we use the open-source Python \texttt{coronagraph} model to degrade the high resolution synthetic spectra and add noise \citep{robinson2016characterizing, lustig2019coronagraph}. \texttt{coronagraph} was originally developed to simulate data from telescopes equipped with a coronagraph and has since been modified to generate simulated data (signal and noise) for a range of exoplanet observing methods using space-based \citep{bolcar2016initial, mennesson2016habitable} and ground-based telescope architectures \citep{meadows2018habitability}. The user provides a high-resolution exoplanet spectrum and the model calculates the data points by degrading the spectrum to the instrument resolution using a boxcar convolution. Though a triangle convolution function is a more realistic simulator of a slit spectrometer, we did not find significant differences when implementing the boxcar filter. \texttt{coronagraph} is then used to calculate the observed photon count rates for a range of observing scenarios based on inputs specified by the user, including telescope diameter and temperature, instrument throughput, and wavelength-dependent spectral resolution. \texttt{coronagraph} calculates noise contributions from zodiacal and exozodiacal dust, telescope thermal emission, coronagraph speckles, dark current, read noise, and clock-induced charge. The model then provides the wavelength-dependent signal-to-noise (S/N) ratios from which the error bars are generated on the synthetic data. Previous applications of this model are detailed in \cite{robinson2016characterizing}, and the inputs to the noise model are described in Table \ref{tab:missions}. We note that the model can be used to simulate data points with randomized noise instances, but this step is unnecessary for this application as shown by \citet{feng2018characterizing}.

\begin{deluxetable*}{ccccccc}
\tablecaption{Summary of the mission concepts used as a template for comparing transmission and direct imaging in this report. We adapt instrument specifications from the \textit{Origins} and LUVOIR final reports, respectively. In the noise parameters column, $R$ stands for resolution and $t$ stands for throughput, with coronagraph throughput values calculated by \cite{kopparapu2021nitrogen} and optical throughput values calculated by \cite{stark2019exoearth} for direct imaging, and throughput values taken from the \textit{Origins} final report for transmission \citep{meixner2019origins}. \label{tab:missions}}
\tablewidth{0pt}
\tablehead{
\colhead{\makecell{Observing \\ Technique}} &
\colhead{\makecell{Mission/\\ Instrument \\ Template}} & 
\colhead{Wavelength [$\mu$m]} & \colhead{Noise Parameters} & \colhead{\makecell{Inscribed \\ Diameter [m]}} & \colhead{Temperature [K]}
}

\startdata
\\
Transmission & \makecell{\textit{Origins} \\ MISC-T} & 2.8~\textendash~20.0 & \makecell{$R = 100$, $t = 0.34$ (2.8 $< \lambda < 10.5$ $\mu$m) \\ $R = 200$, $t = 0.26$ (10.5 $< \lambda < 20.0$ $\mu$m)} & 5.9 & 4.5 \\
\\
Direct Imaging & \makecell{LUVOIR-A (B) \\ ECLIPS} & 0.2~\textendash~2.0 & \makecell{ $R = 7$, $t = 0.05 \times 0.27$ (0.2 $< \lambda < 0.38$ $\mu$m) \\ $R = 140$, $t = 0.2 \times 0.27$ (0.38 $< \lambda < 0.74$ $\mu$m) \\ $R = 100$, $t = 0.3 \times 0.27$ (0.74 $< \lambda < 2.0$ $\mu$m) } & 15 (6.7) & 150 \\
\\
\enddata

\end{deluxetable*}

\subsection{Retrieval Model}

To solve for the atmospheric parameters of the simulated exoplanets, we use the SMART Exoplanet Retrieval (SMARTER) model \citep[][Lustig-Yaeger et al. 2023, submitted]{lustig2020detection, lustig-yaegerHBAR2021}. This model uses SMART to generate synthetic spectra that are then fit to the observed spectrum to infer posterior probability distributions for a given set of atmospheric parameters, e.g., temperature, surface pressure, gas abundances, and albedo. As in most atmospheric retrievals, these posteriors for the atmospheric characteristics are generated by combining a forward model and an inverse model to solve a Bayesian inference problem. 

To solve for the atmospheric parameters that produced an observed or simulated spectrum, we implement Bayesian inference to solve for the posterior probability distributions for atmospheric parameters given the observed spectrum, or the data, $\mathcal{D}$. In this case, the posterior probability refers to the probability that a given atmospheric composition would produce the observed spectrum. The posterior distributions are given by

\begin{equation}
\label{eqn:bayesthm}
    \mathcal{P}(\theta | \mathcal{D}, \mathcal{M}) = \frac{\mathcal{P}(\theta | \mathcal{M})\mathcal{L}(\mathcal{D} | \theta, \mathcal{M})}{\mathcal{Z}(\mathcal{D} | \mathcal{M})}.
\end{equation}

\noindent where $\mathcal{P}(\theta | \mathcal{M})$ is the \textit{prior} probability of the model parameters, $\mathcal{L}(\mathcal{D} | \theta, \mathcal{M})$ is the \textit{likelihood} of the observed spectral data $\mathcal{D}$ being produced by a given model at a particular set of parameters $\mathcal{M}(\theta)$, and $\mathcal{Z}(\mathcal{D}, \mathcal{M})$ is the marginal likelihood or \textit{evidence} normalization factor. In this case, the prior refers to the probability of the model parameters at a particular atmospheric state given previous measurements or observations. Since we currently have no prior observations of terrestrial exoplanet atmospheres relevant to this study, we use uninformative priors for all parameters, where the probability of any value within a finite logarithmic range is given an equal weight. As is common in Bayesian inference problems, we implement the logarithm of Equation \ref{eqn:bayesthm}. For data with uncorrelated errors, we define the log-likelihood function in relation to the traditional $\chi^2$ goodness of fit metric such that $\log\mathcal{L} = -\frac{1}{2}\chi^2$, where this metric describes the goodness of fit that a given atmospheric state, and its corresponding spectrum given by the forward model, provide to the observed spectrum. The log-likelihood function is a sum over $m$ wavelengths in the spectrum such that,
\begin{equation}
\mathcal{L} = -\frac{1}{2} \sum_{l=1}^{m}{    \frac{[{\mathcal{M}({\theta_N})}_l - \mathcal{D}_l]^2}{\sigma_l^2}}.
\end{equation}
Evaluating the likelihood at an atmospheric state $\theta_N$ initiates a run of the forward model $\mathcal{M}$ at this state $\theta_N$, and allows Bayes theorem to be evaluated at this point. The marginal likelihood is defined as 

\begin{equation}
    \mathcal{Z}(\mathcal{D}, \mathcal{M}) = \int \mathcal{P}(\theta, \mathcal{M})\mathcal{L}(\mathcal{D} | \theta, \mathcal{M}) d\theta.
\end{equation}

\noindent This multi-dimensional integral accounts for the likelihood of all atmospheric parameters as well as the number of degrees of freedom allowed in the radiative transfer forward model, where degrees of freedom in this case equals the number of free parameters. While this multi-dimensional integral is generally difficult to calculate, techniques like nested sampling (see below) can more tractably estimate this term and provide a metric for comparing models $\mathcal{M}$ \citep{skilling2006nested} and determining which model best describes the observed spectrum. Next, we describe our nominal forward model as well as vertically-resolved variants that we compare against.

\subsubsection{SMART (Forward Model)}

SMART is the core of the retrieval forward models, $\mathcal{M}$. As described above, SMART calculates transmission, emission, and reflectance spectra for exoplanets using stellar and planetary input parameters combined with absorption coefficients for the given atmospheric gases. We use \texttt{coronagraph} to degrade this high-resolution spectrum to the resolution of the observing instrument. In general, the forward model takes a set of $N$ atmospheric parameters $\theta_N$ to generate a model spectrum, $\mathcal{M}(\theta_N)$, at the same resolution as the data. We now describe in detail the sequence of events that occur inside the forward models.

We make assumptions in LBLABC and SMART to reduce computation time in the retrieval model. 
We use the previously computed rotational-vibrational line absorption coefficients for each gas produced for the synthetic data, as described in Section \ref{sec:firstsmart}, within the retrieval. Although we do not regenerate new absorption coefficient files in successive iterations as the atmospheric state changes due to computational expense, this simplifying assumption does not produce errors in the vicinity of the solution. This necessary assumption is discussed further in Section \ref{sec:discussion}. Additionally, solar zenith angle and the spectral mapping binning criteria are kept fixed for all SMART runs in a given retrieval.

We enable the forward model to construct an atmosphere based on the state vector in one of three modes: evenly-mixed, vertically-resolved with \textbf{fixed} pressure points, and vertically-resolved with \textbf{free} pressure points. In the evenly-mixed mode, each gas profile is fixed throughout the vertical extent of the atmosphere at the abundance sampled from the prior. In the vertically-resolved mode with fixed pressure points, the user specifies a number of pressure points for a selected gas. Due to the additional computational expense of each fixed pressure point, which corresponds to an additional free parameter in gas abundance space, in this study we limit our exploration to 3 or 5 fixed pressure points per gas. Similarly, although multiple gases could be vertically resolved in a given retrieval, in this study we explore one gas to both reduce computational expense and isolate the effect on the fit. All other gases are assumed to have evenly-mixed profiles. Once the gas abundances have been drawn from the prior for each pressure point in the vertically-resolved profile, the forward model performs a linear interpolation  to construct a gas abundance profile. 

\begin{figure*}[htb]
    \includegraphics[scale=0.6]{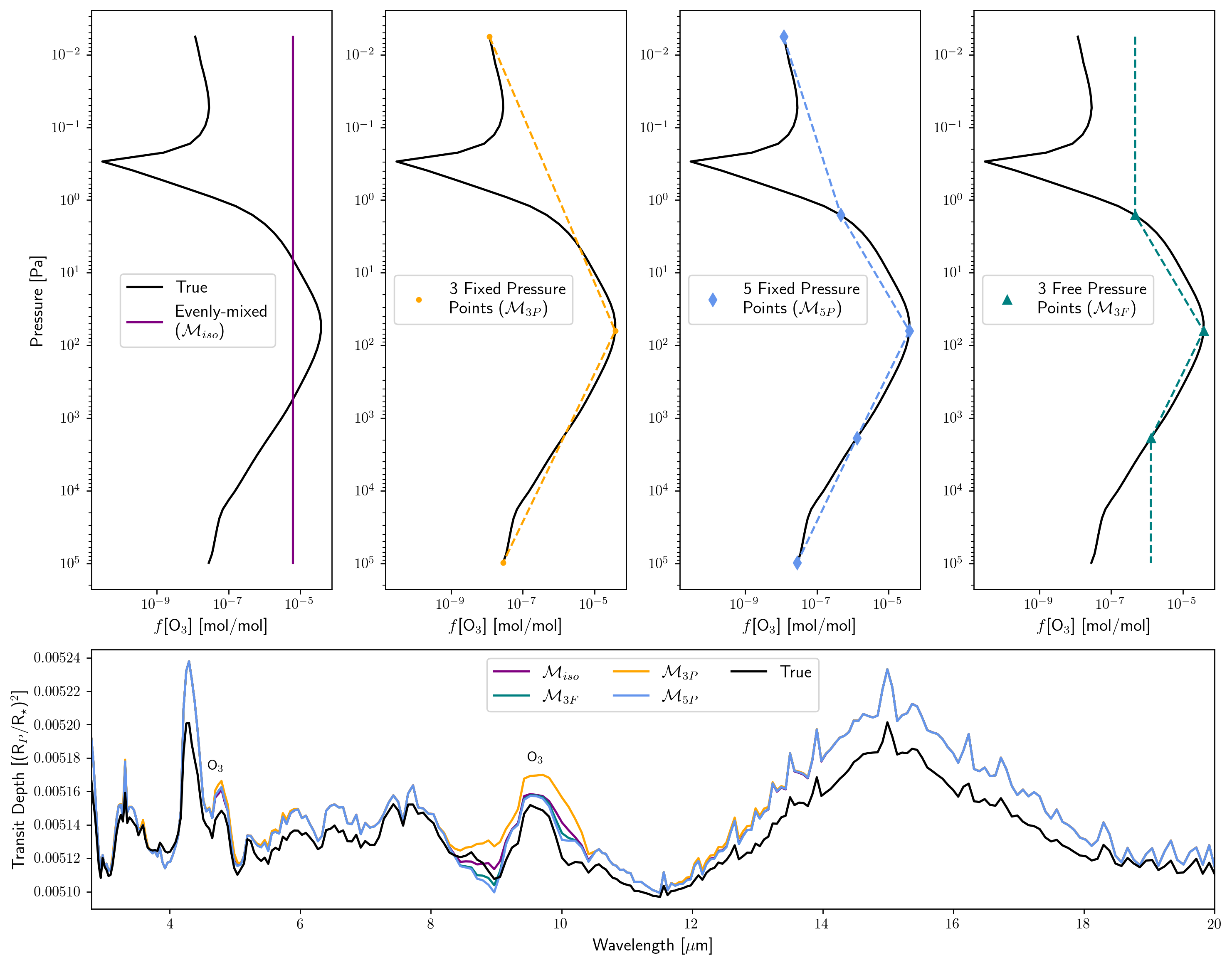}
    \caption{A comparison of the evenly-mixed ($\mathcal{M}_{iso}$), vertically-resolved with 3~\textendash~5 fixed pressure points ($\mathcal{M}_{3P}$, $\mathcal{M}_{5P}$), and vertically-resolved with 3 free pressure points ($\mathcal{M}_{3F}$) forward models for retrieving O$_3$ from a simulated transmission spectrum of a clear-sky, Earth-like TRAPPIST-1 e. The four top panels in the figure show an example of an O$_3$ profile that the forward model may try to produce to generate a fit to the spectral data. The bottom panel compares the corresponding spectra generated by each forward model to the true simulated spectrum. The assumption of isothermal pressure-temperature profiles in each of the forward models also modifies the continuum and the depth of other spectral features relative to the true spectrum.}
    \label{fig:fxcomp}
\end{figure*}

In the vertically-resolved mode with free pressure points, instead of specifying the pressure point values, the user includes three pressure points \textit{within} the state vector, along with three abundance points. This allows the model to freely select the pressures and abundances that provide the best fit to the observed spectrum, and then perform a linear interpolation to construct the vertical profile. We impose a simple prior that the pressures maintain order from smallest to largest to preserve the uniqueness of the points, thereby breaking a labelling degeneracy. As in the fixed points version of the model, all other gases besides the user-selected, vertically-resolved gas are given evenly-mixed profiles. We show a comparison of these four forward model modes with their corresponding spectra to examine the effect on O$_3$, a vertically inhomogeneous gas, in the spectrum of a clear-sky, Earth-like TRAPPIST-1 e in Figure \ref{fig:fxcomp}. We note that the example O$_3$ abundance profile fits generate visible differences in the depth of the O$_3$ spectral absorption features, with changes to the the 9.6 $\mu$m band being more pronounced, relative to the 4.7 $\mu$m O$_3$ band where the difference is much smaller.

For all three modeling modes, the full atmosphere is constructed by assuming an isothermal temperature-pressure (T-P) profile, which is computationally straightforward for the retrieval. Although very few terrestrial planets exhibit isothermal T-P profiles, especially near the surface, isothermal stratospheres are possible, especially for planets orbiting M dwarfs \citep{lincowski2018evolved}.  Since both transmission and reflectance spectra are only weakly sensitive to atmospheric thermal structure \cite{robinson2017characterizing, feng2018characterizing}, the isothermal assumption is common in retrieval modeling work. The assumption of an isothermal temperature structure inside the retrieval is also implemented by \citet{feng2018characterizing}. We discuss this assumption further in Section \ref{sec:discussion}. We also backfill the atmosphere with N$_2$ in all three modes to ensure the total volume mixing ratios of each atmospheric state add to unity so that,

\begin{equation}
\label{eq:n2fill}
 f[N_2] = 1 - \sum^{n_{gas}}_{j=1}{f_j},
\end{equation}

\noindent where $f_j$ is the volume mixing ratio of the $j$-th gas for a total of $n_{gas}$ user-specified gases. N$_2$ is a fill gas in Venus, Mars, and Earth, the terrestrial planets with significant atmospheres in our Solar System, and similar studies also backfill terrestrial exoplanet atmospheres with N$_2$ \citep[e.g.,][]{barstow2016habitable, krissansen2018detectability, changeat2019toward, barstow2020comparison}. However, other fill gases could be used if scientifically warranted, like Ar \citep{lustig2020detection}, or a mixture of H$_2$/He \citep[e.g.,][]{tsiaras2016detection, mollierepetitRADTRANS2019}. 

Once the volume mixing ratio of the filler gas has been calculated, the forward model calculates the mean molecular weight $\mu$ of the atmosphere using the relation
\begin{equation}
    \mu = \frac{\sum^{n_P}_k {P_k(\sum^{n_{gas}}_j m_j f_{j,k})_k}}{\sum^{n_P}_k P_k},
\end{equation}

\noindent where $P_k$ is the $k$-th of $n_P$ total pressure layers in the atmosphere, $m_j$ is the molecular mass of the $j$-th gas out of $n_{gas}$ total gases, and $f_{j,k}$ is the volume mixing ratio of the $j$-th gas at the $k$-th pressure layer. Accordingly, backfilling the atmosphere creates a bias for the filling gas that in turn propagates a bias in mean molecular weight. Since observable molecular spectral features are proportional to the scale height of the atmosphere---which is governed  by atmospheric mean molecular weight and temperature, as well as gravity (planetary mass)---then a bias in the atmospheric mean molecular weight may lead to the propagation of biases in retrievals of the other properties, and any results that depend on those properties.  In this sense, backfilling with N$_2$ can create a false constraint that tends not to bias the retrieval of any of the other gases in the transmission spectrum, but does bias other parameters like the surface temperature and also the bulk gas volume mixing ratio if that bulk gas is in fact not N$_2$. To avoid this bias caused by using filler gases, other retrieval models use centered-log ratio priors \citep{aitchison1982statistical} for the gas abundances \citep{benneke2012atmospheric, damiano2021reflected}. Instead of requiring that all volume mixing ratios sum to one, centered-log ratio priors require that the \textit{$\log$} of all volume mixing ratios sum to zero. Thus, instead of assuming a bulk gas and a flat prior for all trace gases, centered-log ratio priors allow any retrieval gas to be the bulk gas as well without favoring or disfavoring any species. The implementation of centered-log ratio priors in SMARTER is outside the scope of this study, but will be explored in future work.

Once we have constructed an atmosphere, we pass this atmosphere to SMART to generate either a transmission or reflectance spectrum. In transmitted light, SMART calculates the change in flux $\Delta {F}$ during transit as the fractional transit depth,
\begin{equation}
    \Delta{F} = \left ( \frac{R_P}{R_\star} \right )^2,
\end{equation}

 \noindent where $R_p$ is the planet radius and $R_\star$ is the stellar radius. For these transmission spectrum calculations, we use the SMART ray-tracing model described in \cite{robinson2017theory} with refraction turned on. To calculate the transmission spectrum, SMART may either include the solid body or assume a solid body and compute the differential due to the atmosphere, where this differential constitutes the opacities along a line of sight as well as the refraction, which is a function of the viewing geometry of the star-planet system. Consequently, SMART can calculate a number of different transit products, including the total transit depth (atmosphere and solid body), the relative transit depth (atmosphere \textit{above} the solid body), as well as the effective transit height relative to the surface in kilometers. In the transmission retrieval mode, the user may therefore include the solid-body planet radius $R_P$ and calculate the total transit depth of the atmosphere in the transmission forward model, as we do here in this study.  In the absence of transit Jacobians, we use the effective transit height to indicate where in the atmosphere the true spectrum is sensitive to a given gas. The retrieval user may also include continuum pressure $P_0$, and temperature $T$ as free parameters. Using the assumption of an isothermal temperature-pressure profile, the retrieved temperature may also effectively refer to the surface temperature, $T_0$. 
 
 In reflected light, SMART gives the unit-less planet-to-star flux ratio as,
 
 \begin{equation}
     \frac{F_P}{F_\star} = \bigg(\frac{R_P}{a}\bigg)^2 \frac{F_{P, TOA}}{F_{\star, TOA}},
 \end{equation}

\noindent where $a$ is the semi-major axis of the planet's orbit, $F_{P, TOA}$ is the top of atmosphere planet flux, and $F_{\star, TOA}$ is the top of atmosphere stellar flux. In the reflectance retrieval mode, the user may include surface albedo $A_s$ and surface temperature $T_0$ as free parameters. SMARTER produces the reflected light spectrum assuming a uniform or gray surface albedo, where we take the average reflectivity of the Earth as $\bar{A_{e}} = 0.2$ to generate the synthetic spectral data.

Once we have generated the high-resolution spectrum, we use \texttt{coronagraph} to produce the degraded spectrum, $\mathcal{M}(\theta_N)$. \texttt{coronagraph} convolves the high resolution spectrum with a boxcar filter to down-bin the spectrum to the resolution of the data. With more realistic instruments, this down-binning process can be more complicated if there are known instrument characteristics that need to be accounted for, such as non-linearities in pixel resolution. Since we are simulating hypothetical instruments that have not yet been designed, built, or characterized, we implement this simplification in place of a more defined instrument response function. 

Once we have generated the degraded spectrum $\mathcal{M}(\theta_N)$ from the atmospheric state $\theta_N$, we pass the spectral inputs to the likelihood function, $\mathcal{L}$. This in turn triggers the logic of the inverse model. The inverse model maximizes the posterior probability and minimizes the residuals in the fit to the observed spectrum. After assessing the fit to the data $\mathcal{D}$, the inverse model selects the next set of parameters randomly drawn from the priors. Next, we describe the inverse model component of the retrieval framework. 
 
\subsubsection{\texttt{PyMultiNest} (Inverse Model)}

For this work, we employ a nested sampling inverse model to numerically solve Bayes theorem called \texttt{PyMultiNest} \citep{buchner2014x}. \texttt{PyMultiNest} is a Python wrapper for \texttt{MultiNest}, an open-source Fortran program \citep{feroz2008multimodal, feroz2009multinest, feroz2013importance}. Introduced by \cite{skilling2004nested}, nested sampling is commonly used in atmospheric retrievals as a computationally efficient method for solving the Bayesian inference problem \citep{benneke2013distinguish, waldmann2015tau, lavie2017helios, gandhi2018retrieval}. Nested sampling algorithms maintain a user-specified set of live points, each containing a sample of the parameters $\theta$ from the prior distributions $\mathcal{P}(\theta)$. At each successive iteration, nested sampling implements a cost function by removing the live point from the set with the lowest likelihood, and replaces it with a new live point with parameter samples randomly drawn from the prior distributions such that the new live point samples have a higher likelihood than the last. Also at each iteration, \texttt{PyMultiNest} calculates the log evidence, and then compares the change in evidence between successive iterations using the live points. Once the change in evidence is equal to or smaller than a tolerance value, the process is complete. The evidence tolerance is thus the model convergence criterion, where we implement \texttt{PyMultiNest} with 1000 live points and an evidence tolerance of $\le$0.5.

\subsection{Model Comparison}

To compare our models and assess their accuracy, we analyze the model evidence. Nested sampling has the added benefit of concurrently calculating the evidence term by reducing the multidimensional integral $\mathcal{Z}(\mathcal{D})$ described above. 

The evidence metric rewards the model with the best fit to the data while penalizing excess parameter dimensions, so the model with the higher evidence is typically taken to be the preferred model given fits to the same data. Thus, to compare two models $\mathcal{M}_1$ and $\mathcal{M}_{2}$, we merely calculate the difference in their log-evidence terms so that,

\begin{equation}
{B_{1,{2}}} = \ln{Z_1} - \ln{Z_{2}},
\end{equation}

\noindent where $B$ is known as the Bayes factor. In this formulation, a Bayes factor with $B_{1,{2}} > 0$ indicates that $\mathcal{M}_1$ is the favored model, whereas a Bayes factor with $B_{1,{2}} < 0$ indicates that $\mathcal{M}_{2}$ is preferred instead. In this study, we interpret the significance of our Bayes factors based on the empirically-derived odds ratios proposed by  \cite{jeffreys1998theory}. 

To assess the quality of the fits provided to the spectra by our models, we compare the $\chi^2$ for the median spectral solutions produced by the retrieval. To calculate $\chi^2$, we use the fact that $\mathcal{L} = -\frac{1}{2}\chi^2$. The Bayes factor analysis allows us to determine whether additional model complexity is justified by the data, while separately analyzing the $\chi^2$ values allows us to determine which model provided the best fit to the data regardless of model complexity. In combination, these analyses help determine whether improved data resolution or precision may increase the evidence in favor of the more complex model.

\section{Experiments}
\label{sec:experiments}
\begin{figure*}[htb]
    \centering
    \includegraphics[scale=0.65]{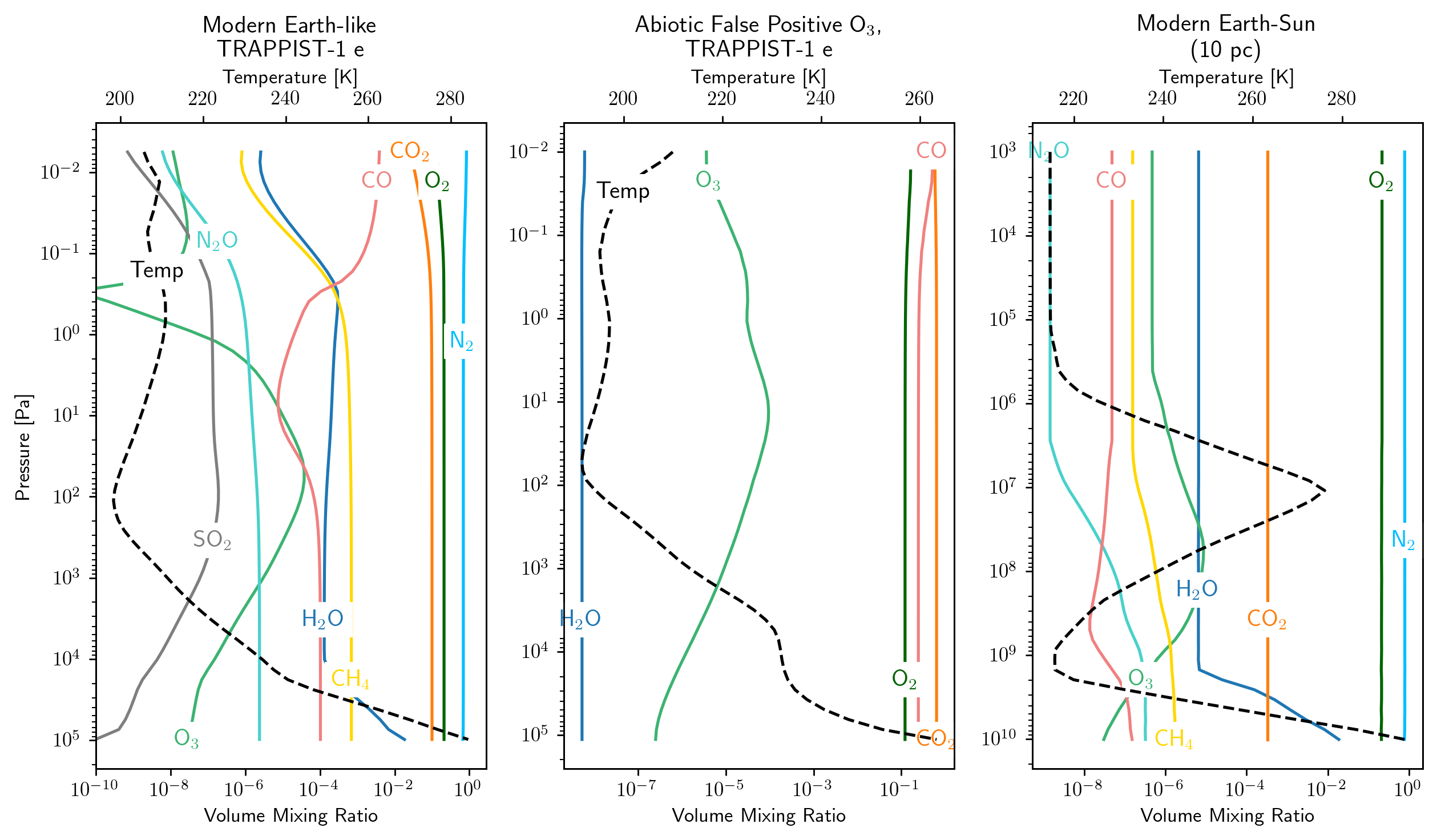}
    \caption{A comparison of the vertically-resolved temperature structure and composition of each atmospheric case addressed in this study. The Earth-like and abiotic, ozone false-positive planets (first and second panels) have similar ozone inventories with visible stratospheric bulges.}
    \label{fig:atmcomp}
\end{figure*}

In this study, we apply both our evenly-mixed and vertically-resolved retrieval models to a set of atmospheric cases in transmitted and reflected light, which we describe in detail here. In transmission, we simulate an Earth-like, inhabited TRAPPIST-1 e and an abiotic, uninhabited TRAPPIST-1 e. Our uninhabited TRAPPIST-1 e represents a possible O$_3$ false-positive case, with a 1-bar, false positive O$_3$, desiccated atmosphere based on the most extreme case described in \cite{gao2015stability}. In reflected light, we simulate an inhabited, clear-sky Pre-industrial Earth at 10 parsecs and the same atmosphere with thermally self-consistent H$_2$O clouds. We describe these cases in detail below, and summarize them in Table \ref{tab:cases} along with our chosen exposure times for both methods. We compare two distinct telescope architectures using two different observation methods that will be optimized for different wavelength ranges and planetary systems. To make the investigation as fair as possible given these intrinsic differences, we run our retrieval model for similar \textit{Origins} and LUVOIR-B exposure times (t = 40 hours). For the larger aperture LUVOIR-A, less exposure time is required to achieve the same quality of data. Choosing similar \textit{Origins} and LUVOIR-B exposure times allows us to draw conclusions about the comparative value of observing optimal transmission and direct imaging targets. Overheads due to, for example, slew, settle, and out-of-transit baseline, are neglected from these exposure calculations. We show a comparison of the structure and composition of each atmospheric case addressed in Figure \ref{fig:atmcomp}. We note that in the case of the clear-sky, false positive O$_3$, H$_2$O-poor TRAPPIST-1 e, the abiotically-generated O$_3$ abundances are comparable to that of the Earth-like TRAPPIST-1 e with an oxygenic biosphere. Both profiles show a stratospheric O$_3$ bulge, though the desiccated planet has a much shallower bulge.

\begin{deluxetable*}{cccccc}
\tablecaption{A summary of the observing cases explored in this work. While \textit{Origins} is also capable of observing some exoplanets in emission and LUVOIR is also capable of transmission observations for a limited number of targets, we focus on each of the observatory's main observing modes to provide instrument templates for comparing transmission and direct imaging. \label{tab:cases}}
\tablewidth{0pt}
\tablehead{
\colhead{Mission} & \colhead{Observing Method} & \colhead{Planet/Star} & \colhead{Biosphere} & \colhead{Comparison Case}
}

\startdata
\\
\textit{Origins} & \makecell{Transmission \\ $t_{exp} = 40$ hours} & TRAPPIST-1 e & \makecell{Clear-Sky Modern Earth-\textbf{like} \\ 1-bar, N$_2$ dominated} & \makecell{Abiotic 1-bar Mars-like \\ CO$_2$ dominated, false positive O$_3$, \\ desiccated  \citep{gao2015stability}} \\
\\
LUVOIR-A (B\tablenotemark{c}) & \makecell{Direct Imaging \\ $t_{exp} = 10$ ($\sim$40\tablenotemark{c}) hours} & Earth-Sun (10 pc) & \makecell{Clear-Sky Modern Earth \\ 1-bar, N$_2$ dominated} & \makecell{Modern Earth with H$_2$O Clouds \\ 1-bar, N$_2$ dominated} \\
\\
\enddata
\tablenotetext{c}{Separate noise models and retrievals are not run for LUVOIR-B, and differences between architecture mirror diameter and throughputs complicate scaling between A and B exposure times.}
\end{deluxetable*}

\subsection{Transmission}

For the transmission experiments, we use spectra generated for an Earth-like TRAPPIST-1 e with a photosynthetic biosphere and an abiotic, desiccated TRAPPIST-1 e with false positive O$_3$ as the synthetic data inputted to the retrieval. Our Earth-like TRAPPIST-1 e transmission spectrum is generated based on temperature and mixing ratio profiles self-consistently calculated for a habitable zone planet around an M8V star (Davis et al., \textit{in prep}). The abiotic case represents a possible O$_3$ false-positive generated by CO$_2$ photolysis in a dry, 1-bar atmosphere \citep{gao2015stability}. In this false positive scenario, the extremely low abundance of H$_2$O prevents CO$_2$ from recombining after photodissociation, thereby allowing enough O$_3$ to accumulate to mimic the O$_3$ levels we would expect on an inhabited planet. This is an important false positive mechanism, as an \textit{Origins}-like mission would observe transiting planets in the mid-IR where O$_3$ must be used as a proxy for O$_2$. 

\begin{figure*}[htb]
    \centering
    \includegraphics[width=0.9\textwidth]{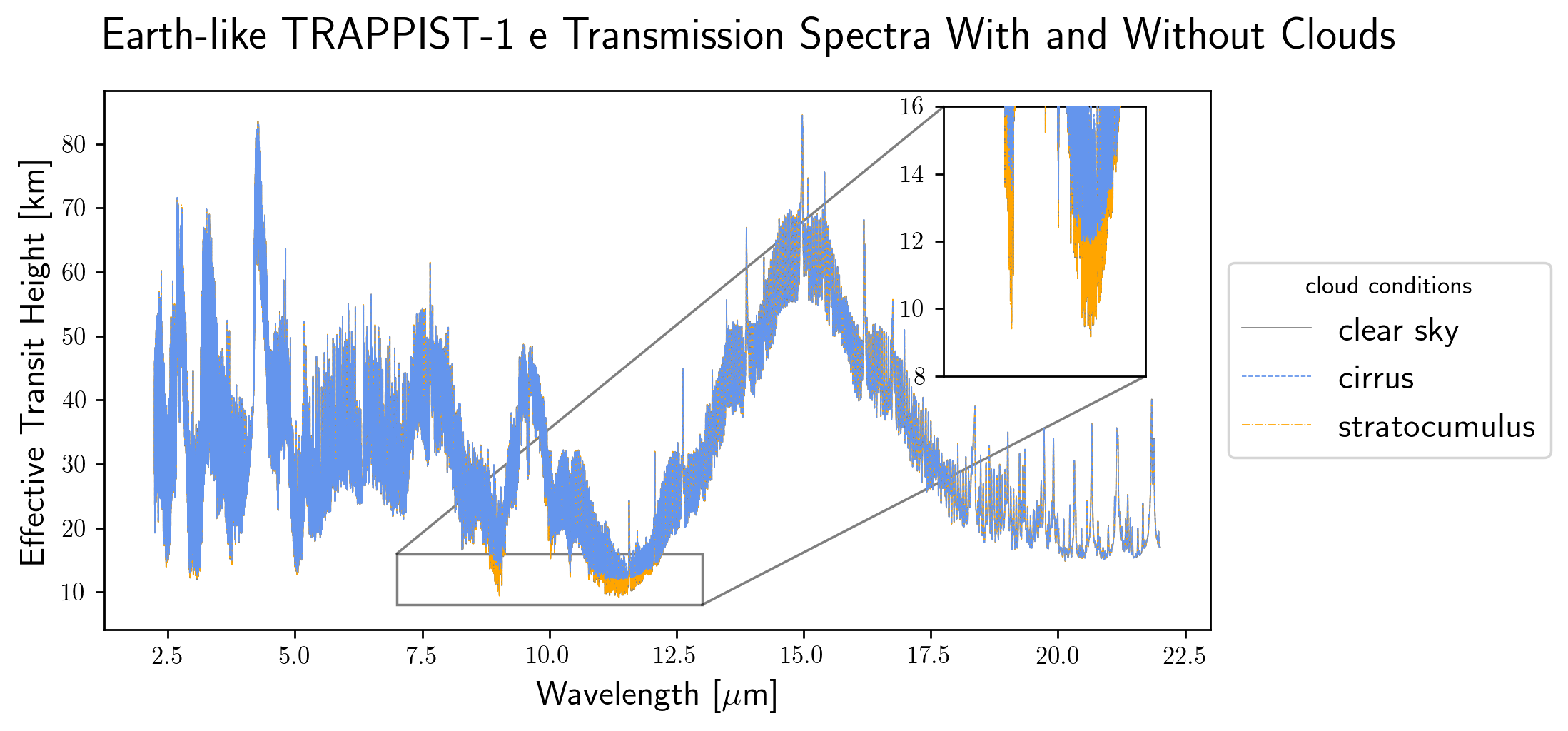}
    \caption{A comparison of simulated transmission spectra for TRAPPIST-1 e modeled with a self-consistent, Earth-like atmosphere. When self-consistent ice and water clouds are included, no significant spectroscopic changes occur to the spectrum. We see small impacts due to clouds in the relatively transparent portion of the atmosphere between 8~\textendash~13 $\mu$m, where the clouds reduce the minimum altitude probed from 10 to 12 km.}
    \label{fig:earth_t1e_clouds}
\end{figure*}

The Earth-like TRAPPIST-1 e atmosphere is also modeled with thermally self-consistent water (stratocumulus) and ice (cirrus) clouds. The vertical cloud distributions and ice cloud optical properties are self-consistently calculated with a single scattering model for a variety of particle types as described in \citet{Meadows2023}. However, when we simulate the spectroscopic effects of these clouds on the planet's transmission spectrum, we see no significant changes to the spectrum compared to the clear-sky case, except for minimal changes in transit height probed by the 8~\textendash~13 $\mu$m window region, as shown in Figure \ref{fig:earth_t1e_clouds}, which effectively raises the minimum altitude probed from 10km to 12km. Therefore, we conclude that Earth-like water and ice clouds are unlikely to have a significant impact on our transmission  retrieval experiments and we omit them here. These results are consistent with \citet{Meadows2023}, who show that clouds do not significantly impact the transit depths of spectroscopic features for an Earth-like TRAPPIST-1 e for wavelengths greater than 2.8 $\mu$m. As a caveat, we note that 3-D, General Circulation Models (GCMs) include more sophisticated dynamics than 1-D models, and can predict different cloud altitudes as well as a distinct day-side cloud accumulation pattern for tidally locked planets like those in the TRAPPIST-1 system \citep{komacek2020clouds}, which may or may not be synchronously rotating. However, the 1-D results used here predict clouds that are at altitudes comparable to, if not slightly higher than those predicted for TRAPPIST-1 e by 3-D models \citep{fauchez2019impact, Meadows2023}. Though 3-D models are currently computationally expensive and thus difficult to implement within a spectroscopic retrieval framework, current work is exploring how to make the implementation of 3-D models in atmospheric retrievals more tractable by speeding up radiative transfer calculations \citep{macdonald2021trident}.

For the 1-bar false positive O$_3$ planet, we do not consider or model aerosols, because we concluded that neither clouds, hazes, or dust would likely affect the observations. Since the abundance of H$_2$O is well below the saturation vapor pressure required for water to condense, we conclude that water clouds may not form. Additionally, since this is a methane-free, highly-oxidizing Mars-like atmosphere, we also rule out the presence of organic hazes since \citet{arney2016pale} showed that a CH$_4$/CO$_2$ ratio of $\geq$0.2 is required for haze to form (and become spectrally significant) in an Earth-like atmosphere. However, given a desiccated Mars-like atmospheric composition, it is reasonable that global dust storms could occur on such a planet, and that these dust particles may impact the opacity of the atmosphere. Global dust storms on Mars are reasonably well-studied, with observations reporting ``rocket storms'' that may loft dust as high as 70 km \citep{wang2018parameterization}. However, the Martian atmosphere is considerably thinner than the 1-bar atmosphere simulated here, so dust storms on Earth represent a better analogue to our Earth-like planet. On Earth, the atmospheric transport of Saharan dust is a well-studied phenomenon. Thus, we can look to these studies as an analogue for dust transport within a 1-bar atmosphere to understand how dust particles may impact the transmission spectra of the 1-bar, H$_2$O-desiccated planet in our study. Lidar observations of the Earth taken from space over a 5 year period have shown that the peak vertical distribution  of the dust does not exceed 5 km \citep{tsamalis2013seasonal}. We therefore find that dust particles in this atmosphere are unlikely to impact observations in transmission. 

In transmitted light, we attempt to retrieve the planet radius, continuum pressure, surface temperature, and gas abundances. A summary of the parameters and their associated priors are described in Table \ref{tab:param_trnst}.

\begin{deluxetable*}{ccccc}
\tablecaption{A summary of the parameters and priors explored in the two transmission cases of this study (TRAPPIST-1 e Earth-like and false positive O$_3$ planet). In the non-vertically-resolved retrievals, all gas profiles from the forward model are generated assuming evenly-mixed abundances. $R_p$ is planet radius, $P_0$ is continuum pressure, and $T_0$ is surface temperature. Finally, $f$ indicates the gas mixing ratio of a given molecule (e.g., H$_2$O, CH$_4$, CO$_2$, O$_2$, O$_3$, CO).
\label{tab:param_trnst}}
\tablewidth{0pt}
\tablehead{
\colhead{\makecell{Parameter \\ $\theta$}} & \colhead{} &  \colhead{\makecell{Prior \\ $\mathcal{P}(\theta)$}} & \colhead{Lower Bound} & \colhead{Upper Bound}
}

\startdata
\\
Planetary radius & $R_p$ & Uniform & 0.85$R_\oplus$ & 0.95$R_\oplus$ \\
\\
Continuum pressure & P$_0$ & Uniform in log-space & $10^2$ Pa & $10^6$ Pa  \\
\\
Surface temperature & $T_0$ & Uniform & 100 K & 400 K  \\
\\
Gas abundance & $f$[$x$] & Uniform in log-space & $10^{-10}$ mol/mol & 1 mol/mol \\
\\
$^{a,b}$Vertically-resolved O$_3$ abundance & $f$[O$_3$]$_{n}$ ($n$ = 3, 5)& Uniform in log-space & $10^{-10}$ mol/mol & 1 mol/mol \\
\\
$^{b}$Vertically-resolved O$_3$ pressure & $P$(O$_3$)$_{n}$ ($n$ = 3, 5) & Uniform in log-space & $10^{-2}$ Pa & $10^{5}$ Pa \\
\\
\enddata
\footnotesize{$^a$Parameters considered in the fixed points models ($\mathcal{M}_{3P}$, $\mathcal{M}_{5P}$) and $^b$those considered in the free points model ($\mathcal{M}_{3F}$).} \\ 
\end{deluxetable*}

\subsection{Direct Imaging}
For the two direct imaging experiments, we use spectra generated for an Earth-like environment with and without clouds as the synthetic data input to the retrieval. Our Earth spectrum is generated based on temperature and mixing ratio profiles from case 62 of the Intercomparison of Radiation Codes in Climate Models (ICRCCM), representing the averaged midlatitude Earth during the summer months, as shown in \citet{lincowski2018evolved}. Comparison of spectra generated using ICRCCM and full-disk-averaged spectra of the Northern hemisphere in the spring \citep{robinson2011earth} showed  only minor differences in the depths of water vapor bands, which are known to be highly spatially and temporally variable on the Earth. Unlike in the M dwarf transmission cases, the known potential false positives for abundant O$_2$ for planets orbiting G dwarfs are difficult to generate and difficult to distinguish from true biospheres \citep{wordsworth2014abiotic, meadows2017reflections}. In particular, G-dwarf planets with low abundances of non-condensable gases (such as N$_2$) may produce inefficient cold traps that allow water into higher levels of the atmosphere, where it is more readily photolyzed to produce O$_2$ \citep{wordsworth2014abiotic}. Therefore, for our comparison case we instead consider the impact of realistic clouds on our ability to retrieve planetary properties in direct imaging.  For noise calculations, we assume the observed planets are orbiting a G dwarf, at a distance of 10 pc.  

  While cloud decks have been shown to effectively truncate the atmospheric scale height and thus reduce the size of spectral features in transmission \citep{lincowski2018evolved, fauchez2019impact, lustig2019coronagraph, komacek2020clouds}, the behavior of atmospheric clouds in reflected light is more complex and may actually enhance spectral features in some cases \citep[e.g.,][]{rugheimer2013spectral}. For our synthetic spectral observations, the vertical cloud distributions and ice cloud optical properties are again self-consistently calculated with a single scattering model for a variety of particle types as described in \citet{Meadows2023}. We simulate Earth with realistic patchy clouds by linearly combining the water (25\%), ice (25\%), and clear-sky (50\%) spectra to produce the cloudy spectrum. Finally, we assume a gray (constant with wavelength) surface albedo of 0.2 as in \cite{feng2018characterizing} and \citet{damiano2022reflected}. Wavelength dependent changes in surface albedo expected from a realistic planetary surface, which likely contribute significantly to an observed spectrum, are beyond the scope of this study and will be the subject of future work. A summary of the parameters and their associated priors are described in Table \ref{tab:param_refl}.

\begin{deluxetable*}{ccccc}
\tablecaption{A summary of the parameters and priors explored in the direct imaging (clear and cloudy Earth) cases of this study. In the non-vertically-resolved retrievals, all gases profiles from the forward model are generated assuming evenly-mixed abundances. $A_s$ is a uniform surface albedo of 0.2 reported as a unitless ratio and $T_0$ is surface temperature. Finally, $f$ indicates the gas mixing ratio of a given molecule (e.g., H$_2$O, CH$_4$, CO$_2$, O$_2$, O$_3$, CO).  \label{tab:param_refl}}
\tablewidth{0pt}
\tablehead{
\colhead{\makecell{Parameter \\ $\theta$}} & \colhead{} &  \colhead{\makecell{Prior \\ $\mathcal{P}(\theta)$}} & \colhead{Lower Bound} & \colhead{Upper Bound}
}

\startdata
\\
Surface albedo & $A_s$ & Uniform in log-space & 10$^{-2}$ & 1 \\
\\
Surface temperature & $T_0$ & Uniform & 100 K & 400 K  \\
\\
Gas abundance & $f[x]$ & Uniform in log-space & $10^{-12}$ mol/mol & 1 mol/mol \\
\\
$^{a,b}$Vertically-resolved O$_3$ abundance & $f$[O$_3$]$_{n}$ ($n$ = 3, 5)& Uniform in log-space & $10^{-12}$ mol/mol & 1 mol/mol \\
\\
$^b$Vertically-resolved O$_3$ pressure & $P$(O$_3$)$_{n}$ ($n$ = 3, 5) & Uniform in log-space & $10^{-2}$ Pa & $10^{5}$ Pa \\
\\
\enddata
\footnotesize{$^a$Parameters considered in the fixed points models ($\mathcal{M}_{3P}$, $\mathcal{M}_{5P}$) and $^b$those considered in the free points model ($\mathcal{M}_{3F}$).} \\ 
\end{deluxetable*} 

\section{Results}
\label{sec:results}
The results of our study consist of two main components: posterior distributions generated by the retrieval, which provide the distribution of atmospheric properties that best match the spectrum, and the fits to the spectrum generated by running SMART for atmospheric states sampled from these posteriors. We compare the posterior distributions to the true profiles to assess the accuracy of our retrieval model and our ability to constrain parameters from the data. We supply \textit{all} posterior distributions with covariances in Appendix \ref{sec:ap:corners}. In the main text of this paper, we show the marginal posterior distributions for the evenly-mixed forward model cases. All non-O$_3$ posteriors for the vertically-resolved cases are within 1-$\sigma$ of their evenly-mixed counterparts, so we omit them in the main text. We show the posterior distributions for the significant vertically-resolved O$_3$ profiles in the main text.  The calculation of the posteriors includes the calculation of the evidence, $\log\mathcal{Z}$. We use the evidence term as a metric for determining the relative likelihood of a given model, and we compare model evidences via Bayes factors to perform model comparisons. Finally, we use the $\chi^2$ values for the median spectral fit to assess the goodness-of-fit of each model. We first individually compare our pairs of transmission and direct imaging cases, and then compare the results of our transmission cases to the results of our direct imaging cases.

\subsection{Transmission}
\label{sec:res_trnst}
We first compare our retrieval cases in transmitted light. We compare the posterior distributions from the evenly-mixed cases for the Earth-like and abiotic false positive O$_3$  TRAPPIST-1 e in Figure \ref{fig:t1e_hist} below. To indicate the atmospheric layers to which the true, noiseless spectrum is most sensitive, we rerun the input spectrum without a given gas and determine the wavelengths over which the gas is spectrally active. Then, we calculate the size of these absorption features in the true spectrum in terms of the effective transit height, and convert this to the corresponding range of pressures in the atmosphere that are displayed in Figure \ref{fig:t1e_hist}.
 
 \begin{figure*}[htb]
    \centering
    \setcounter{subfigure}{0}
    \subfigure{\includegraphics[scale=0.4]{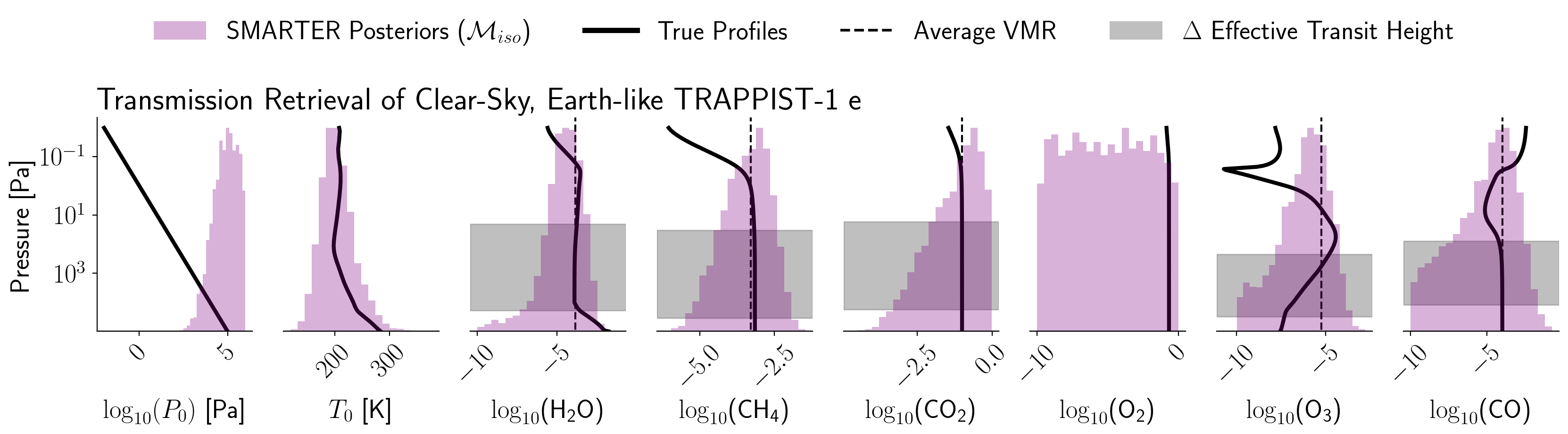}} \\
    \subfigure{\includegraphics[scale=0.4]{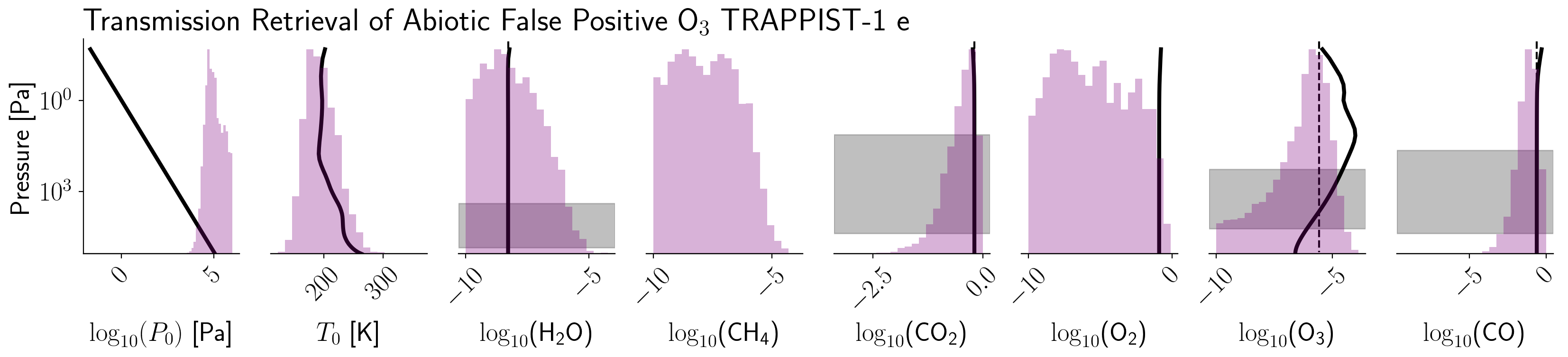}} \\
    \caption{A comparison of the posterior distributions for the Earth-like and abiotic false positive O$_3$ TRAPPIST-1 e using the evenly-mixed forward model ($\mathcal{M}_{iso}$). The black lines represent the true profiles used to generated the observed spectrum, and the dotted black lines indicate the corresponding average volume mixing ratios (VMR) of the true gas abundance profiles. We calculate the change in effective transit height for the spectral features of each gas in the noiseless input spectrum, to indicate the atmospheric layers to which the spectrum is sensitive (gray shading). For both atmospheres, there are no O$_2$ absorption features in the 2.8~\textendash~20.0 $\mu$m mid-IR wavelength range we considered, so we are unable to perform this analysis. For the abiotic planet, there is no CH$_4$ in the atmosphere, so we do not have true values to compare to.}
    \label{fig:t1e_hist}
\end{figure*}
 
 For the Earth-like TRAPPIST-1 e case, we first assess the retrieval of the surface pressure and the (nominally) evenly-mixed gases in the true atmosphere, which include CH$_4$, CO$_2$, O$_2$, and CO. We retrieve a median continuum pressure of $8 \times 10^{4}$ Pa, which is lower by a factor of 80\% from the true surface pressure ($\pm 1$ dex). This median retrieved pressure is consistent with the pressure found at an altitude of $\sim$2 km in the true atmosphere. However, the broad 1-$\sigma$ distribution of pressures we retrieve span $10^4$~\textendash~$4.1\times 10^5$ Pa, or 0.1~\textendash~4.1 bars. The lower bound of this range corresponds to an altitude of 23 km in the true atmosphere, while the 4.1-bar upper bound exceeds the true surface pressure of 1 bar and therefore has no altitude analog in the true atmosphere. For CH$_4$ and CO$_2$, we retrieve median abundances that are within a factor of 2 of the true mixing ratios. We note that at 1-$\sigma$, the retrieved CH$_4$ abundance may be as low as 43 ppm, but at 2- and 3-$\sigma$ the abundance may be as low as 8 ppm and 2 ppm, respectively. The retrieved O$_2$ volume mixing ratio is not constrained at all, with a posterior that provides no more information than the prior. Lastly, the CO volume mixing ratio is not precisely constrained, with a distribution peak that aligns with the evenly-mixed portion of the CO profile to which the spectrum is sensitive, but a 1-$\sigma$ interval spanning $5$ orders of magnitude in volume mixing ratio. 
 
 Assessing the retrieval of parameters that change with pressure in the true atmosphere is more complex, and requires us to consider a number of reference points along the true profile to which the retrieval may be sensitive. We note that the retrieved isothermal temperature distribution is consistent with the temperature in the upper atmosphere $\sim$50~\textendash~70 km ($\sim$10$^2$ Pa), but biased low by 30\% relative to the true surface temperature, $T_0$. This suggests that the transmission spectrum is not probing the surface environment despite the lack of clouds in this atmosphere. Similarly, the retrieved H$_2$O volume mixing ratio is underestimated by 80\% relative to the true surface water vapor abundance, but the change in effective transit height for the H$_2$O features imply that the true spectrum should be sensitive primarily to the lower stratospheric abundance as shown in Figure \ref{fig:t1e_hist}, which is evenly mixed. However, even if we adjust our comparison to account for this spectral sensitivity, the retrieved H$_2$O volume mixing ratio distribution is within a factor of 5 lower than this isothermal, evenly-mixed region of the true water profile. Finally, O$_3$ is not precisely constrained with the evenly-mixed model, with a broad posterior distribution that spans the entire width of the prior. Furthermore, the retrieved O$_3$ abundance is underestimated relative to the bulge abundance, but the upper bound of the distribution coincides with this maximum O$_3$ abundance. The retrieved O$_3$ abundance is consistent with both the true abundance at $\sim$$10^3$ Pa, which is within the atmospheric column probed by the spectrum, and the average volume mixing ratio of the true ozone profile within 1-$\sigma$ (albeit underestimated within a factor of $7$). 

For the abiotic false positive O$_3$ TRAPPIST-1 e, the evenly-mixed gases now include H$_2$O, as the planet is sufficiently desiccated so as to inhibit condensation. We retrieve a median continuum pressure of $1.0 \times 10^{5}$ Pa, which is within a factor of $0.9$ of the true surface pressure ($\pm 1$ dex). This median retrieved pressure is consistent with the pressure found at an altitude of $\sim$1 km in the true atmosphere. However, similar to the Earth-like TRAPPIST-1 e, we find that the 1-$\sigma$ pressure distribution we retrieve spans a broad range of pressures from $3 \times 10^4$~\textendash~$4 \times 10^5$ Pa (0.3~\textendash~4 bars), where the lower bound corresponds to an altitude of 11 km in the true atmosphere, and the upper bound does not correspond to any altitude in the true atmosphere. We retrieve an upper limit on the H$_2$O volume mixing ratio, which is within a factor of $33$ of the true volume mixing ratio, and the spectrum appears to be sensitive to the optically thin H$_2$O features closer to the surface than in the case of the Earth-like atmosphere. At 1-$\sigma$, the upper limit on the water vapor abundance is 167 ppb, while at 2-$\sigma$ it is 5 ppm and at 3-$\sigma$ it is 13 ppm. Similarly, we retrieve a 1-$\sigma$ upper limit of ${\sim}70$ ppb on CH$_4$, a very low abundance that is consistent with the absence of CH$_4$ in this atmosphere. However, the 2- and 3-$\sigma$ upper limits are much higher at 5 ppm and 27 ppm, respectively. For CO$_2$, we retrieve a median volume mixing ratio of $40$\%, which is underestimated by a factor of $1.5$ relative to the truth for this CO$_2$-dominated atmosphere. Similar to the Earth-like case, the O$_2$ volume mixing ratio is not constrained. Lastly, the CO volume mixing ratio is underestimated by a factor of $4$.

For this desiccated planet, the pressure-dependent parameters include temperature and O$_3$, which shows a stratospheric bulge akin to the Earth-like atmosphere. The retrieved isothermal temperature distribution is biased low by 3\% from the coolest point ($\sim$200 K) in the true temperature profile. Finally, the broader peak of the retrieved O$_3$ distribution for the evenly-mixed forward model ($\mathcal{M}_{iso}$) more closely aligns with the average volume mixing ratio than in the case of the Earth-like TRAPPIST-1 e, but it is still slightly underestimated. As in the case of the Earth-like atmosphere, the retrieved O$_3$ abundance is consistent with the true abundance in one of the atmospheric layers probed by the spectrum $\sim$$10^4$ Pa, but the retrieved distribution is very broad and spans the entire width of the prior. We summarize and compare all retrieved values from both of our transmission cases to their corresponding truths in Table \ref{tab:results_summ_trnst_lin}.

\begin{figure*}[htb]
  \centering
  \includegraphics[scale=0.9]{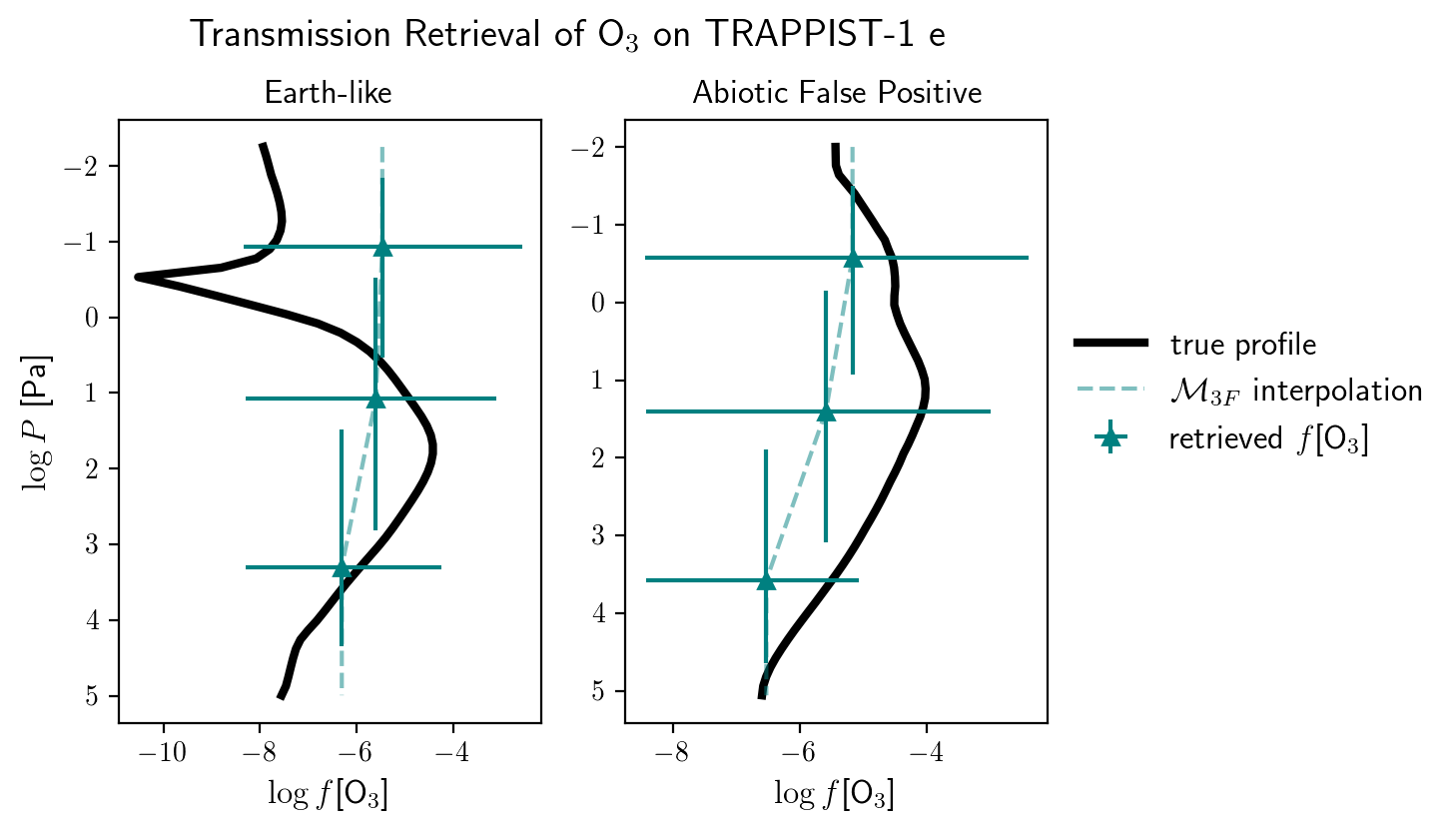}
  \caption{A comparison of the vertically-resolved O$_3$ profiles retrieved by the $\mathcal{M}_{3F}$ forward model for the Earth-like and abiotic TRAPPIST-1 e cases. We note that, for both atmospheres, the retrieval effectively reproduced an evenly-mixed O$_3$ profile despite being allowed to move freely in pressure and abundance space for 3 points along the profile.} 
  \label{fig:t1e_o3_vert_comp}
\end{figure*}

\begin{figure*}[htb]
  \centering
  \includegraphics[scale=0.9]{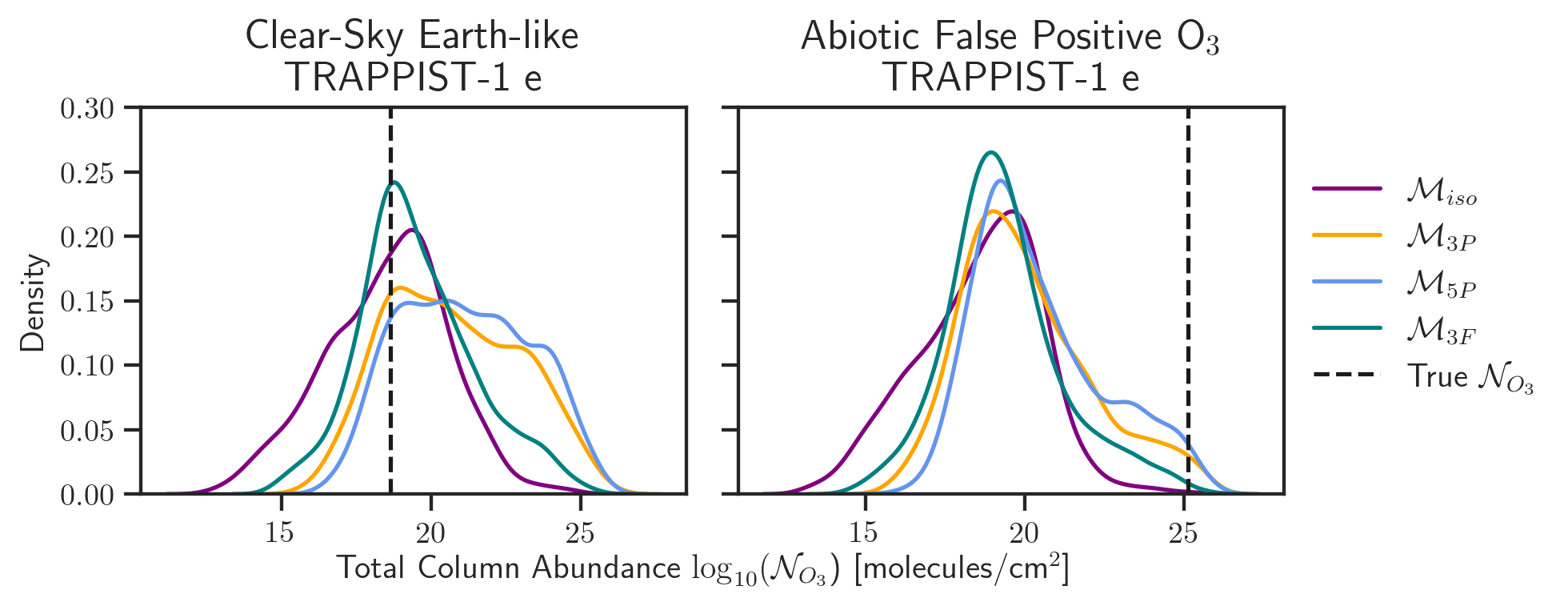}
  \caption{A comparison of O$_3$ total column abundance distributions calculated for retrieval models with and without vertical resolution for both transmission cases. For both cases and across all models, we find that the total column abundance of O$_3$ is broadly distributed and unconstrained, suggesting that the transmission spectra are relatively insensitive to the column even in the absence of aerosols.} 
  \label{fig:o3_trns_col_abs}
\end{figure*}

The vertically-resolved forward models ($\mathcal{M}_{3P}$, $\mathcal{M}_{5P}$, $\mathcal{M}_{3F}$) are unable to retrieve an O$_3$ abundance profile with the characteristic stratospheric bulge for both the Earth-like and abiotic TRAPPIST-1 e, from the transmission spectra. In fact, for both atmospheric cases, all three vertically-resolved models effectively reproduce an evenly-mixed O$_3$ volume mixing ratio profile due to the lack of constraints in abundance space (abundance and pressure space) for $\mathcal{M}_{3P}$ and $\mathcal{M}_{5P}$ ($\mathcal{M}_{3F}$). We show a comparison of the vertically-resolved O$_3$ profiles for the Earth-like and abiotic atmospheric cases using the forward model with 3 free pressure points ($\mathcal{M}_{3F}$) in Figure \ref{fig:t1e_o3_vert_comp}. In Figure \ref{fig:o3_trns_col_abs}, we show secondary total column abundance posteriors calculated by sampling the pressure, temperature, and O$_3$ abundance profiles from the retrieval posteriors, generating the putative atmosphere using the relevant forward model (evenly-mixed or vertically-resolved), and solving for the column abundance using the ideal gas and hydrostatic equilibrium equations. We compare the size and location of these distributions to the total column abundances calculated for the ``true'' atmospheres used to generate the synthetic data. For both the Earth-like and abiotic cases, and across all models, we find that the retrieval of the total column abundance is highly unconstrained from these transmission spectra. In the case of the Earth-like planet, no total column abundance distribution is biased higher from the truth by more than 1.1-$\sigma$, but the peak column abundances from these distributions vary from being a factor of 1.5 to 200 times larger than truth. For the abiotic planet, these peaks vary from being a factor of 0.4 to 5 times larger than truth. In both cases, the 5-fixed points model ($\mathcal{M}_{5P}$) shows the largest bias in total column abundance higher than the truth. For the Earth-like planet, the 3-free points model produces the total column abundance that is least biased relative to the true value, while for the abiotic planet no model cases are particularly biased, but the evenly-mixed model produces the least accurate distribution. 

For both the Earth-like and abiotic TRAPPIST-1 e cases, we find that the evenly-mixed model and the vertically-resolved models provide plausible fits to the data, so the evidence values effectively favor the least complex, evenly-mixed model. All $\chi^2$, evidence, and Bayes factor values are reported in Table \ref{tab:summ}. Though the more flexible models produce smaller $\chi^2$ values, the model evidence demonstrates that these more complex models are not statistically favored at the signal-to-noise level explored here. In fact, in the case of the 3-free points model $\mathcal{M}_{3F}$, the Bayes factor $B=-1.8$ suggests weak evidence in favor of the least flexible, evenly-mixed model $\mathcal{M}_{iso}$. This is because the model evidence penalizes models with additional parameters that do not sufficiently improve the fit to the data. This indicates that we find no statistical evidence in favor of vertically parameterized ozone profiles for the transmission cases. 

\begin{figure*}[htb]
    \centering
    \setcounter{subfigure}{0}
    \subfigure{\includegraphics[scale=0.5]{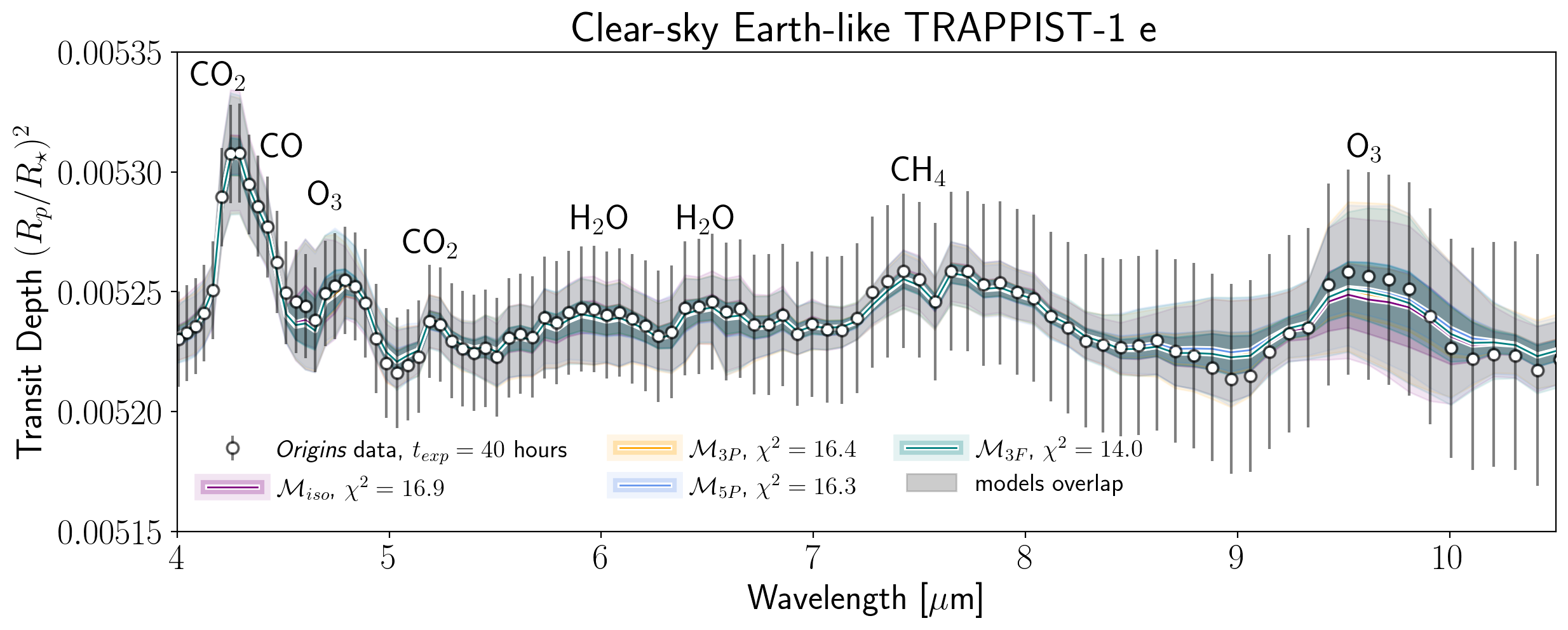}} 
    \subfigure{\includegraphics[scale=0.5]
    {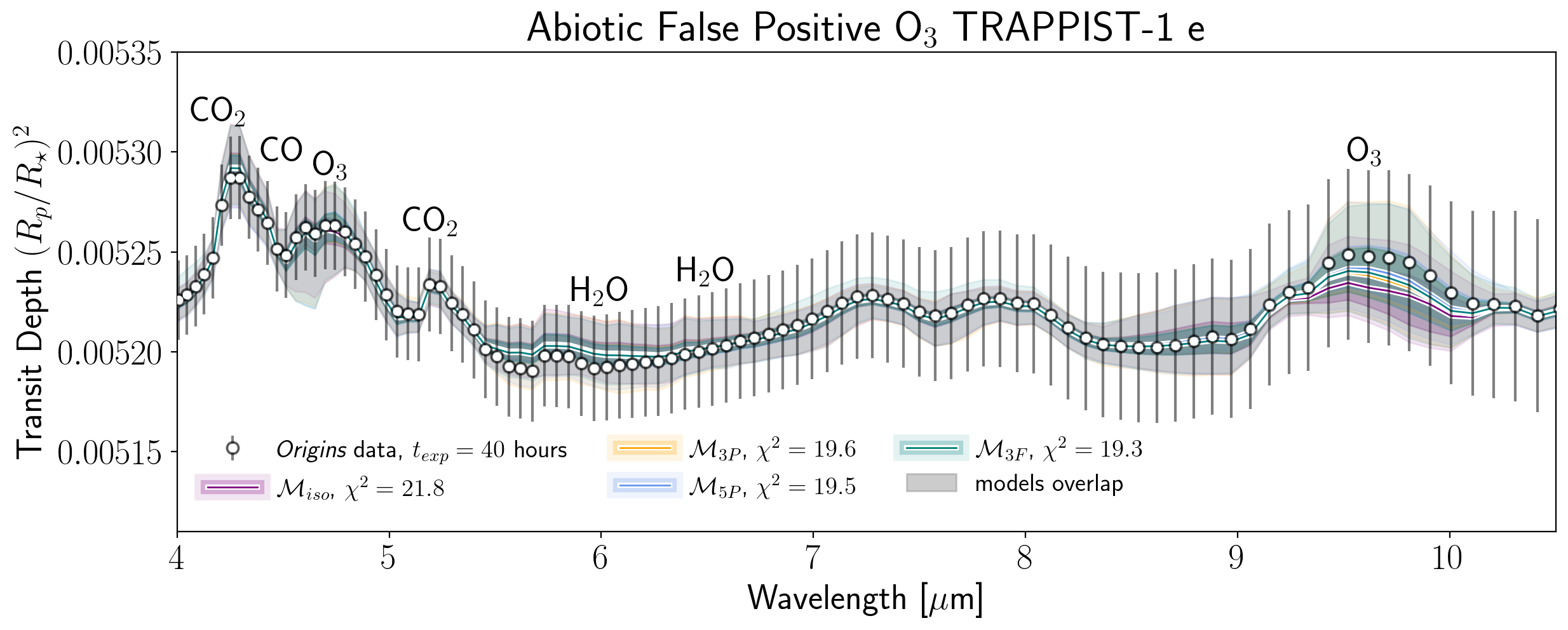}}
  \caption{A comparison of the median spectral fits with $\pm$1-$\sigma$, 3-$\sigma$ envelopes for the Earth-like and abiotic false positive O$_3$ TRAPPIST-1 e. On each plot, we show the fit to the spectrum supplied by each version of the forward model and report its associated $\chi^2$ value.  Note that the lines and envelopes overlap significantly and appear gray when they do so. All models provide plausible fits to all spectral features for both transmission cases. We zoom in on the 4-12 $\mu$m range of the spectrum, as this is where the signal-to-noise is largest.} 
  \label{fig:t1e_spec_comp}
\end{figure*}

\subsection{Direct Imaging}

\begin{figure*}[htb]
    \centering
    \setcounter{subfigure}{0}
    \subfigure{\includegraphics[scale=0.4]{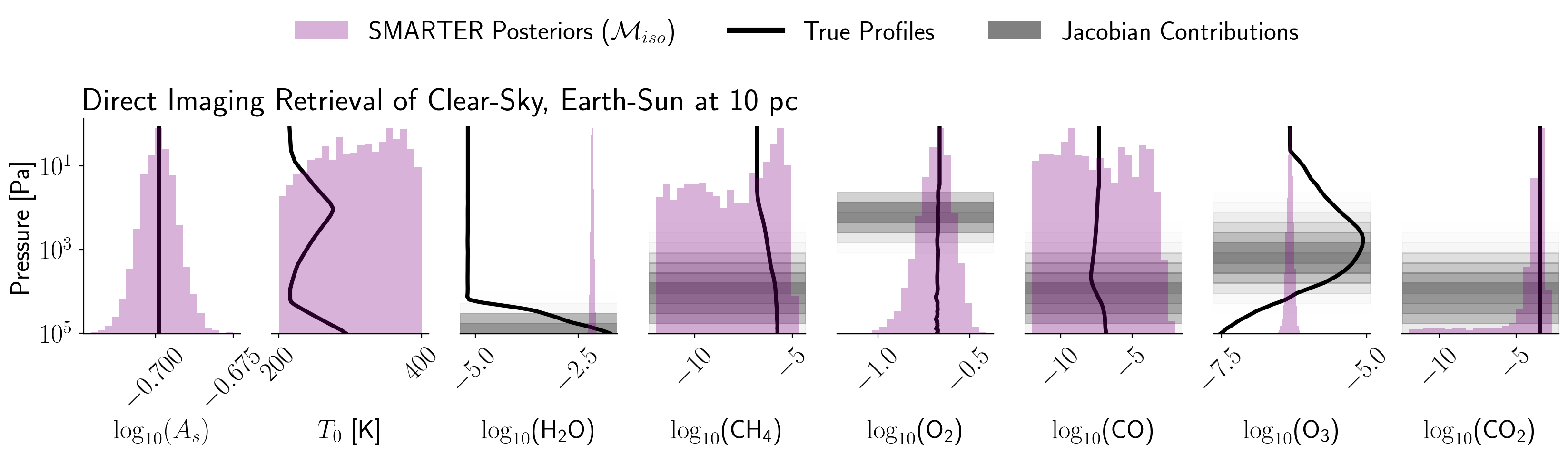}} \\
    \subfigure{\includegraphics[scale=0.4]{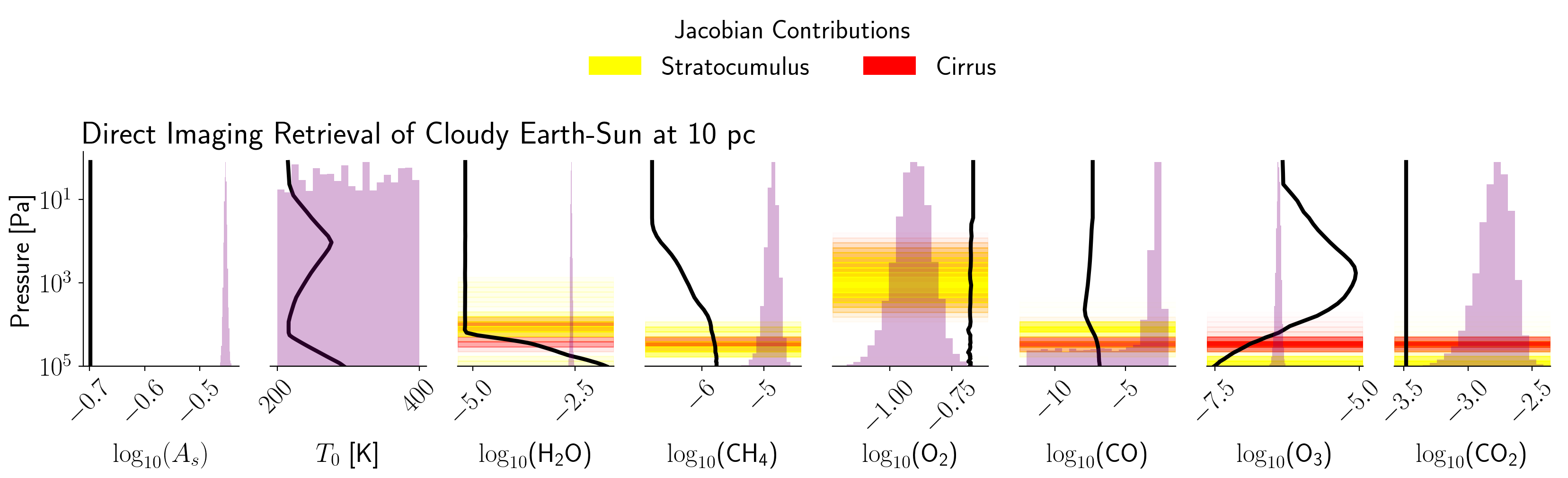}}
    \caption{A comparison of the posterior distributions for the clear-sky and cloudy Earth using the evenly-mixed forward model ($\mathcal{M}_{iso}$). Note that the y-axis on the albedo subplot provides a reference axis for interpreting the pressure-dependent profiles of all other parameters in the plot; we do not retrieve albedo as a function of pressure. The black lines represent the true profiles used to generated the observed spectrum. For the clear-sky case, the gray shaded regions indicate the Jacobian contributions to the true spectrum, where the opacity of these regions scales with their relative weight. Thus, a darker gray region indicates the atmospheric layer(s) to which each spectral feature is most sensitive. For the cloudy case, we show the Jacobian contributions to the true spectrum for an atmosphere with either 100\% stratocumulus (yellow) or 100\% cirrus clouds (red). The opacity of these regions also scales with their relative weight, and regions with significant overlap between stratocumulus and cirrus clouds Jacobian contributions take on an orange hue.}
    \label{fig:earth_hist}
\end{figure*}

We now compare our retrieval cases in reflected light, beginning with the posterior distributions from the evenly-mixed cases for the clear-sky and cloudy Earth in Figure \ref{fig:earth_hist}. To show the atmospheric layers to which the true spectrum is sensitive, we compute the radiance Jacobian as a function of gas abundance in each atmospheric layer, to calculate wavelength dependent spectral sensitivity to each gas, vertically-resolved throughout the atmospheric column. We then integrate the Jacobians over all wavelengths in the simulated data range for each gas, and report the result as a pressure-dependent Jacobian contribution. 

 For the clear-sky Earth case, beginning with the largely pressure-independent parameters, we observe that the average surface albedo is accurately retrieved and well constrained. In contrast, both the retrieved CH$_4$ and CO mixing ratio distributions are not constrained, though we gain valuable upper limits on the abundances of both gases. For CH$_4$, we retrieve a 1-$\sigma$ upper limit of 1.3 ppm, and a 3-$\sigma$ upper limit of 1.86\% VMR. The retrieved O$_2$ volume mixing ratio is confidently constrained at $21$\%, which matches the true Earth oxygen abundance of $21$\%. Lastly, the CO$_2$ volume mixing ratio appears to be accurately retrieved within a factor of 2 ($<$1-$\sigma$) of the true value, but the distribution contains a tail trailing towards lower abundances. The 1-$\sigma$ upper bound of the retrieved CO$_2$ distribution is 489 ppm, and the 3-$\sigma$ upper bound is 91\% VMR.
 
 For the pressure-dependent parameters, we observe that the retrieved isothermal temperature distribution is completely unconstrained, but the retrieved H$_2$O volume mixing ratio distribution is confidently constrained within 10\% at 1-$\sigma$ near the true surface value, to which the spectrum is sensitive. Specifically, our median retrieved H$_2$O volume mixing ratio is within a factor of 3 of the true surface value, and is consistent with the tropospheric abundance at $0.7$ bars ($7\times10^{4}$ Pa, $\sim$2.5 km). Finally, the retrieved O$_3$ volume mixing ratio for the evenly-mixed forward model ($\mathcal{M}_{iso}$) is  underestimated relative to both the average volume mixing ratio of the true profile ($<$2-$\sigma$) and the bulge abundance ($<$3-$\sigma$). The spectrum is most sensitive to O$_3$ in the stratospheric bulge layers, but the retrieved abundance does not overlap with the true profile in these layers of the atmosphere.
 
 For the cloudy Earth case (50\% clear, 25\% stratocumulus, 25\% cirrus), we show the Jacobian contributions to the true spectrum in Figure \ref{fig:earth_hist} for each gas for an atmosphere with 100\% stratocumulus clouds and an atmosphere with 100\% cirrus clouds to visualize the impact of water and ice clouds on the spectral features. The water Jacobians for each cloud type show the altitudes at which these clouds are distributed in our model ($\sim$18 km, $\sim10^{4}$ Pa). Importantly, we restate that clouds are \textit{not} considered in the retrieval model, but are included in the observed spectrum. For the pressure-independent parameters, we note that the retrieved average surface albedo is twice as bright as the truth ($>$20-$\sigma$) due to the presence of clouds in the data and the omission of clouds in the retrieval model. Interestingly, the retrieved CH$_4$, CO, and CO$_2$ abundances are all confidently overestimated, and CO has a long tail trailing towards lower abundances. CH$_4$ is overestimated by a factor of $10$, while CO$_2$ is overestimated by a factor of $50$. For CO, the distribution is non-normal and nearly bimodal, with a well-defined peak and a long tail trailing towards smaller abundances. Though this peak in the CO posterior gives a much larger abundance than the true profile, approximately 40\% of the distribution lies \textit{below} abundances of $10^{-5}$, which suggests that the retrieval has not confidently ruled out these lower abundances. For CH$_4$ and CO, cirrus clouds appear to increase the sensitivity of the spectral features to lower altitudes compared to the clear-sky case. For CO$_2$, both cloud types appear to increase the sensitivity of the spectrum to lower atmospheric layers in the true spectrum. We also note that the tail of the CO$_2$ posterior distribution has been reduced compared to its clear-sky counterpart. Lastly, the spectrum's sensitivity to O$_2$ has not changed with altitude due to the addition of clouds, but the median retrieved O$_2$ abundance (13\% VMR) is underestimated by a factor of 2 compared to the true profile. 
 
 For the pressure-dependent parameters, the retrieved isothermal temperature distribution is completely unconstrained as in the clear-sky case. Also compared to the clear-sky case, the retrieved H$_2$O mixing ratio is offset lower from the true surface abundance within a factor of $7$, and would be consistent with the true profile in the atmosphere around $0.55$ bars ($5.5\times10^{4}$ Pa, $\sim$5 km). The cirrus clouds appear to make the H$_2$O spectral features sensitive closer to the tropopause, helping to explain the lower retrieved H$_2$O abundance relative to the clear-sky case. 
 
 Finally, the retrieved O$_3$ abundance is now underestimated by a more significant margin relative to both the true average volume mixing ratio ($<$2.5-$\sigma$) and the true bulge abundance ($<$5-$\sigma$). However, the clouds appear to increase the number of atmospheric layers to which the spectrum is sensitive, with stratocumulus clouds pushing the spectral sensitivity to higher altitudes. We summarize and compare all retrieved values from both our clear-sky and cloudy direct imaging cases to their corresponding truths in Table \ref{tab:results_summ_refl_lin}.
 
\begin{figure*}[htb]
  \centering
  \includegraphics[scale=1]{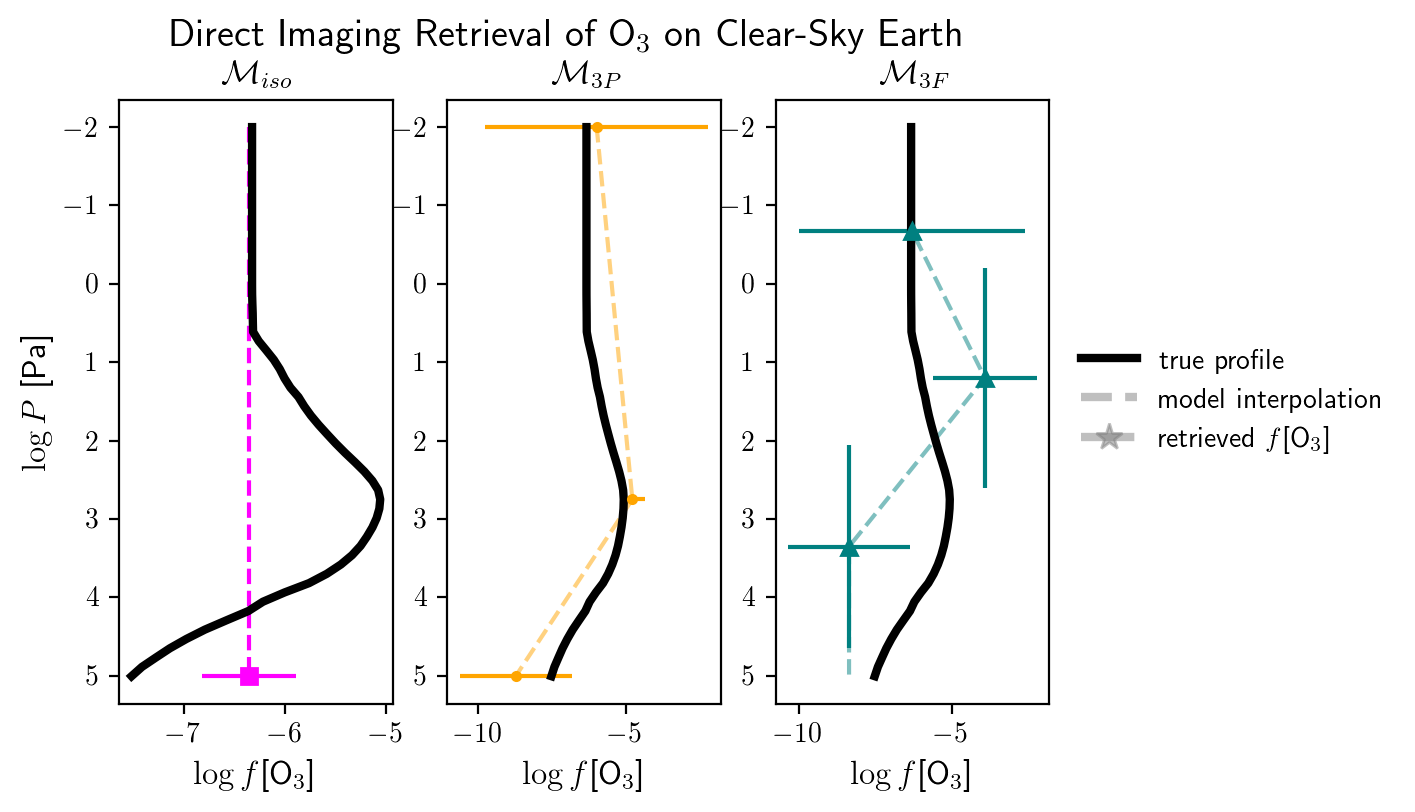}
  \caption{A comparison of the vertically-resolved O$_3$ profiles retrieved by the $\mathcal{M}_{3P}$ and $\mathcal{M}_{3F}$ forward model for the clear-sky Earth. The Bayes factors weakly favor the 3-fixed points model ($\mathcal{M}_{3P}$) over the evenly-mixed case ($\mathcal{M}_{iso}$) and the 3-free points model ($\mathcal{M}_{3F}$), which has more free parameters.} 
  \label{fig:earth_o3_refl_comp}
\end{figure*}

\begin{figure*}[htb]
  \centering
  \includegraphics[scale=0.9]{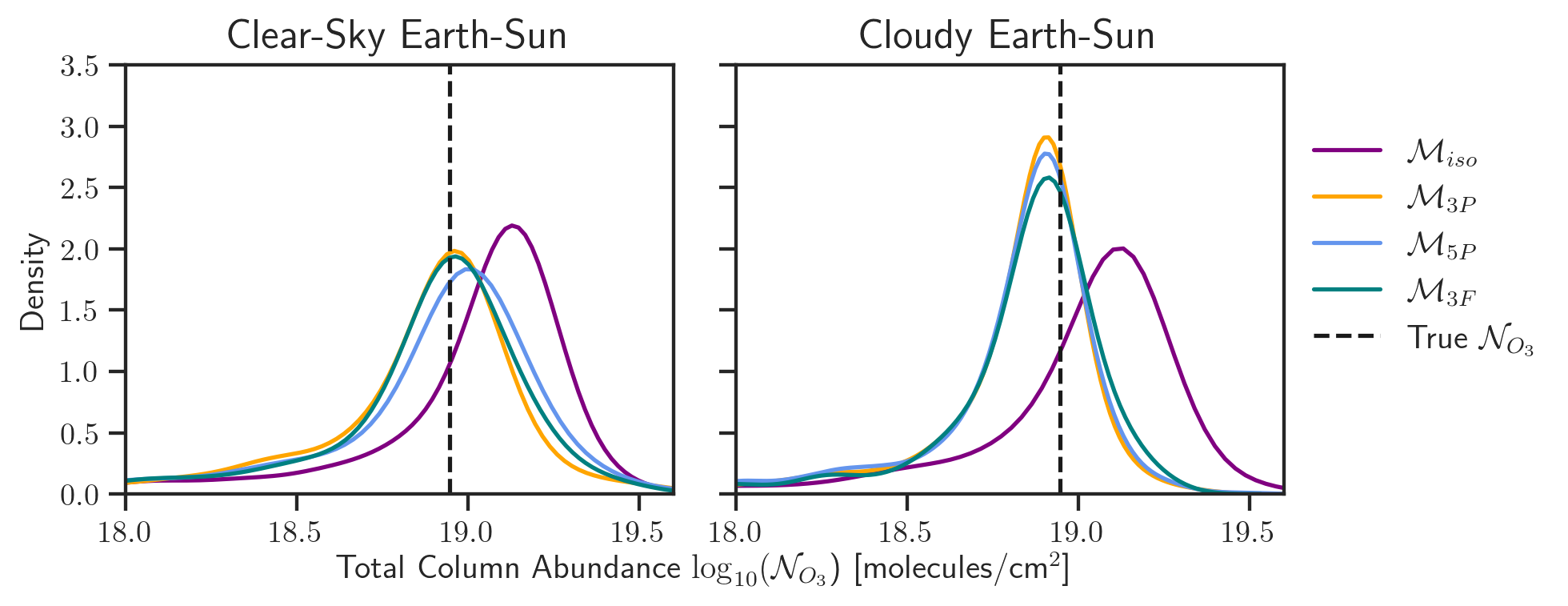}
  \caption{A comparison of O$_3$ total column abundance distributions calculated for retrieval models with and without vertical resolution for both the clear-sky and cloudy direct imaging cases. For both cases, we find that the total column abundance of O$_3$ is well-constrained using any of the vertically-resolved models, while the evenly-mixed model shows a small bias towards higher column abundances that worsens in the presence of clouds.} 
  \label{fig:o3_refl_col_abs}
\end{figure*}

 The $\mathcal{M}_{3P}$ and $\mathcal{M}_{3F}$ vertically-resolved forward models retrieve an O$_3$ abundance profile with a characteristic stratospheric bulge for both the clear-sky and cloudy Earth. The $\mathcal{M}_{5P}$ forward model is not able to constrain the vertical structure of the profile, effectively returning an evenly-mixed profile. We show a comparison of the vertically-resolved O$_3$ profiles for the clear-sky Earth using the $\mathcal{M}_{3P}$ and $\mathcal{M}_{3F}$ forward models in Figure \ref{fig:earth_o3_refl_comp}. In Figure \ref{fig:o3_refl_col_abs}, we show secondary total column abundance posteriors calculated using the same method described in section \ref{sec:res_trnst}. We find that, in both the clear-sky and cloudy cases, the evenly-mixed model ($\mathcal{M}_{iso}$) shows the largest bias to higher column abundances than the truth when compared to any of the vertically-resolved models ($\mathcal{M}_{3P}$, $\mathcal{M}_{5P}$, $\mathcal{M}_{3F}$). The peaks of the evenly-mixed total column abundance distributions for both the clear-sky and cloudy cases are overestimated by 40\% relative to the truth. The vertically-resolved models perform better, producing total column abundance distributions with peaks that slightly underestimate the true value by only 5\% in the clear-sky case and 15\% in the cloudy case. 

\begin{figure*}[htb]
  \centering
  \setcounter{subfigure}{0}
  \subfigure{\includegraphics[scale=0.5]{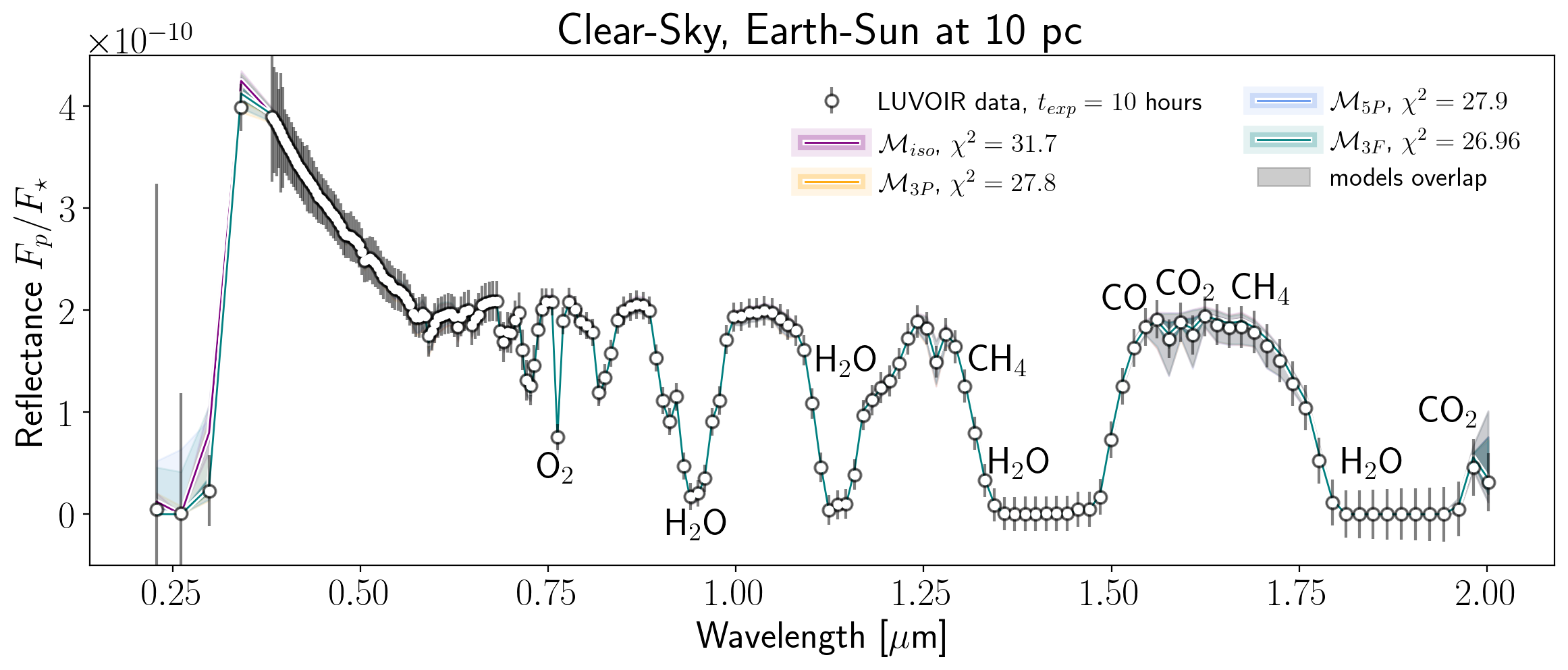} \label{fig:a}} \\
  \subfigure{\includegraphics[scale=0.5]{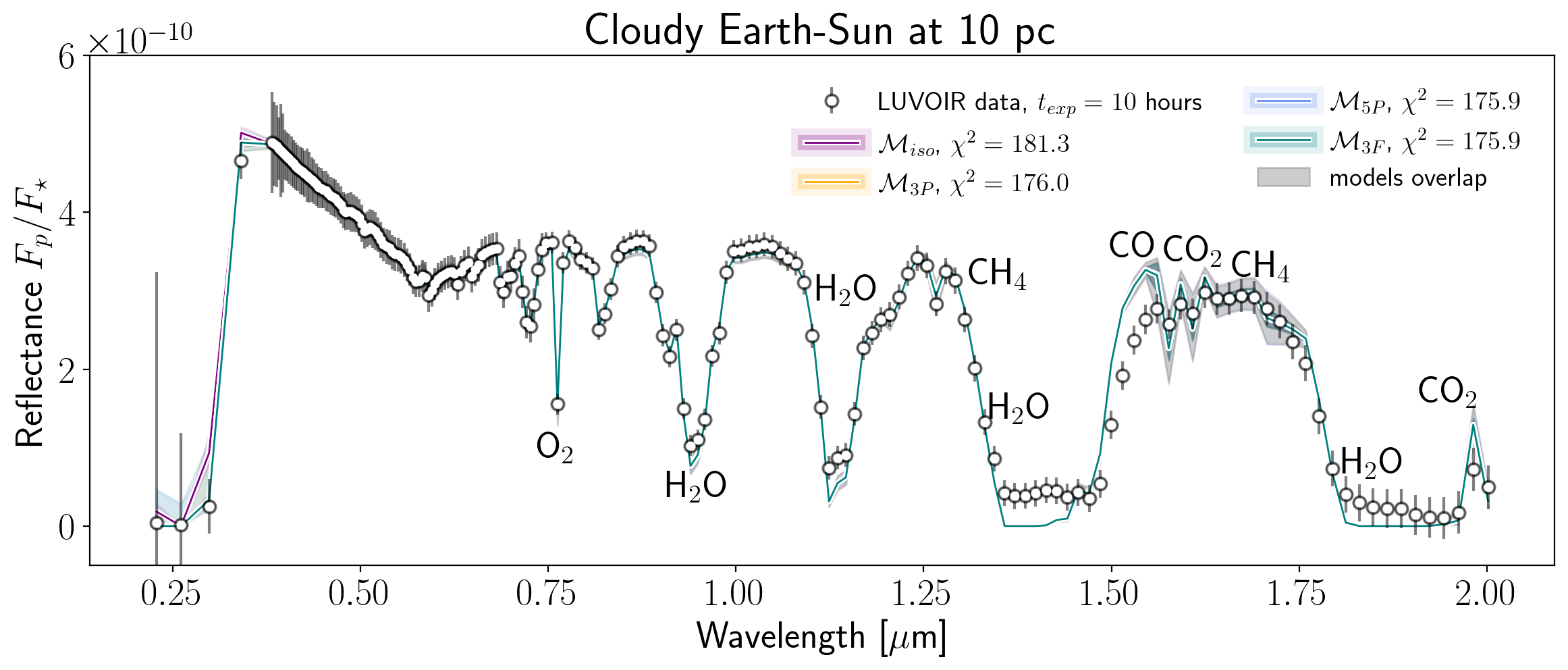} \label{fig:b}}
  \caption{A comparison of the median spectral fits with $1$- and 3-$\sigma$ posterior distributions shown as shaded envelopes (dark and pale, respectively) for the clear-sky and cloudy Earth. Note that the lines and envelopes corresponding to the different models overlap significantly and appear gray when they do so. On each plot, we show the fit to the spectrum supplied by each version of the forward model and report its associated $\chi^2$ value. For the clear-sky Earth, the evenly-mixed model visibly provides a better fit to the 0.3 $\mathrm{\mu}$m feature. In the upper panel, note the larger distribution on the CO posteriors, and that vertically resolving the O$_3$ profile results in the most significant changes to the fit to the spectrum in the region of the UV Hartley band. In the lower panel, note the large residuals in the water bands and in the CO/CO$_2$ features $\sim$1.6 $\mu$m when wavelength-dependent clouds are present in the data but not included in the retrieval.} 
  \label{fig:earth_spec_comp}
\end{figure*}

Comparing the evidence for the models in both the clear-sky and cloudy cases, we find that there is weak evidence in favor of the $\mathcal{M}_{3P}$ forward model when compared to $\mathcal{M}_{iso}$ and $\mathcal{M}_{3F}$ ($0.6 < {B} < 1.7$). All model evidences ($\log\mathcal{Z}$) are reported in Table \ref{tab:summ}. We also compare the forward model fits to the spectrum for both the clear-sky and cloudy Earth cases in Figure \ref{fig:earth_spec_comp}. For the fits to the simulated spectral data of the clear-sky Earth, all four versions of the forward model provide a plausible fit to the spectrum with the given error bars, but the most flexible model with 3 free points ($\mathcal{M}_{3F}$) has the smallest $\chi^2$ of 26.9.  In particular, we observe that the $\mathcal{M}_{3F}$ model produces the best fit to the 0.30 $\mu$m O$_3$ feature, and produces a better fit to the true O$_3$ profile than the evenly-mixed model. Like the clear-sky case, the cloudy direct imaging case also shows weak evidence in favor of the 3-fixed points ($\mathcal{M}_{3P}$) model as well as the 5-fixed points ($\mathcal{M}_{5P}$) model, but we note that the fit to the spectrum is visibly poor in the bottoms of the water vapor bands and at the 1.6 $\mu$m CO/CO$_2$ feature. All $\chi^2$ values are reported in Table \ref{tab:summ}.

 \begin{deluxetable*}{l|rrr|rrr}
\tablewidth{0.98\textwidth}
\tabletypesize{\normalsize}
\tablecaption{Retrieved constraints on Earth-like and abiotic TRAPPIST-1 e transmission spectra generated with the evenly-mixed forward model. As noted in the text, the vertically-resolved models only significantly impacted the retrieved distributions of O$_3$. The results of the vertically-resolved comparisons are summarized in Table \ref{tab:summ}. For O$_3$, and CO, which are not evenly-mixed gases in the Earth-like atmosphere, we use the average volume mixing ratio of the true profiles as our true values. Similarly, for H$_2$O, we designate the isothermal region of the true profile to be our comparison value given the apparent sensitivity of the water absorption features in the transmission spectrum to this region of the atmosphere. Finally, ``UL'' denotes parameters for which we obtained only an upper limit, and parameters that are not constrained are marked ``Unconstrained''.
\label{tab:results_summ_trnst_lin}}
\tablehead{\colhead{} & \multicolumn{3}{c}{Earth-like T-1 e} & \multicolumn{3}{c}{Abiotic O$_3$ T-1 e} \\
           \colhead{Parameters} & \colhead{Truths} & \colhead{Retrieved} & \colhead{Ratio} &  \colhead{Truths} & \colhead{Retrieved} &  \colhead{Ratio} }
\startdata
$R_p$ [$R_{\oplus}$] & 0.91                   & $0.920^{+0.002}_{-0.002}$        & $1.010^{+0.002}_{-0.002}$     & 0.91                  & $0.9200^{+0.0009}_{-0.0020}$   & $1.0000^{+0.0009}_{-0.0020}$ \\ 
$P_0$ [Pa]           & 9.8 $\times$ 10$^{4}$  & $8^{+33}_{-7} \times 10^{4}$  & $0.8^{+3.4}_{-0.7}$          & 1.2 $\times$ 10$^{5}$ & $1.0^{+3.0}_{-0.7} \times 10^{5}$ & $0.9^{+3.0}_{-0.6}$ \\ 
$T_0$ [K]            & 284                    & $202^{+34}_{-30}$               & $0.7^{+0.1}_{-0.1}$          & 263                   & $194^{+30}_{-25}$              & $0.74^{+0.01}_{-0.09}$ \\ 
$f$[H$_2$O]          & 158 ppm                & $32^{+330}_{-30}$ ppm           & $0.2^{+2.0}_{-0.2}$            & 5 ppb             & $7^{+160}$ ppb (UL)   & $1.3^{+31.0}$ (UL) \\ 
$f$[CH$_4$]          & 696 ppm                & $403^{+1900}_{-360}$ ppm        & $0.6^{+3.0}_{-0.5}$            & 0                     & $16^{+55}$ ppb (UL)       & $160000^{+600000}$ (UL) \\ 
$f$[CO$_2$]          & 0.10                   & $0.07^{+0.30}_{-0.06}$           & $0.7^{+3.0}_{-0.6}$           & 0.63                  & $0.4^{+0.3}_{-0.2}$           & $0.6^{+0.5}_{-0.3}$ \\ 
$f$[O$_2$]           & 0.20                   & Unconstrained                   & -                            & 0.12                  & Unconstrained                             & - \\ 
$f$[O$_3$]           & 6 ppm                  & $0.84^{+13.00}_{-0.80}$ ppm         & $0.2^{+2.0}_{-0.2}$           & 7.3 ppm               & $0.73^{+5.00}_{-0.72}$ ppm        & $0.100^{+0.700}_{-0.098}$ \\ 
$f$[CO]              & 606 ppm                & $12.4^{+650.0}_{-12.0}$ ppm         & $0.02^{+1.00}_{-0.02}$          & 0.24                  & $0.064^{+0.200}_{-0.054}$          & $0.3^{+0.8}_{-0.2}$ \\ 
\enddata
\end{deluxetable*}

\begin{deluxetable*}{l|rrr|rrr}
\tablewidth{0.98\textwidth}
\tabletypesize{\normalsize}
\tablecaption{Retrieved constraints on clear-sky and cloudy Earth-twin reflected light spectra generated with the evenly-mixed forward model. As noted in the text, the vertically-resolved models only significantly impacted the retrieved distributions of O$_3$. The results of the vertically-resolved comparisons are summarized in Table \ref{tab:summ}. For H$_2$O, we use the surface abundance of water in the true atmosphere as our comparison value. For O$_3$, we use the bulge abundance in the true atmosphere as our comparison value given the apparent sensitivity of the ozone absorption features in the spectrum to this region of the atmosphere. As in Table 
 \ref{tab:results_summ_trnst_lin}, ``UL'' denotes parameters for which we obtained only an upper limit, and parameters that are not constrained are marked ``Unconstrained''. \label{tab:results_summ_refl_lin}} 
\tablehead{\colhead{} & \multicolumn{3}{c}{Clear-sky Earth} & \multicolumn{3}{c}{Cloudy Earth} \\
           \colhead{Parameters} & \colhead{Truths} & \colhead{Retrieved} & \colhead{Ratio} & \colhead{Truths} & \colhead{Retrieved} &  \colhead{Ratio} }
\startdata
$A_s$        & 0.2      & $0.200^{+0.003}_{-0.003}$    & $1.0^{+0.01}_{-0.01}$  & 0.2     & $0.350^{+0.003}_{-0.003}$     & $1.75^{+0.02}_{-0.02}$ \\
$T_0$ [K]    & 294      & Unconstrained                           & -                      & 294      & Unconstrained                           & - \\
$f$[H$_2$O]  & 0.02     & $0.0070^{+0.0006}_{-0.0005}$ & $0.95^{+0.08}_{-0.08}$ & 0.02     & $0.0030^{+0.0002}_{-0.0002}$ & $0.35^{+0.02}_{-0.02}$ \\
$f$[CH$_4$]  & 1.50 ppm & $6^{+1300}$ ppb (UL)        & $0.004^{+900}$         & 1.50 ppm & $14^{+4}_{-3}$ ppm            & $9^{+2}_{-2}$ \\
$f$[O$_2$]   & 0.21     & $0.21^{+0.05}_{-0.05}$       & $0.99^{+0.2}_{-0.2}$   & 0.21     & $0.13^{+0.02}_{-0.02}$       & $0.6^{+0.1}_{-0.1}$ \\
$f$[CO]      & 90 ppb & $30^{+60000}$ ppb (UL)        & $0.32^{+609}$          & 90 ppb & $528^{+2000}_{-528}$ ppm       & $5600^{+22000}_{-5600}$ \\
$f$[O$_3$]   & 4.2 ppm  & $0.50^{+0.07}_{-0.06}$ ppm   & $0.11^{+0.02}_{-0.02}$ & 4.2 ppm  & $0.40^{+0.04}_{-0.04}$ ppm   & $0.094^{+0.009}_{-0.009}$ \\
$f$[CO$_2$]  & 330 ppm  & $187^{+302}_{-170}$ ppm      & $0.57^{+0.92}_{-0.52}$ & 330 ppm  & $2000^{+500}_{-500}$ ppm & $5.0^{+1.6}_{-1.4}$ \\
\enddata
\end{deluxetable*}

\begin{deluxetable*}{c|c|c|c|c|c|c}
\tablecaption{Summary of the results of the retrieval tests for our observing cases and their atmospheric subcases with evenly-mixed ($\mathcal{M}_{iso}$) and vertically-resolved ($\mathcal{M}_{3P}$, $\mathcal{M}_{5P}$, $\mathcal{M}_{3F}$) forward models. The model that minimizes $\chi^2$ provides the best fit to the data. The model with the largest evidence ($\log \mathcal{Z}$) is the preferred model, but our comparison of the effectiveness of the models depends on the significance \citep{jeffreys1998theory} of the difference in their evidences compared to $\mathcal{M}_{iso}$, shown in the Bayes Factor column of the table below. Here, Bayes factors  $|B| < 1$ are considered to provide inconclusive evidence, while negative Bayes factors $-2 < B \leq -1$ suggest weak evidence in favor of $\mathcal{M}_{iso}$, and positive Bayes factors $1 \leq B < 2$ suggest weak evidence in favor of a given vertically-resolved forward model. Dashes are reported in this column for $\mathcal{M}_{iso}$ in each case given this is the model used for comparison. For both the Earth-like TRAPPIST-1 e transmission case and the false-positive O$_3$ transmission case, we find that there is either no significant difference in the model evidence between $\mathcal{M}_{iso}$ and $\mathcal{M}_{3P}$ and $\mathcal{M}_{5P}$, and that $\mathcal{M}_{iso}$ is weakly preferred when compared to $\mathcal{M}_{3F}$. For the clear-sky Earth-twin direct imaging case, we find weak evidence in favor of the 3-fixed points model ($\mathcal{M}_{3P}$). Like the clear-sky case, the cloudy direct imaging case also shows weak evidence in favor of the 3-fixed points ($\mathcal{M}_{3P}$) as well as the 5-fixed points ($\mathcal{M}_{5P}$) models, but we note that the fit to the spectrum is visibly poor in the bottoms of the water vapor bands and at 1.6 $\mu$m. \label{tab:summ}}
\tablewidth{0.98\textwidth}
\tabletypesize{\normalsize}
\tablehead{{Observing Method} & {Atmosphere} & {Model ($\mathcal{M}$)} & \# Params. ($N$) & Goodness of Fit ($\chi^2$) & Evidence ($\log{\mathcal{Z}}$) & Bayes Factor ($B$)}
\startdata
\multirow[c]{8}{*}{Transmission}           
            & \multirow[c]{4}{*}{Earth-like T-1 e}
                        & $\mathcal{M}_{iso}$   & 9     & 16.9      & -15.9 $\pm$ 0.1  & - \\  \cline{3-7}
            &           & $\mathcal{M}_{3P}$    & 11    & 16.4      & -15.7 $\pm$ 0.3  & 0.2 \\ \cline{3-7}
            &           & $\mathcal{M}_{5P}$    & 13    & 16.3      & -16.3 $\pm$ 0.1  & -0.4 \\ \cline{3-7}
            &           & $\mathcal{M}_{3F}$    & 14    & 14.0      & -17.7 $\pm$ 0.0  & -1.8 \\  \clineB{2-7}{3.0}
            & \multirow[c]{4}{*}{Abiotic O$_3$ T-1 e}
                        & $\mathcal{M}_{iso}$  & 9      & 21.8      &  -16.5 $\pm$ 0.0  & - \\ \cline{3-7}
            &           & $\mathcal{M}_{3P}$   & 11     & 19.6      & -15.9 $\pm$ 0.1   & 0.6 \\  \cline{3-7}
            &           & $\mathcal{M}_{5P}$   & 13     & 19.5      & -17.0 $\pm$ 0.1   & -0.5 \\ \cline{3-7}
            &           & $\mathcal{M}_{3F}$   & 14     & 19.3      & -18.0 $\pm$ 0.0   & -1.5 \\
    \hlineB{5.0}
\multirow[c]{8}{*}{Direct Imaging}           
            & \multirow[c]{4}{*}{Clear-sky Earth}
                        & $\mathcal{M}_{iso}$  & 8      & 31.7      & -25.5 $\pm$ 0.1  & - \\  \cline{3-7}
            &           & $\mathcal{M}_{3P}$   & 10     & 27.8      & -24.2 $\pm$ 0.1  & 1.3 \\  \cline{3-7}
            &           & $\mathcal{M}_{5P}$   & 12     & 27.9      & -24.8 $\pm$ 0.1  & 0.7 \\  \cline{3-7}
            &           & $\mathcal{M}_{3F}$   & 13     & 26.9      & -25.9 $\pm$ 0.1  & -0.4 \\  \clineB{2-7}{3.0}
            & \multirow[c]{4}{*}{Cloudy Earth}
                        & $\mathcal{M}_{iso}$  & 8      & 181.3     & -120.2 $\pm$ 0.0  & - \\  \cline{3-7}
            &           & $\mathcal{M}_{3P}$   & 10     & 176.0     & -118.3 $\pm$ 0.2  & 1.9 \\  \cline{3-7}
            &           & $\mathcal{M}_{5P}$   & 12     & 175.9     & -118.4 $\pm$ 0.1  & 1.8 \\  \cline{3-7}
            &           & $\mathcal{M}_{3F}$   & 13     & 175.9     & -119.8 $\pm$ 0.1  & 0.4 \\
\enddata
\end{deluxetable*}

\section{Discussion}
\label{sec:discussion}
We used a novel terrestrial exoplanet retrieval code to compare the accuracy of retrievals for transmission and direct imaging spectroscopy, to assess both retrieved abundances and our ability to infer the characteristics of an exoplanet's surface environment. In transmitted light, we performed this comparison for both a modern Earth-like and an abiotic TRAPPIST-1 e with false positive O$_3$.  In reflected light, we compared a clear-sky and cloudy Earth-twin at 10 pc. 

For the cases considered, we found that we were able to detect and constrain the abundances of numerous key species. For transmission observations with exposure time $t = 40$ hours for an \textit{Origins}-like mid-IR probe, which is assumed to be nearly twice as sensitive as JWST according to the instrument parameters described by \citet{meixner2019origins} over the wavelength range 2.8~\textendash~20.0 $\mu$m, our model detects and retrieves 20\% of the true H$_2$O, 60\% of the true CH$_4$, and 20\% of the true O$_3$ abundances for the modern Earth-like atmosphere. For the abiotic false positive planet, we detect and retrieve 130\% of the true H$_2$O, 10\% of the true O$_3$, and a useful upper limit on the CH$_4$.  We also accurately constrain CO$_2$ abundances for the inhabited planet to 70\% of the true profile, and retrieve 60\% of the true CO$_2$ abundance for the abiotic planet. 
For the modern Earth-like TRAPPIST-1 e observed in transmission, we obtained good fits to the simulated data, but found that our retrieved water abundance coincides with the true stratospheric abundance of this planet rather than the surface abundance.  For this case, we also retrieved a median temperature of 202 K, which, given our calculations of transmission's sensitivity to that region, is likely to be that of stratospheric altitudes between 50 and 70 km.  

For our direct imaging cases, we found that the retrieval produced relatively high accuracy constraints on  O$_3$ with inferred abundances within 20\% of the true value, and a near-surface abundance of H$_2$O in a clear-sky, Earth-twin atmosphere retrieved to within 10\%. Similarly, O$_2$ is constrained with 1-$\sigma$ errors of $\pm$5\% VMR for the clear-sky case, and within a factor of 2 with clouds. Additionally,  when we implement a vertically-resolved forward model in these cases, we obtain weak sensitivity to the presence of a stratospheric O$_3$ bulge for both the clear-sky and cloudy Earth based on improved fits to the Hartley band in the ultraviolet. In addition, our vertically resolved model more accurately retrieves the true averaged column abundance compared to the evenly mixed assumption. The evenly-mixed model retrieves the true value within $40\%$ errors, while the vertically resolved models retrieve the true value within $\sim$${10}\%$ errors in the clear-sky case and within $\sim$${20}\%$ errors in the cloudy case (\ref{fig:o3_refl_col_abs}). For the clear-sky Earth, our inferred abundances fit the reflected light observations over the entire wavelength range within 1-$\sigma$ error bars.

\subsection{Comparison to Previous Work}

There are benchmark retrieval studies in the literature that we can compare to, although subtle differences in the synthetic observations and retrieval forward models limit direct comparisons. The retrieval results we present for the photochemically self-consistent clear-sky, Earth-like TRAPPIST-1 e in transmitted light may be compared to the transmission retrieval of  \citet{tremblaydetectability2020}, for their non-photochemically-self-consistent Earth orbiting TRAPPIST-1, which has the gas mixing ratios and atmospheric structure of the true Earth.  Consequently, although similar, the atmospheric compositions and vertical distributions for the two cases are not identical. Similarly, the  results we present for the cloudy Earth-twin in reflected light may be compared to the Earth twin planets modeled by \citet{feng2018characterizing} and \citet{damiano2022reflected}. In all three of these studies, the authors use a forward model to produce synthetic data by assuming evenly-mixed gas and isothermal temperature profiles (except in the case of H$_2$O, which is the condensing gas, in \citeauthor{damiano2022reflected} [\citeyear{damiano2022reflected}]). All three studies use the same forward model to generate their data and perform the retrieval, in which cloud properties are also included as free parameters. In particular,  \citet{damiano2022reflected} implement a cloud parameterization scheme that vertically resolves the profile of the condensing gas at the top/bottom of the cloud deck and at the planet surface, allowing exploration of cloud behavior and cloud-gas absorption degeneracies. All three studies show potential degeneracies between cloud parameters, surface albedo, and gas absorption. 

In comparison, our study used the vertically-resolved atmospheric structure of a 1-D-modeled planet to generate the synthetic data, allowing us to explore the biases introduced by using a  simplified forward model to fit spectra  of complex planetary environments. Our study lacks a cloud parameterization for the retrieval model, but instead includes additional free parameters to vertically resolve non-condensing, photochemically-active gases.  Our study shows the biases that are introduced when a cloud-free model is used to fit a cloudy atmosphere, and how vertical parameterization impacts the retrieval and interpretation of non-condensing, photochemically-active gases. 

Comparing our results to the retrieval of synthetic transmission spectra at $R$=100 and $\lambda$ = 3.0~\textendash~30.0 [$\mu$m] for Earth-like planets orbiting TRAPPIST-1 from \citet{tremblaydetectability2020}, we more precisely constrain H$_2$O and CH$_4$, while \citet{tremblaydetectability2020} put more precise constraints on CO$_2$ and O$_3$. However, this difference in retrieval sensitivity is most likely due to compositional differences in our model atmospheres. \citet{tremblaydetectability2020} did not generate synthetic data for a climatically and photochemically self-consistent Earth-like planet around an M dwarf, and our photochemical models (Figure \ref{fig:atmcomp}) show that the spectrum of the M dwarf star would have suppressed O$_3$ and enhanced CH$_4$ abundance  \citep{segura2005biosignatures} compared to modern Earth.  Consequently,  \citet{tremblaydetectability2020} find an O$_3$ column abundance that is enhanced compared to ours, while their CH$_4$ is significantly reduced. Furthermore, their paper includes data generated with more transits than this study, and they scale TRAPPIST-1's stellar spectrum to a more optimistic $K$-band magnitude of 8 (100 times brighter than TRAPPIST-1's observed brightness) to mimic the average brightness of nearby M dwarf stars.  This also serves to increase the SNR of the spectrum particularly at the longer wavelengths where prominent O$_3$ and CO$_2$ absorption features appear.

In contrast, the synthetic observations in reflected light all assume a planetary atmospheric composition close to true Earth and so provide a better comparison with our results.  In particular, the S/N=$15$, R=$140$ simulations of \citet{feng2018characterizing} and the S/N=$10$, R=$140$ simulations of \citet{damiano2022reflected} make a better comparison with our S/N=$15$, R=$100$ cloudy Earth-twin retrieval results. For H$_2$O in the cloudy case (50\% clear/50\% cloudy), we are able to more precisely constrain the near-surface water abundance (7\% errors) than both \citet{feng2018characterizing} (100\%~\textendash~200\% errors) and \citet{damiano2022reflected} (200~\textendash~700\% errors), and we appear to retrieve a near-surface tropospheric water abundance consistent with the true profile at 5 km (5.6$\times 10^4$ Pa), between the surface and cloud deck. Similar to \citet{feng2018characterizing} and \citet{damiano2022reflected}, O$_2$ and O$_3$ are well constrained, as we retrieve 60\% of the true O$_2$ and 10\% of the O$_3$ bulge abundance, and we are able to rule out abundances lower than 9\% for O$_2$ and abundances lower than 0.36 ppm for O$_3$. However, un-modeled, wavelength-dependent cloud absorption \citep{robinson2011earth} causes our retrieval to overestimate the abundance of CH$_4$ and CO$_2$ as it increases their VMRs in an attempt to match the additional cloud absorption. By contrast, \citet{damiano2022reflected} obtain more accurate but highly imprecise retrieved abundances and \citet{feng2018characterizing} did not model or retrieve these gases in their study. We note that this inaccuracy of our retrieval model will likely be remedied by the inclusion of cloud parameters in the retrieval forward model, which we intend to explore in future studies.

\subsection{Comparison of Transmission and Direct Imaging Capabilities}

Our results have broader implications for the differing capabilities of transmission and direct imaging spectroscopy to extract accurate information regarding a terrestrial planet’s environmental properties. For the two observing modes below  we compare the sensitivity to cloud absorption, the ability to assess the surface habitability of an exoplanet, and the interpretation of biosignature pairs and their discriminants in the context of the planetary environment. 

\subsubsection{Sensitivity to Clouds}
\label{sec:clouds}

Clouds are likely to be a significant feature of the planetary atmospheres explored here, whether they are impediments to observables \citep[e.g.,][]{lincowski2018evolved, fauchez2019impact, suissa2020dim, komacek2020clouds} or observables themselves \citep{kreidberg2014clouds, wakeford2015transmission, arney2018organic, benneke2019water}. While the SMART forward model is capable of modeling aerosols, and we use this capability in creating the simulated observations, the aerosols must be rigorously defined in terms of particle size distributions and wavelength-dependent optical properties \citep{robinson2011earth}, which are typically much more complex than the simpler parameterizations used in existing retrieval models \citep[e.g.,][]{feng2018characterizing, tremblaydetectability2020, damiano2022reflected}, and therefore computationally expensive to use in a retrieval model. As a result, we focus our study on the outcomes we obtain when realistic aerosols are present in exoplanet measurements but not fitted for and retrieved on.

We saw relatively little impact between clear and cloudy spectra in our \textit{Origins}-like mid-IR transmission simulations of an Earth-like TRAPPIST-1 e with an exposure time of 40 hours. Similar results were found in the retrieval study of \cite{mayicecloud2021}, who showed that water ice cloud variability did not affect retrieved abundances at detectable levels for synthetic observations of TRAPPIST-1 e with JWST's NIRSpec PRISM. This is despite previous evidence that clouds can potentially truncate transmission spectra in observations of hotter exoplanets \citep{kreidberg2014clouds} and models of Earth-like habitable-zone terrestrial planets \citep{fauchez2019impact, suissa2020dim, komacek2020clouds,lincowski2018evolved}, and can also affect the retrieval of atmospheric properties \citep{Line2016}.

Our lack of cloud impact in the transmission case compared to previous results is due to two factors: 1) primarily that our photochemically-self-consistent TRAPPIST-1 e atmosphere contains higher abundances of both CH$_4$ and CO$_2$ than the true Earth cases explored previously \citep[e.g.,][]{fauchez2019impact}, thereby providing greater atmospheric opacity with a pseudo-continuum of absorption and 2) to a lesser extent, that the two studies take different approaches to modeling the ice and water clouds included in the modern Earth-like atmospheres.  Our inferred lack of spectral truncation by clouds is consistent with the cloud altitudes---ice clouds at 300 mbar ($3\times10^{4}$ Pa) and water clouds at 900 mbar ($9\times10^{4}$ Pa)---which are below the regions of the atmosphere where the retrieval is sensitive (c.f. Figure \ref{fig:t1e_hist}).  Although it has been suggested that 3-D GCMs may form clouds higher than predicted by 1-D climate models \citep{fauchez2019impact}, clouds in our model are in fact located a few kilometers higher at the terminator, where transmission probes, compared to \citet{fauchez2019impact}. Instead, truncation of the spectrum occurs because our higher trace gas abundances produce stronger bands with extended wings (especially CO$_2$ and CH$_4$) that form a pseudo-continuum of opacity for most wavelengths $>$2 $\mu$m. This can be seen in Figure \ref{fig:earth_t1e_clouds} in this work and Figure 11 in \citet{Meadows2023}, when compared to the atmospheric spectra with lower CH$_4$ and CO$_2$ abundances in Figure 11 in \citet{lincowski2018evolved}, and Figure 5a in \citet{fauchez2019impact}. The additional spectrally-broad opacity from these higher gas abundances raises the minimum atmospheric level probed for even the clear-sky case, thereby minimizing the difference between the atmospheric layers probed in the clear and cloudy cases, and effectively reducing the impact of the addition of clouds. The effect of this additional opacity can explain much, but not all, of the reduced spectral truncation due to clouds found in this study relative to \citet{fauchez2019impact}; cloud modeling differences are therefore also needed to fully explain this spectral truncation gap. Even though our cloud altitudes are similar, clouds may produce a second order effect on the minimum transmission altitude between these two sets of results, likely due to differences in cloud optical depths and particle distributions in the two models.

For the clear and cloudy direct imaging comparison, clouds do truncate the probed atmospheric region, and we retrieve 40\% less O$_2$, 20\% less O$_3$, and 60\% less near-surface water abundance in the cloudy case when compared to the clear sky case. We also find that the presence of wavelength-dependent cloud absorption in the data (especially near 1.6 $\mu$m) causes us to overestimate the abundances of CO$_2$, CH$_4$, and CO, which all overlap these cloud absorption features. This overestimate occurs because the model currently cannot fit for cloud-induced, wavelength-dependent reflectivity changes, and so attempts to compensate by enhancing the wavelength-independent surface albedo as well as increasing or decreasing the inferred abundances  of gases with overlapping absorption features to better fit the spectrum. 

However, the surface albedo can only affect the spectral continuum, and not the bottoms of saturated absorption bands where photons do not reach the surface, while true clouds, which are partially gray scatterers and higher in the atmosphere, may increase the reflectivity in the bottoms of the bands. While the near-surface water abundance we retrieve assuming evenly-mixed vertical profiles provides a reasonable fit to the shape of the water bands, it will overestimate the water abundance in the cloud-truncated regions of the atmosphere, and this manifests as a much deeper core of the band than seen in the synthetic data. We find that, when assuming evenly-mixed profiles, the near-surface abundance we retrieve provides a good fit to the overall shapes of the water bands, but that only the cloud-top abundance can provide a satisfactory fit to the core of the bands. One such data-model mismatch is particularly noticeable in the poor fit to the H$_2$O bands ${>}1$ $\mu$m (c.f. Figure \ref{fig:earth_spec_comp}). We can quantify the size of the large residuals in the bottoms of these near-IR H$_2$O bands by calculating the quadrature sum of the S/N at each wavelength of a given absorption feature. Using this method, we find that the residuals in the near-IR H$_2$O bands have S/N as large as 6.5. Another visible data-model mismatch occurs in the CO band at $\sim$1.55~\textendash~1.6 $\mu$m (c.f. Figure \ref{fig:earth_spec_comp}). 
The CO, CO$_2$, and CH$_4$ absorption features near 1.6 $\mu$m overlap with a large ice cloud absorption feature from 1.5~\textendash~1.6 $\mu$m, and since the ice cloud feature is not present in the model, this results in an overestimate of the inferred abundances of these gases.

\subsubsection{Habitability Assessment}
\label{sec:habass}

Constraining surface temperature, surface pressure, the presence and abundance of water vapor \citep{robinson2018characterizing, meadows2018habitability}, and the abundance of atmospheric CO$_2$ \citep{catlingdavid2018exoplanet} are key environmental parameters needed to constrain a climate model and thereby assess the likelihood that a terrestrial planet is habitable. Atmospheric temperature and pressure are fundamental properties that can reveal the structure of the atmosphere, and their surface values can help determine whether liquid water can be sustained on the planet's surface. The abundance of greenhouse gases, such as water vapor and CO$_2$, especially when combined with any knowledge of atmospheric pressure and temperature, provide further constraints for climate modeling of the surface temperature. The detection of atmospheric water vapor near the surface of the planet can also help assess climate, and may also increase the likelihood that surface liquid water is present. Below we describe and compare the strengths and weaknesses of the transmission or direct imaging technique for obtaining the information needed to constrain habitability assessment via a climate model.  

\paragraph{Characterizing habitability using transmission observations}
\label{par:trans_hab}

Transmission observations are not sensitive to the planetary surface \citep{lincowski2018evolved}, and our results did not provide informative constraints on planetary surface temperature, or near-surface water abundance for both the Earth-like planet and the abiotic planet studied here. Furthermore, given our isothermal temperature model for the Earth-like and abiotic planets, our 1-$\sigma$ retrieved pressure ranges of 0.1~\textendash~4.1 bars ($1\times10^4$~\textendash~$4.1\times10^5$ Pa) and 0.3~\textendash~4 bars ($3\times10^4$~\textendash~$4\times10^5$ Pa) translate to altitudes at or below $\sim$23 km and $\sim$11 km, respectively. The high-pressure $\sim$4 bar upper bound for both planets indicates an insensitivity to how much additional atmosphere may lie below the region probed by the transmission spectrum. The retrieved median isothermal temperature agrees well with the stratosphere ($\sim$10$^2$ Pa, 50~\textendash~70 km) of the Earth-like planet within 10 K, and that of the abiotic planet within several K, although both temperatures are below the freezing point of water. We were also not able to retrieve precise constraints on the atmospheric CO$_2$ abundances for either planet, making it challenging to model the planetary climate. 

We retrieve median pressures for both the living and abiotic TRAPPIST-1 e that are near that of the true surface pressure, but in an agnostic setting it will be difficult to know whether we were truly seeing the surface of the planet or the top of a cloud deck given the nature of transmission observations \citep{benneke2012atmospheric, lustig2019mirage}. In transmitted light, the retrieved maximum pressure corresponds to the spectral continuum, which probes to lower altitudes than molecular absorption.  However, for a habitable planet this level is still likely to be above the surface, due to the intrinsic opacity of molecules in the atmosphere (\citet{meadows2018habitability} and Figure \ref{fig:t1e_hist}) as well as clouds and hazes \citep{fortney2005effect, kreidberg2014clouds, arney2017pale}.  Depending on the host star and planetary distance, refraction could also limit access to the near-surface atmosphere  \citep{misra2014effects, betremieux2014impact, lincowski2018evolved, lustig2019detectability}.

For the spectrum of the clear-sky Earth-like TRAPPIST-1 e (which is indistinguishable from the cloudy case at almost all wavelengths, c.f., Figure \ref{fig:earth_t1e_clouds}) we retrieved a median pressure from the continuum that is consistent with the true atmosphere's  $\sim$2 km altitude  pressure---which is 80\% of the true surface pressure ($\pm 1$ dex) for the Earth-like planet.  For the cloud-free abiotic planet, we retrieve a median continuum pressure consistent with 1 km altitude,  or 90\% of the true surface pressure.  These differences in median retrieved pressure reflect differences in scale height between the two atmospheres. Although these median retrieved pressures are close to the true surface values, the retrieved distribution suggests a broader range of pressures, and without knowledge of the ground truth, we would not be able to rule out that we had probed to a cloud-top, rather than close to the surface, without additional information. However, the relative sensitivity of transmission spectra to stratospheric temperature in combination with tighter constraints on the near-surface pressure may potentially be used with a climate model to infer the atmospheric temperature structure if a cold trap is present \citep{robinson2014common}.

CO$_2$ is a key greenhouse gas, and while the peaks of our retrieved distributions for stratospheric CO$_2$ are relatively accurate (70\% and 60\% of the true value for the Earth-like and abiotic planets, respectively), we retrieve imprecise distributions spanning $\sim$3 orders of magnitude for both cases. Since CO$_2$ is commonly evenly mixed in terrestrial planetary atmospheres \citep{hu2012photochemistry}, the stratospheric abundance we retrieve is likely a good proxy for the abundance throughout the atmosphere, which can help constrain surface temperature. However, the broad distributions we retrieve correspond to a 1-$\sigma$ range of CO$_2$ abundances of approximately 1\%-40\% for the Earth-like planet and 1\%-70\% of the abiotic planet. Given that TRAPPIST-1 e has $\sim$60\% Earth's insolation \citep{gillon2017seven}, this broad range of CO$_2$ abundances could give both habitable and uninhabitable scenarios assuming Earth-like fluxes \citep{turbet2020review, Meadows2023}. This insensitivity to CO$_2$ abundance is in part due to the strong absorption of CO$_2$, which produces spectral features that saturate at low abundances. Additionally, CO$_2$ absorption in the MIR is a product of both abundance and atmospheric thermal structure, and an assumed isothermal temperature structure could compromise the ability to fit CO$_2$ features in the spectrum. The correlation between temperature and thermally-dependent CO$_2$ absorption features (e.g., the 5.3 $\mu$m CO$_2$ feature) could be used in future retrievals to constrain vertical temperature structure for uninhabitable planets with strong CO$_2$ absorption. Additionally, future work could propagate broad posteriors through a climate model to fully depict the range of planetary environments that are possible given the fits to the observed spectra. 

Transmission is also sensitive to stratospheric water vapor, and will allow us to use order of magnitude constraints to identify or rule out extremely desiccated planets, but will likely not allow us to discriminate between Venus-like and Earth-like environments using the retrieved water vapor abundance. For the Earth-like TRAPPIST-1 e planet, we retrieve peak stratospheric H$_2$O abundances of 32 ppm, with a 1-$\sigma$ distribution spread of 2~\textendash~362 ppm, corresponding to altitudes of 17~\textendash~88 km in the true atmosphere, to which the spectrum is indeed sensitive (c.f., Figure \ref{fig:t1e_hist}). In comparison, measurements at the terminator of Venus's night-side estimated the water vapor abundance to be 0.56~\textendash~2.45 ppm \citep{chamberlain2020soir}, while the stratospheric water vapor abundance of the true Earth is $\sim$6 ppm. Thus, our broad distribution for the Earth-like planet overlaps with both a Venus-like and an Earth-like scenario. For the desiccated planet corresponding to the most extreme case modeled by \citet{gao2015stability}, we retrieve a 1-$\sigma$ upper limit on the H$_2$O abundance on the order of 167 ppb, which is approximately 12 times smaller than Venus' maximal stratospheric water abundance. Although the 2- and 3-$\sigma$ upper limits are higher abundances of 5 and 13 ppm, these stricter upper limits are still orders of magnitude smaller than the stratospheric water vapor abundances (130-302 ppm, c.f., Figure \ref{fig:atmcomp}) that we predict for the self-consistent Earth-like TRAPPIST-1 e. While this upper limit would enable us to distinguish such an extremely dry planet from an Earth-like planet around an M-dwarf, planets that are less desiccated may be more ambiguous.

\paragraph{Characterizing habitability using direct imaging observations}

Despite relative insensitivity to temperature at the spectral resolutions considered, our results suggest that direct imaging observations are remarkably sensitive to the near-surface water vapor abundance of a 50\% cloudy Earth-twin at 10 pc, and can confidently constrain near-surface water abundances to within 0.1\% (at 1-$\sigma$).

In contrast to the transmission cases, we were unable to constrain temperature in any part of the atmosphere at the S/N and spectral resolutions considered. The insensitivity of direct imaging to temperature poses a challenge to future habitability assessments, as it is a key climate modeling parameter, and should be investigated in greater detail. Although direct imaging has been shown to be sensitive to surface pressure \citep{feng2018characterizing, damiano2022reflected}, we did not fit for surface pressure for the direct imaging cases, and will explore this in future work.  A number of techniques have been proposed to determine surface pressure and temperature, including dimer strengths compared to rho-vibrational or electronic states \citep{misra2014using}, broadening of individual molecular bands \citep{palle2009earth, schwieterman2015detecting}, and fitting the Rayleigh scattering slope \citep{benneke2012atmospheric}. Recent work also considers how we may be able to constrain global temperature from the thermally-dependent behavior of water vapor absorption \citep{young2024retrievals}. However, all of the inference techniques listed above will be impacted by truncation of the atmospheric path length due to clouds \citep[e.g., Earth;][]{misra2014using} and haze \citep[e.g., Archean Earth or Venus;][]{arney2016pale, Crisp1997}. In addition, absorption from the planetary surface, or clouds and hazes, can also obliterate the pressure signal \citep{arney2017pale}.  

Planetary albedo could help identify the presence of clouds or help discriminate between ocean and ice surfaces, and although we obtained compelling albedo constraints by assuming strong prior knowledge about the target, multiple factors complicate albedo retrievals. Our cloud-free case yielded a robust albedo constraint (0.200 $\pm 0.003$), but we retrieved a planetary albedo that is 1.8 times the value of the 50\% cloud-cover Earth, as another effect of the surface albedo being used by the retrieval to account for clouds. These results assumed tight priors constraints on planetary radius and phase, which are both degenerate with albedo \citep{nayak2017atmospheric, guimond2018direct}, and therefore propagate greater uncertainty into the retrieved surface albedo if unknown \citep{feng2018characterizing, Carrion-Gonzalez2020}. This effect has long been appreciated for observations of small bodies in the solar system \citep{davies2009solar}, and poses a challenge for future exoplanet direct imaging \citep{fujii2017rotational}, especially since uncertainties in the radius may propagate to larger errors in surface gravity, surface pressure, and molecular abundances \citep{feng2018characterizing}. 

We find that direct imaging observations cannot confidently retrieve CO$_2$ in the atmosphere of a clear-sky Earth-twin. We retrieve a 1-$\sigma$ distribution of abundances spanning 17~\textendash~489 ppm with a long tail trailing to longer abundances. We conservatively interpret this distribution as providing a 3-$\sigma$ upper limit of 91\% VMR on the retrieved abundance of CO$_2$ in the clear-sky case. In the atmosphere of the Earth-twin with 50\% cloud coverage, although we detect CO$_2$ more confidently, we greatly overestimate the abundance of CO$_2$, and this is again due to the impact of the synthetic data including cloud extinction, which is not in the retrieval forward model. The peak of the retrieved CO$_2$ distribution is consistent with 5 times more carbon dioxide than is present in the true atmosphere (330 ppm), with a 1-$\sigma$ distribution spanning 1500~\textendash~2500 ppm.  As described above in Section \ref{sec:clouds}, this bias towards much higher abundances of CO$_2$ occurs due to overlapping ice cloud and CO$_2$ absorption features at $\sim$1.6 $\mu$m and  $\sim$1.95 $\mu$m in the synthetic data. Overestimating the CO$_2$ abundance in this atmosphere as 3~\textendash~7 times the true value would imply a much stronger greenhouse effect than on the true Earth, and depending on other planetary characteristics, may increase the probability that an environment is incompatible with life, even if the inferred higher abundance does not lead to a runaway greenhouse \citep{goldblatt2013low}. While we retrieved a more accurate CO$_2$ abundance for the Earth-like TRAPPIST-1 e, we found a broad distribution of abundances that may also be compatible with uninhabitable scenarios, suggesting that both transmission and direct imaging may provide ambiguous assessments of CO$_2$ for Earth-like planets.

Finally, while our transmission retrievals were not sensitive to near-surface water abundances, our direct imaging retrievals returned 35\% of the true surface water abundance for the cloudy Earth-twin with a 1-$\sigma$ distribution spanning abundances of 0.28~\textendash~0.32\%. Though the retrieved distribution is 65\% lower than the true surface abundance (consistent with the true atmosphere's abundance at $\sim$5 km or $\sim$${10^4}$ Pa, c.f. Figure \ref{fig:earth_hist}), the narrow range of abundances would confidently distinguish this planet from the few ppm expected above planet-wide clouds for a Venus-like environment \citep{chamberlain2020soir}. This result complements previous work that shows that reflectance spectra could also be sensitive to surface liquid water via ocean glint at visible and near-IR wavelengths \citep{williams2008detecting, robinson2010detecting, lustig2018detecting}. 

However, our retrieved distribution for water vapor will have been impacted by the presence of cloud extinction in the synthetic data that was excluded in the retrieval forward model. As described above in Section \ref{sec:clouds}, because the model must assume that water vapor is evenly-mixed in the atmosphere, it overestimates the water vapor abundance relative to the distribution of the cloud decks, but underestimates it relative to the surface. It is unclear whether the lower abundance we retrieve is due primarily to the significantly lower water vapor abundance at the cloud-tops, or the bias due to the lack of clouds in the retrieval. However, the inclusion of clouds in future retrievals would likely mitigate the bias towards lower water vapor abundances by allowing the model to correctly infer the near-surface water abundance from the width of the bands, and the cloud-top pressure from the band depths. We note that vertically parameterizing the water vapor abundance in the retrieval model \citep{damiano2022reflected} may further constrain this problem.

\subsubsection{Biosignature Pairs and Discriminants}

Below we describe in detail what we were able to conclude about the presence of biosignature gases and false positives in the planets we simulated, and the strengths and weaknesses of using either transmission or direct imaging to assess the presence of life in its environmental context.

\paragraph{Assessing biosignatures using transmission observations}

Our results suggest that a mid-IR transmission telescope with twice the sensitivity of JWST will be adept at detecting and placing constraints on the abundances of the CO$_2$/CH$_4$ biosignature pair for a self-consistent M dwarf Earth-like planet, although it may struggle to meaningfully constrain O$_3$. False positive discriminants may be present and may be at least, if not more detectable than the putative biosignatures, helping to discriminate between an abiotic or biological origin for these gases.  

\textbf{CO$_2$/CH$_4$}: 
Our results are consistent with previous work that shows carbon dioxide and methane may be a particularly accessible biosignature pair for transmission observations \citep[e.g.,][]{krissansen2018disequilibrium, lin2021differentiating, mikal2022detecting, rotman2022general, Meadows2023}. 

Using the detectability method described by \citet{lustig2019detectability}, we find that for observations ($t=40$ hours) of an Earth-like TRAPPIST-1 e with an \textit{Origins}-like telescope, CO$_2$ absorption bands can be detected with a confidence level of 11-$\sigma$ and CH$_4$ absorption bands can be detected at 9.4-$\sigma$. Together, these high-significance detections of the individual gases imply a strong detection of the pair. For the clear-sky, Earth-like TRAPPIST-1 e transmission case, we retrieve a broad CO$_2$ 1-$\sigma$ distribution with abundances spanning 1\% ~\textendash~37\%. For the same planet, we retrieve a CH$_4$ 1-$\sigma$ distribution spanning 43 ~\textendash~2303 ppm. When assessing this biosignature pair, the abundance of methane may provide additional evidence to discriminate between geological and biological sources of methane, by determining the inferred surface flux rate needed to maintain the observed abundance against photochemical destruction.  A theoretical upper limit for abiotic methane production can be derived from serpentinization, water interacting with ultramafic rock, which is the most efficient known means of producing abiotic methane \citep{arney2016pale}.   \citet{guzman2013abiotic} calculated that methane abundances exceeding the serpentinization upper limit of 2.5 ppm on an Earth-like planet around a G star require a biological origin. For a planet around an M dwarf, the abundance of methane from serpentinization may be photochemically enhanced by up to $\sim$3$\times$ \citep{domagal2014abiotic}, giving a theoretical upper limit of 7.5 ppm. 

We can therefore interpret our results assuming that higher abundances of CH$_4$ combined with low abundances of CO will increase the likelihood that the methane has a biological origin  \citep{krissansen2018disequilibrium,wogan2020abundant}. Our minimum retrieved CH$_4$ abundance at 1-$\sigma$ (43 ppm) is nearly 6$\times$ larger than the potential M-dwarf-enhanced serpentinization upper limit of 7.5 ppm described above. However, the minimum retrieved abundance at 2-$\sigma$ (8 ppm) roughly equals this serpentinization upper limit, and the minimum retrieved abundance at 3-$\sigma$ (2 ppm) is less than a third of this theoretical abiotic methane upper limit. This suggests that, although biogenic CH$_4$ is still likely, a conclusive ($>3\sigma$) interpretation of methane’s origin may be difficult without the simultaneous detection of CO$_2$ found here. Similarly, the relatively poor constraint on CO, which encompasses hundreds of ppm (0.4~\textendash~662.4 ppm at 1-$\sigma$) could suggest that we cannot rule out abiotic methane produced by volcanism from a reduced mantle \citep{krissansen2018disequilibrium,wogan2020abundant}.  Tighter constraints, or a stricter upper limit on CO, would be needed to conclusively discriminate between the abiotic and biological source hypotheses, barring the detection of the CO$_2$/CH$_4$ pair.

\textbf{O$_3$/CH$_4$}: Both O$_3$ and CH$_4$ were detected for the inhabited TRAPPIST-1 e, which, given the short photochemical lifetime of methane in an oxidizing atmosphere, implies a chemical disequilibrium produced by an active flux of methane \citep{hitchcock1967life}. The indication of an active methane flux on the Earth-like TRAPPIST-1 e planet implies a highly-reducing atmosphere, which decreases the likelihood that the detected O$_3$ was generated abiotically \citep{domagal2014abiotic}. Conversely, the fact that no H-bearing species were detected in the false positive O$_3$ atmosphere implies that this atmosphere may be in the ultra-oxidizing regime required to generate abundant abiotic O$_2$/O$_3$ \citep{gao2015stability}.  However, the broad O$_3$ distributions we retrieve for both the living and the dead planets overlap at 1-$\sigma$ (c.f., the O$_3$ panels in Figure \ref{fig:t1e_hist}), implying that the biological O$_3$ and the abiotic O$_3$ cannot be confidently distinguished without additional environmental context.

To strengthen the case for biologically-mediated CH$_4$ on the inhabited planet, we can use serpentinization as an upper bound on the abiotic production to assess whether the retrieved abundances are consistent with life. However, as described above, we find that our 3-$\sigma$ minimum retrieved CH$_4$ abundance for the inhabited TRAPPIST-1 e is smaller than the potential 7.5 ppm serpentinization upper limit estimated for M-dwarf stars, suggesting that the methane observed in the atmosphere could be consistent with abiogenesis. Likewise, the 3-$\sigma$ upper limit of 27 ppm we obtain for the CH$_4$ non-detection in the false positive O$_3$ TRAPPIST-1 e exceeds the theoretical serpentinization upper limit. Although an abiotic origin is weakly favored ($\sim$2-$\sigma$) by the non-detection, we cannot confidently conclude ($>$3-$\sigma$) that traces of methane do not still exist below our detection sensitivity. Future work can explore planets with differing abiotic methane fluxes to further assess the viability of constraining biological CH$_4$ abundances with transmission observations.

The inherent similarity between the Earth-like and false positive O$_3$ profiles in the transmission cases presents a challenge for accurate characterization despite relatively accurate retrievals. This implies that additional atmospheric context, such as CO and H$_2$O abundances, may be required to distinguish the living planet from the lifeless one when CH$_4$ is not present.  
 
Large abundances of CO have previously been proposed as an abiotic O$_3$ discriminant \citep{hu2012photochemistry, domagal2014abiotic, gao2015stability, schwieterman2018exoplanet, meadows2018exoplanet}. We find that the false positive O$_3$ planet is characterized by a CO abundance that is 20 to potentially 400$\times$ greater than the nominal CO abundance of the living planet. High CO abundances have previously been argued to be possible discriminants for abiotic production of O$_2$ via photolysis of CO$_2$ \citep{domagal2014abiotic, gao2015stability,schwieterman2016identifying}, or for abiotic CH$_4$ via volcanic outgassing from a more reducing mantle \citep{krissansen2018detectability}. The potentially very high CO abundances (1~\textendash~26\%) retrieved in the abiotic case make an abiotic source far more likely. Though previous work showed that living M dwarf planets with photosynthetic biospheres and both reducing or oxidizing atmospheres may also photochemically generate high abundances (100 ppm to several percent) of CO \citep{schwieterman2019rethinking}, this runaway effect has since been shown to be caused by a model artifact \citep{ranjan2023importance}.

Alternatively, strict upper limits on atmospheric water vapor may provide additional context to help identify O$_3$ produced from abiotic O$_2$, as extremely low water vapor abundances inhibit the recombination of photolyzed CO$_2$, leading to a buildup of abiotic  O$_2$, and O$_3$, in desiccated CO$_2$-dominated atmospheres \citep{gao2015stability}. 

While we showed above that the retrieved distribution of water vapor abundances for the Earth-like TRAPPIST-1 e would not allow us to confidently discriminate it from a Venus-like planet, the abiotic planet is extremely desiccated, with a 3-$\sigma$ upper limit of 13 ppm H$_2$O, suggesting that it would be easily distinguished from a self-consistent Earth-like planet around an M dwarf with stratospheric water vapor abundances on the order of hundreds of ppm. However, we note that the false positive O$_3$ atmosphere used in this study is based on the most desiccated case explored by \citet{gao2015stability}, and abiotic planets with higher water vapor abundances may be more difficult to discriminate from a living planet.

Finally, when considering O$_3$ as a biosignature in transmitted light, we rely on O$_3$ to act as a proxy for O$_2$, but it may be difficult to accurately determine the abundance of the source oxygen, and so whether it is more likely to have a biological or abiotic origin. 
The conversion is complicated because O$_2$ and O$_3$ abundances are not linearly related above O$_2$ abundances of $\sim$0.2\% \citep[][Fig. 2]{kastingozone}. Above this limit, a strong degeneracy between ozone and O$_2$ abundances means that the same O$_3$ column depth could be generated in atmospheres with O$_2$ abundances that span orders of magnitude \citep{kastingozone}, depending on stellar type \citep{segura2003ozone, kozakis2022ozone}.  The behavior implies that even highly precise O$_3$ column abundance constraints may yield imprecise O$_2$ constraints. In the linear regime, precise conversions to O$_2$ may still be possible, but the degenerate region spans O$_2$ abundances that are consistent with photosynthetic production \citep{kastingozone}, complicating biosignature interpretation. 
While gaining insight into the vertical structure of O$_3$ in the atmosphere may allow photochemical models to better constrain the abundance of the source O$_2$, and perhaps help to break the O$_2$-O$_3$ degeneracy, this is unlikely to be effective for transmission observations. By modeling the vertical O$_3$ structure with photochemistry, it may be possible to back out the abundance of the source O$_2$, and thereby deduce whether or not the O$_2$ is more or less likely to be biological. Unfortunately, we found that transmission is insensitive to the true vertical structure of O$_3$ for both the Earth-like and the desiccated planets in our experiments. 
Future transmission observations of atmospheres containing O$_3$ will require further lines of evidence to discriminate between biological and abiotic oxygen \citep{domagal2014abiotic}, such as constraints on the stellar UV flux \citep{segura2003ozone, kozakis2022ozone}, and a retrieval model that includes photochemistry.

\paragraph{Assessing biosignatures using direct imaging observations}

Given the caveats that we are using a cloud-free retrieval model to analyze cloudy simulated data, our study shows that a LUVOIR-like direct imaging mission will excel at constraining O$_2$ and O$_3$, but will face challenges constraining modern Earth-like CH$_4$ fluxes around a G-dwarf. For a full description of the biases that result in overestimated abundances of CO, CO$_2$, and CH$_4$ when wavelength-dependent clouds are included in the data but not in the retrieval, see Section \ref{sec:clouds}. Although hazes were not considered in our study, scattering slopes and absorption features from hazes would be expected to produce similar biases if present in the data but unaccounted for in the retrieval. We also find that direct imaging will have the advantage of providing precise constraints on O$_3$ in an oxygenic biosphere.  Direct imaging may even be weakly sensitive to the ozone's vertical structure via the UV Hartley band, potentially providing further context for an O$_2$ detection. Though we only find an upper limit on CH$_4$ for our Earth-twin, more informative methane distributions may be retrieved for planets with higher CH$_4$ abundances.  Higher abundances may be due to larger methane surface fluxes, as may be the case for Archean \citep{arney2016pale} and Proterozoic \citep{segura2003ozone, young2023inferring} Earth, or longer methane lifetimes in the planetary atmosphere, as may be the case for modern methane fluxes around K, rather than G dwarfs \citep{arney2019k}.  However, we also note that CH$_4$ may not have risen much above 10 ppm in the Proterozoic due to rapid oxidation of CH$_4$ in a sulfate rich ocean \cite{olson2016limited}. Abundant O$_2$ may also be a strong biosignature on its own, without the presence of CH$_4$ from a methanogenic biosphere, as long as false-positive mechanisms for the generation of abundant O$_2$ can be ruled out  \citep{meadows2017reflections}.

\textbf{O$_2$/CH$_4$:} 
Although our retrieval places tight constraints on O$_2$ for the Earth-like G dwarf planet, we do not detect CH$_4$ (1.86\% VMR 3-$\sigma$ upper limit), and arguments for the probability of the  biological origin of the observed O$_2$ must therefore be based on its abundance alone \citep{meadows2018exoplanet}. For the clear-sky Earth, we retrieve an accurate O$_2$ abundance of $21 \pm 5\%$ VMR with a significance of 11-$\sigma$. For our cloudy Earth-twin, we are able to constrain 60\% of the true abundance of O$_2$ with a 1-$\sigma$ distribution spanning abundances of 11~\textendash~15\% VMR. Regardless of whether clouds are present in the synthetic data, our results allow us to confidently rule out some ocean-loss false-positive atmospheres \citep{luger2015extreme, schaefer2016predictions} containing $>$30\% VMR O$_2$, as well as abiotic mechanisms that predict limited oxygen accumulation ($<$6\% VMR) \citep{walker1977evolution, selsis2002signature, segura2007abiotic, harman2015abiotic, meadows2017reflections}.

Though our study simulates LUVOIR A (B) observations with 10 (40) hours of exposure time, the LUVOIR final report simulates longer standard integration times of 100 (230) hours for the average target \citep[][their Figure 1-14]{luvoir2019luvoir}. However, even these more generous exposure times are not sufficient to detect CH$_4$ at a 5-$\sigma$ confidence level on a modern Earth twin at 10 pc. Using the detectability method described in \citet{lustig2019detectability}, we determine that absorption features from modern Earth methane abundances only become detectable at 5- (3-)$\sigma$ in 650 (240) hours with LUVOIR-A and approximately 2600 (960) hours with LUVOIR-B for the direct imaging target studied here.  For modern-Earth twins observed with direct imaging, confirmation of the biological origin for any detected O$_2$ may require ruling out the presence of potential false positive discriminants, like CO, rather than confirming the presence of CH$_4$. These results are consistent with \citet{young2023inferring}, who show that narrow bandpass observations with the Habitable Worlds Observatory (HWO) of a modern Earth at SNR=10 may only rule out higher Archean methane abundances.

While direct imaging may not be sensitive to modern Earth fluxes of CH$_4$ around a sun-like star, either larger surface fluxes of CH$_4$, or longer photochemical lifetimes, could result in higher and potentially more detectable abundances of CH$_4$. For example, geological data suggesting that Earth once had an organic haze \citep{zerkle2012bistable}, imply that the Archean required a  higher methane surface flux than the modern Earth to sustain the production of this haze \citep{arney2016pale}. Archean-like exoplanets \citep{arney2016pale} may also have higher fluxes \citep{arney2017pale}, as may Proterozoic exoplanets \citep{segura2003ozone, young2023inferring}.  Even for similar surface fluxes to modern Earth, the incident spectrum of the parent star reduces the destruction rate of methane, potentially allowing it to build up to abundances 100 times greater than modern-Earth levels around a K6V star \citep{arney2019k}. Promisingly, the planned HWO will likely target planets around K and M stars \citep{habworldstargets}.

Constraints on O$_3$ could also provide important additional context for interpreting the planetary environment and an oxygenic biosphere. We retrieve an O$_3$ abundance that is precise ($\pm 70$ ppb), significant (19-$\sigma$), and consistent within the range of O$_3$ VMRs in the true vertical structure (see Fig. \ref{fig:earth_o3_refl_comp}), but this abundance is  only about half the value of the peak abundance in the stratospheric bulge. When we included vertical O$_3$ structure in the retrieval, we achieved an improved fit, but with weak statistical evidence in favor of the more complex model (see Table \ref{tab:summ}). Since O$_3$ is a photochemical byproduct of O$_2$ \citep{ratner1972atmospheric}, accessing its vertical structure could help reveal indirect constraints on the atmospheric chemistry and potentially the biological productivity of the atmosphere through additional modeling steps. In addition, since direct imaging is particularly sensitive to total column abundance \citep{robinson2017characterizing}, we find that retrieving the vertical O$_3$ structure helps to mitigate bias that can occur when assuming an evenly mixed O$_3$ profile, and enable total column abundance inferences that are closer to the true value (c.f., Figure \ref{fig:o3_refl_col_abs}). 

Finally, we note that while an accurate O$_2$ abundance can rule out false positive processes that generate significantly larger and smaller abundances of abiotic O$_2$, for habitable-zone planets around G-dwarf stars, identifying false positive generation of Earth-like O$_2$ amounts ($\sim$20\% VMR) could be quite difficult.  For G dwarf planets, the most likely known abiotic mechanism for generating Earth-like abundances of oxygen is the photolysis of stratospheric H$_2$O in atmospheres with low abundances of non-condensable gases \citep{wordsworth2014abiotic}. To discriminate between abiotic and biological O$_2$ in these scenarios, we require constraints on the abundance of non-condensable gases in the planetary atmosphere \citep{meadows2018exoplanet}. However, for a nitrogen-dominated atmosphere, the N$_2$-N$_2$ collision-induced absorption feature occurs at longer wavelengths than those accessible to HWO \citep{schwieterman2015detecting, meadows2018exoplanet}, and will likely not be accessible in future direct imaging observations. Instead, it may be possible to constrain the bulk gas by fitting for the mean molecular weight of the atmosphere, and retrieving the abundances of other non-condensable, spectrally active gases \citep{lustig2020detection}.

\textbf{CO$_2$/CH$_4$:} 
The CO$_2$/CH$_4$ disequilibrium pair was first proposed as a biosignature for the Earth's Archean and for planets orbiting M dwarfs \citep{krissansen2018disequilibrium, krissansen2016detecting}, and we confirm that it is unlikely to be detectable for modern-Earth-like planets orbiting G dwarfs observed using direct imaging, due primarily to the difficulty in detecting modern-Earth-like quantities of CH$_4$. For our clear-sky modern Earth, we obtain a modest detection of CO$_2$, retrieving a 1-$\sigma$ distribution of CO$_2$ abundances spanning 17~\textendash~489 ppm in the clear-sky case, which we conservatively interpret as providing a 3-$\sigma$ upper limit of 91\% VMR. Similarly, we do not detect CH$_4$, obtaining only a 3-$\sigma$ upper limit of 1.86\% VMR. 

Furthermore, high abundances of CO may indicate that an abiotic volcanic source for CH$_4$ is more likely \citep{wogan2020abundant}. However, our retrievals lacked the sensitivity to rule out high CO abundances, as we retrieve a 1-$\sigma$ upper limit of 60 ppm on the abundance of CO for the clear-sky case, which is 600 times larger than our true value, and 600 times larger than the average CO abundance measured in the modern Earth's atmosphere \citep{badr1994carbon}. 
Consequently we confirm that even though the CO$_2$/CH$_4$ biosignature pair is a high priority biosignature for transmission observations of M-dwarf targets, and will likely be observable even for a planet with modern-Earth-like fluxes orbiting an M  \citep{segura2005biosignatures,Meadows2023} or K star \citep{arney2019k}, it will be challenging to detect for a modern Earth-Sun observed in reflected light.

\subsection{Model Limitations and Future Work}
\label{sec:modellimits}

To simplify our atmospheric retrieval model, we made assumptions related to atmospheric composition and vertical structure, and neglected spectral contamination effects, which may have impacted our results. These model limitations included assuming N$_2$ as the backfill gas for the atmospheric bulk composition, implementing an isothermal temperature profile, using a single set of absorption coefficients for calculating radiative transfer, excluding clouds from the forward model, and omitting spectral contamination effects from the synthetic data.  We discuss impacts and future work for each of these limitations below.

Our model adds N$_2$ as the bulk component of the retrieved atmospheric composition to achieve the inferred pressure of the terrestrial atmosphere in all cases (Equation \ref{eq:n2fill}). In atmospheres with  small amounts of trace gases, this results in the assumption that N$_2$ is the predominant bulk gas. The practice of backfilling a terrestrial atmosphere with N$_2$ is well-established in the retrieval community \citep[e.g.,][]{barstow2016habitable, krissansen2018detectability, changeat2019toward, barstow2020comparison} as N$_2$ is a common additional bulk gas in the Solar System's O$_2$ and CO$_2$-dominated terrestrial atmospheres, as it comprises more than 1\% of the atmosphere. 

However, assuming N$_2$ as the filling gas in our retrieval imposes a strong prior that N$_2$ is the bulk gas in the atmosphere due to our use of uniform log priors \citep[][their Figure 1]{benneke2012atmospheric}. For our Earth-like cases where N$_2$ is in fact the dominant gas in the atmosphere, this assumption of N$_2$ as the bulk composition does not bias our inferred abundances. However, for the abiotic false positive O$_3$ TRAPPIST-1 e scenario, we find that this strong prior on N$_2$ as the bulk gas likely causes us to underestimate the CO$_2$ abundance by a factor of $\sim$2 in this CO$_2$-dominated atmosphere. This assumption also causes us to underestimate the mean molecular weight of the atmosphere, since  N$_2$ (28 g/mol) is comparatively lighter than CO$_2$ (44 g/mol). This bias in the mean molecular weight enhances the inferred scale height of the atmosphere, and the retrieval compensates by decreasing the temperature of the atmosphere, and in some cases, the CO$_2$ abundance. The pitfalls of assuming a filling gas could be avoided by implementing centered-log ratio (CLR) priors, which allow any gas included in the retrieval to take on high (bulk gas) abundances at equal probability \citep[e.g.,][]{benneke2012atmospheric, damiano2021reflected}, or by directly retrieving on the atmospheric mean molecular weight \citep{tremblaydetectability2020}. These methods will be implemented in subsequent versions of our retrieval code. 

Across all of our cases, our retrieval forward model also assumed an isothermal temperature profile to fit data generated with a realistic profile (see Figure \ref{fig:atmcomp}). The use of an isothermal temperature profile did not impact our retrieval of temperature in reflected light, given our inability to constrain it from the spectrum at the spectral resolution investigated. This is likely a reasonable assumption going forward for studies of future direct imaging observations with comparable spectral resolution. In contrast, transmission spectra are more sensitive to temperature, given the dependence of spectral feature size on temperature via the atmospheric scale height \citep{seager2011exoplanets}. Our constraints on the retrieved temperature for the modern Earth-like planet are consistent with the stratospheric temperature, suggesting that assuming an isothermal temperature profile did not significantly impact our results. This may be an adequate approximation for stratospheres in Earth-like M-dwarf planet atmospheres, which do not show temperature inversions \citep{lincowski2018evolved}. Future work could investigate whether or not transmission spectroscopy provides insight to the temperature structure of secondary atmospheres by implementing a more complicated parameterization of temperature in the forward model \citep[e.g.,][]{madhusudhan2009temperature, macdonald2020so}.

To increase computational speed, our retrievals did not recalculate absorption coefficients during the retrieval exploration, relying instead on using a single set of absorption coefficients calculated from the realistic simulated atmosphere. While the optical depth in the radiative transfer translates changes in the abundance of each species into observables, the change in the intrinsic line shape and strength are fixed at the assumed line-by-line absorption coefficients throughout the retrieval. This assumption may have contributed to some biases in our CO$_2$ and H$_2$O retrieval results, but it does not affect our overall conclusions. 
    
To test the impact of this model assumption, we re-ran absorption coefficients for every state in the transmission retrieval.  These tests returned identical posteriors to those using the single set of absorption files generated with the input spectrum, with the exception of CO$_2$, when it was a bulk component of the atmosphere ($\geq$ 10\%). In this case, the CO$_2$ posterior differed by about 20\%, and provided better agreement with the true CO$_2$ abundance. The only significant effect of the single set of absorption coefficients in this test was a small underestimation of the CO$_2$ abundance (within 1-$\sigma$ at the simulated precision) in our abiotic false positive O$_3$ TRAPPIST-1 e case. We additionally tested transmission retrievals run with a single set of absorption coefficients for the input spectrum atmosphere, and retrievals ran with absorption coefficients generated at isothermal/evenly-mixed approximations of the true answer, and these also returned equivalent posteriors, given the finite sampling precision. 

Lastly, we quantify the impact of neglecting to recalculate absorption coefficients at each forward model instance by performing a perturbation sensitivity test. We take our best-fit retrieval parameters and perturb the abundances of the critical gases (CO$_2$ and CH$_4$ for transmission, and O$_2$ and H$_2$O for direct imaging). We then calculate the perturbed spectrum using the unperturbed best-fit absorption coefficients, and calculate it again using recalculated absorption coefficients consistent with the perturbed abundances. Finally, we compare the spectra generated at both sets of absorption coefficients within the simulated noise. We find that for \textit{Origins}-like transmission observations, when CH$_4$ and CO$_2$ abundances are perturbed by up to a factor of 10 without recalculating the absorption coefficients, we observe changes in the spectrum no larger than 1 ppm, which are insignificant ($<$ 0.5-$\sigma$) for the SNR associated with the $t=40$ hours exposure time explored here. For LUVOIR-A-like direct imaging observations, when oxygen and water abundance is doubled we observe changes to the spectrum on the order of 1/100th the planet-star contrast, which are insignificant given 10 hours of exposure time. However, when water abundance is enhanced by a factor of 10 without recalculating the absorption coefficients, we observe changes to the spectrum on the order of 1/10th the planet-star contrast, which are significant at 4-$\sigma$. This suggests that our retrieved water distributions with high-SNR LUVOIR observations may be weakly sensitive to our simplifying line shape assumptions.

Although these model simplifications have not significantly impacted our results, identifying an accurate and computationally efficient means to treat the rotational-vibrational absorption coefficients in terrestrial atmospheres remains important work on the path to performing biosignature assessment with future telescopes. Recent work has shown that differing opacity tables and assumptions may lead to biases in the results of thermal emission retrievals of Earth-twins \citep{Alei_2022} and JWST retrievals of Super-Earth and Warm-Jupiter atmospheres \citep{Niraula2022}. Similarly, the community has identified the growing need for additional laboratory measurements to better inform the opacities that support these critical exoplanet science applications \citep{fortney2019need}. 

Another limitation of our retrieval model is that it does not simulate the effect of clouds in the forward model, as discussed in more detail in Section \ref{sec:clouds}, which can result in overestimates of the surface albedo and retrieved gas abundances. Since the simulated observations had realistic water and ice clouds, which produce wavelength dependent absorption features in the observed spectrum, we find that the cloud-free model attempts to compensate for cloud scattering by enhancing the reflectivity of the planet surface and increasing the absorption of CH$_4$, CO, and CO$_2$, which have features in the near-IR where the ice clouds also absorb.  The cloud-free simplification consequently results in overestimates of the retrieved CH$_4$ (9$\times$), CO (5600$\times$), and CO$_2$ (5$\times$) abundances, compared to their true values,  as well as a large overestimate of the grey surface albedo (1.8$\times$) for direct imaging observations of the cloudy Earth-twin.  Other retrieval models have successfully included parameterized clouds for transmission \citep{benneke2012atmospheric} and direct imaging cases \citep{feng2018characterizing, damiano2022reflected}, and this capability will also be included in subsequent versions of our retrieval code.

Similar to cloud and condensation treatment in this study, while our simulated data implicitly includes the effect of photochemistry by virtue of our photochemically self-consistent atmospheres, the retrieval itself does not. Though this model simplification is currently the standard in the retrieval community, there is abundant work predicting the detectable effects of photochemistry on observations of exoplanet atmospheres \citep[e.g.,][]{segura2003ozone, segura2005biosignatures, hu2012photochemistry, miguel2013exploring, rugheimer2015effect, arney2017pale, meadows2017proxima, lincowski2018evolved, schwieterman2018exoplanet, arney2019k}. Furthermore, \citet{tsai2023photochemically} recently showed that separate photochemical forward models could be used to explain the retrieval of disequilibrium SO$_2$ in hot Jupiter atmospheres.  In this case, standard retrieval models without photochemistry detected SO$_2$, but separate photochemical modeling was required to explain its presence. For rocky planets, \citet{krissansen2018detectability} compared molecular abundance posteriors to surface fluxes as an example of connecting retrieval results to photochemistry. Notably, none of this recent work fully folds photochemistry into the retrieval forward model. We take a step in this direction with this study by attempting to vertically resolve the photochemically-mediated structure of the O$_3$ profile in an Earth-like atmosphere. Future work could explore methods for explicitly including photochemistry within a retrieval framework,  such as implementing photochemistry into the forward model or post-processing retrieval posteriors with a separate photochemistry retrieval.

Finally, we note that we have not included stellar and instrumental spectral contamination in this study, but previous work has shown that contamination sources may have a significant impact on our ability to characterize exoplanets in transmitted and reflected light. For example, in transmitted light, stellar variability and contamination due to unocculted star spots and faculae may produce a signal that is more than 10 times greater than the spectral transit depths anticipated for terrestrial exoplanets \citep{rackham2018transit}. Initial JWST results of TRAPPIST-1 and other M dwarfs suggest that the effects of stellar activity (e.g., transit light source effect, flares, etc.) need to be mitigated in order to access secondary atmospheres for rocky exoplanets transiting M dwarfs \citep{lim2023atmospheric, moran2023high, may2023double}. In this study, we assume this mitigation has been applied prior to our analysis, although work is ongoing to develop such methods. In reflected light, faint background stars may be a significant source of contamination for planets located at low galactic latitudes \citep{cracraft2021characterizing}, and scattered light may be a significant background and noise signal source for large aperture telescopes like HWO \citep{pfisterer2018role}. The impact of stellar activity on reflected light observations has not been extensively studied and must be the subject of future investigations. While outside the scope of this work, future model development is necessary to incorporate more realistic contamination sources into forward-looking retrieval investigations for both transmission and direct imaging observations.

\section{Conclusion}
\label{sec:conclusion}

In this study, we used a new terrestrial exoplanet retrieval model to compare the capabilities of transmission and direct imaging spectroscopy for habitability assessment and biosignature pair detection in rocky, habitable zone exoplanet atmospheres. 

Transmission spectroscopy with a 5.9-m \textit{Origins}-like telescope will not be able to probe the surface and near-surface environment of TRAPPIST-1 e, hampering habitability assessment, but will likely have the sensitivity to detect the CO$_2$/CH$_4$ biosignature pair and discriminate an extremely desiccated O$_3$ false positive case. By fitting a simplified temperature profile to the transmission data, for both the biotic and abiotic planets we find that the 1-$\sigma$ lower bound on the continuum pressure of $\sim$0.1-bars ($1\times10^4$ Pa) corresponds to altitudes tens of kilometers above the surface in the true atmosphere, but the high-pressure upper bound of $\sim$4 bars ($4\times10^5$ Pa) has no altitude counterpart in the true atmosphere. The broad pressure distributions we retrieved exemplify that the transmission continuum does not disambiguate a solid surface from atmospheric opacity. We find that the inferred isothermal temperature is most representative of the input temperature profile at levels corresponding to the stratosphere (50-70 km). We therefore cannot use surface temperature or near-surface water abundance to inform habitability assessment of the Earth-like TRAPPIST-1 e since transmission does not sufficiently probe these phenomena. Despite these limitations for characterizing the habitability of the self-consistent Earth-like TRAPPIST-1 e, we find that a mid-IR transmission telescope with twice the sensitivity of JWST can detect CO$_2$ absorption features at 11-$\sigma$ and CH$_4$ absorption features at 9.4-$\sigma$ in 40 hours of exposure, providing a high-confidence detection of the biosignature pair. Finally, transmission probes broad distributions of O$_3$, which overlap for the biotic and abiotic planets and thus cannot be used to discriminate between a photosynthetic biosphere and a CO$_2$-dominated abiotic atmosphere. However, in the context of an O$_3$ detection, we find that the non-detection of water vapor, with a 3-$\sigma$ upper limit of 13 ppm, coupled to the detection of abundant CO (1-$\sigma$ range of 1-26\%), could be used to distinguish between O$_3$ sourced from CO$_2$ photolysis and biogenic O$_3$ sourced from abundant O$_2$.

For future direct imaging missions with a 15-m LUVOIR-A like telescope, we cannot constrain surface temperature at the spectral resolution explored here, and although our model was relatively sensitive to planetary albedo, this parameter is degenerate with planetary radius. While insensitivity to temperature and the size/albedo degeneracy complicate our ability to assess habitability, our model was able to constrain the near-surface water abundance with 50\% or 0\% cloud coverage in the simulated data to within 0.1\% of the true value at high significance ($>$10-$\sigma$) even for our assumed exposure time of only 10 hours. We detect O$_2$ and O$_3$ with high detection significances of 11-$\sigma$ and 19-$\sigma$, respectively. We also precisely constrain their abundances with (1-$\sigma$) errors of $\pm$5\% VMR for O$_2$ and $\pm$70 ppb VMR for O$_3$. However, we fail to detect CO$_2$ and CH$_4$, retrieving 3-$\sigma$ upper limits on molecular abundances of 91\% and 1.83\%, respectively, on the modern Earth-twin.  Given our finding that short ($\sim$10 hour) exposure times with a 15-m direct imaging telescope are insensitive to modern Earth-like CH$_4$ abundances on planets around G stars, the CO$_2$/CH$_4$ and O$_2$/CH$_4$ biosignature pairs are difficult to assess for these planetary systems. However, the predicted photochemical enhancement of CH$_4$ on M- \citep{segura2005biosignatures, rugheimer2015effect} and K-dwarf \citep{arney2019k} planets may make the retrieval of the O$_2$/CH$_4$ pair more feasible in these systems, and HWO will target 3 M dwarfs and 27 K dwarfs as part of its planned target list for discovering and characterizing Earth-like planets \citep{luvoir2019luvoir, habworldstargets}. For exposure times of up to 10 ($\sim$40) hours with a 15-m LUVOIR-A-like (6.7-m LUVOIR-B-like) telescope of a planet at 10 parsecs, we find that direct imaging is weakly sensitive to the vertical structure of the stratospheric ozone bulge via the shape of the Hartley band (0.2-0.3 $\mu$m) when comparing an evenly-mixed and vertically-resolved retrieval model {(Bayes factor $\sim$$1.3$)}. Finally, when clouds are present in the observed spectrum, we see evidence for them in the large residuals at the bottoms of the water-bands (S/N up to 6.5) and from NIR ice absorption features, which may be used to uniquely identify the presence of clouds in the reflected light spectra of Earth-like planets even when not explicitly included in the retrieval model. The profound science objectives of future exoplanet missions---to constrain habitable and inhabited environments---will require further retrieval model developments to fully assess in advance.

\software{Astropy \citep{Astropy2013, Astropy2018, price2022astropy}, coronagraph \citep{lustig2019coronagraph}, LBLABC \citep{meadows1996ground}, Matplotlib \citep{Hunter2007}, MultiNest \citep{feroz2008multimodal, feroz2009multinest}, NumPy \citep{Walt2011, NumPy2020}, SciPy \citep{Virtanen2019scipy, SciPy2020}, SMART \citep{meadows1996ground}, SMARTER \citep{lustigyaeger2023earth}}

\begin{acknowledgments}

This work was performed by the Virtual Planetary Laboratory Team, which is a member of the NASA Nexus for Exoplanet System Science, and funded via NASA Astrobiology Program Grant 80NSSC18K0829. J.L.Y acknowledges support from JHU APL’s Independent Research And Development program. This work made use of the advanced computational, storage, and networking infrastructure provided by the Hyak supercomputer system at the University of Washington. We thank two anonymous reviewers for their thorough reviews of our manuscript and thoughtful comments that helped to improve the quality, clarity, and reproducibility of our work. 

\end{acknowledgments} 

\newpage
\bibliography{retrieval}{}

\begin{thebibliography}{}
\expandafter\ifx\csname natexlab\endcsname\relax\def\natexlab#1{#1}\fi
\providecommand{\url}[1]{\href{#1}{#1}}
\providecommand{\dodoi}[1]{doi:~\href{http://doi.org/#1}{\nolinkurl{#1}}}
\providecommand{\doeprint}[1]{\href{http://ascl.net/#1}{\nolinkurl{http://ascl.net/#1}}}
\providecommand{\doarXiv}[1]{\href{https://arxiv.org/abs/#1}{\nolinkurl{https://arxiv.org/abs/#1}}}

\bibitem[{Aitchison(1982)}]{aitchison1982statistical}
Aitchison, J. 1982, Journal of the Royal Statistical Society: Series B (Methodological), 44, 139

\bibitem[{Alei {et~al.}(2022)Alei, Konrad, Molli{\`{e}}re, Quanz, Angerhausen, \& Ranganathan}]{Alei_2022}
Alei, E., Konrad, B., Molli{\`{e}}re, P., {et~al.} 2022, in Space Telescopes and Instrumentation 2022: Optical, Infrared, and Millimeter Wave, ed. L.~E. Coyle, M.~D. Perrin, \& S.~Matsuura ({SPIE}), \dodoi{10.1117/12.2631692}

\bibitem[{Anglada-Escud{\'e} {et~al.}(2016)Anglada-Escud{\'e}, Amado, Barnes, Berdi{\~n}as, Butler, Coleman, de~La~Cueva, Dreizler, Endl, Giesers, {et~al.}}]{anglada2016terrestrial}
Anglada-Escud{\'e}, G., Amado, P.~J., Barnes, J., {et~al.} 2016, Nature, 536, 437

\bibitem[{Arney {et~al.}(2018)Arney, Domagal-Goldman, \& Meadows}]{arney2018organic}
Arney, G., Domagal-Goldman, S.~D., \& Meadows, V.~S. 2018, Astrobiology, 18, 311

\bibitem[{{Arney} {et~al.}(2014){Arney}, {Meadows}, {Crisp}, {Schmidt}, {Bailey}, \& {Robinson}}]{Arney2014}
{Arney}, G., {Meadows}, V., {Crisp}, D., {et~al.} 2014, Journal of Geophysical Research (Planets), 119, 1860, \dodoi{10.1002/2014JE004662}

\bibitem[{Arney {et~al.}(2016)Arney, Domagal-Goldman, Meadows, Wolf, Schwieterman, Charnay, Claire, H{\'e}brard, \& Trainer}]{arney2016pale}
Arney, G., Domagal-Goldman, S.~D., Meadows, V.~S., {et~al.} 2016, Astrobiology, 16, 873

\bibitem[{Arney(2019)}]{arney2019k}
Arney, G.~N. 2019, The Astrophysical Journal Letters, 873, L7

\bibitem[{Arney {et~al.}(2017)Arney, Meadows, Domagal-Goldman, Deming, Robinson, Tovar, Wolf, \& Schwieterman}]{arney2017pale}
Arney, G.~N., Meadows, V.~S., Domagal-Goldman, S.~D., {et~al.} 2017, The Astrophysical Journal, 836, 49

\bibitem[{{Astropy Collaboration} {et~al.}(2013){Astropy Collaboration}, {Robitaille}, {Tollerud}, {Greenfield}, {Droettboom}, {Bray}, {Aldcroft}, {Davis}, {Ginsburg}, {Price-Whelan}, {Kerzendorf}, {Conley}, {Crighton}, {Barbary}, {Muna}, {Ferguson}, {Grollier}, {Parikh}, {Nair}, {Unther}, {Deil}, {Woillez}, {Conseil}, {Kramer}, {Turner}, {Singer}, {Fox}, {Weaver}, {Zabalza}, {Edwards}, {Azalee Bostroem}, {Burke}, {Casey}, {Crawford}, {Dencheva}, {Ely}, {Jenness}, {Labrie}, {Lim}, {Pierfederici}, {Pontzen}, {Ptak}, {Refsdal}, {Servillat}, \& {Streicher}}]{Astropy2013}
{Astropy Collaboration}, {Robitaille}, T.~P., {Tollerud}, E.~J., {et~al.} 2013, \aap, 558, A33, \dodoi{10.1051/0004-6361/201322068}

\bibitem[{Badr \& Probert(1994)}]{badr1994carbon}
Badr, O., \& Probert, S. 1994, Applied Energy, 49, 99

\bibitem[{Barstow {et~al.}(2020)Barstow, Changeat, Garland, Line, Rocchetto, \& Waldmann}]{barstow2020comparison}
Barstow, J.~K., Changeat, Q., Garland, R., {et~al.} 2020, Monthly Notices of the Royal Astronomical Society, 493, 4884

\bibitem[{Barstow \& Irwin(2016)}]{barstow2016habitable}
Barstow, J.~K., \& Irwin, P.~G. 2016, Monthly Notices of the Royal Astronomical Society: Letters, 461, L92

\bibitem[{Batalha {et~al.}(2018)Batalha, Lewis, Line, Valenti, \& Stevenson}]{batalha2018strategies}
Batalha, N.~E., Lewis, N.~K., Line, M.~R., Valenti, J., \& Stevenson, K. 2018, The Astrophysical Journal Letters, 856, L34

\bibitem[{Benneke \& Seager(2012)}]{benneke2012atmospheric}
Benneke, B., \& Seager, S. 2012, The Astrophysical Journal, 753, 100

\bibitem[{Benneke \& Seager(2013)}]{benneke2013distinguish}
---. 2013, The Astrophysical Journal, 778, 153

\bibitem[{Benneke {et~al.}(2019)Benneke, Wong, Piaulet, Knutson, Lothringer, Morley, Crossfield, Gao, Greene, Dressing, {et~al.}}]{benneke2019water}
Benneke, B., Wong, I., Piaulet, C., {et~al.} 2019, The Astrophysical Journal Letters, 887, L14

\bibitem[{B{\'e}tr{\'e}mieux \& Kaltenegger(2014)}]{betremieux2014impact}
B{\'e}tr{\'e}mieux, Y., \& Kaltenegger, L. 2014, The Astrophysical Journal, 791, 7

\bibitem[{Bolcar {et~al.}(2016)Bolcar, Feinberg, France, Rauscher, Redding, \& Schiminovich}]{bolcar2016initial}
Bolcar, M.~R., Feinberg, L., France, K., {et~al.} 2016, in Space telescopes and instrumentation 2016: Optical, infrared, and millimeter wave, Vol. 9904, SPIE, 192--203

\bibitem[{Buchner {et~al.}(2014)Buchner, Georgakakis, Nandra, Hsu, Rangel, Brightman, Merloni, Salvato, Donley, \& Kocevski}]{buchner2014x}
Buchner, J., Georgakakis, A., Nandra, K., {et~al.} 2014, Astronomy \& Astrophysics, 564, A125

\bibitem[{Caldas {et~al.}(2019)Caldas, Leconte, Selsis, Waldmann, Bord{\'e}, Rocchetto, \& Charnay}]{caldas2019effects}
Caldas, A., Leconte, J., Selsis, F., {et~al.} 2019, Astronomy \& Astrophysics, 623, A161

\bibitem[{{Carri{\'o}n-Gonz{\'a}lez} {et~al.}(2020){Carri{\'o}n-Gonz{\'a}lez}, {Garc{\'\i}a Mu{\~n}oz}, {Cabrera}, {Csizmadia}, {Santos}, \& {Rauer}}]{Carrion-Gonzalez2020}
{Carri{\'o}n-Gonz{\'a}lez}, {\'O}., {Garc{\'\i}a Mu{\~n}oz}, A., {Cabrera}, J., {et~al.} 2020, \aap, 640, A136, \dodoi{10.1051/0004-6361/202038101}

\bibitem[{Catling {et~al.}(2018)Catling, Kiang~Nancy, Robinson~Tyler, Rushby~Andrew, Genio~Anthony, {et~al.}}]{catlingdavid2018exoplanet}
Catling, D.~C., Kiang~Nancy, Y., Robinson~Tyler, D., {et~al.} 2018, Astrobiology

\bibitem[{Chamberlain {et~al.}(2020)Chamberlain, Mahieux, Robert, Piccialli, Trompet, Vandaele, \& Wilquet}]{chamberlain2020soir}
Chamberlain, S., Mahieux, A., Robert, S., {et~al.} 2020, Icarus, 346, 113819

\bibitem[{Changeat {et~al.}(2019)Changeat, Edwards, Waldmann, \& Tinetti}]{changeat2019toward}
Changeat, Q., Edwards, B., Waldmann, I., \& Tinetti, G. 2019, The Astrophysical Journal, 886, 39

\bibitem[{Charnay {et~al.}(2015)Charnay, Meadows, Misra, Leconte, \& Arney}]{charnay20153d}
Charnay, B., Meadows, V., Misra, A., Leconte, J., \& Arney, G. 2015, The Astrophysical Journal Letters, 813, L1

\bibitem[{Cracraft {et~al.}(2021)Cracraft, De~Rosa, Sparks, Bailey, \& Turnbull}]{cracraft2021characterizing}
Cracraft, M., De~Rosa, R., Sparks, W., Bailey, V., \& Turnbull, M. 2021, arXiv preprint arXiv:2110.08097

\bibitem[{{Crisp} \& {Titov}(1997)}]{Crisp1997}
{Crisp}, D., \& {Titov}, D. 1997, in Venus II: Geology, Geophysics, Atmosphere, and Solar Wind Environment, ed. S.~W. {Bougher}, D.~M. {Hunten}, \& R.~J. {Phillips}, 353

\bibitem[{Damiano \& Hu(2021)}]{damiano2021reflected}
Damiano, M., \& Hu, R. 2021, The Astronomical Journal, 162, 200

\bibitem[{Damiano \& Hu(2022)}]{damiano2022reflected}
---. 2022, The Astronomical Journal, 163, 299

\bibitem[{Davies {et~al.}(2009)Davies, Wright, \& Glasse}]{davies2009solar}
Davies, J.~K., Wright, G.~S., \& Glasse, A.~C. 2009, Earth, Moon, and Planets, 105, 73

\bibitem[{Des~Marais {et~al.}(2002)Des~Marais, Harwit, Jucks, Kasting, Lin, Lunine, Schneider, Seager, Traub, \& Woolf}]{des2002remote}
Des~Marais, D.~J., Harwit, M.~O., Jucks, K.~W., {et~al.} 2002, Astrobiology, 2, 153

\bibitem[{Dittmann {et~al.}(2017)Dittmann, Irwin, Charbonneau, Bonfils, Astudillo-Defru, Haywood, Berta-Thompson, Newton, Rodriguez, Winters, {et~al.}}]{dittmann2017temperate}
Dittmann, J.~A., Irwin, J.~M., Charbonneau, D., {et~al.} 2017, Nature, 544, 333

\bibitem[{Domagal-Goldman {et~al.}(2014)Domagal-Goldman, Segura, Claire, Robinson, \& Meadows}]{domagal2014abiotic}
Domagal-Goldman, S.~D., Segura, A., Claire, M.~W., Robinson, T.~D., \& Meadows, V.~S. 2014, The Astrophysical Journal, 792, 90

\bibitem[{Fauchez {et~al.}(2019)Fauchez, Turbet, Villanueva, Wolf, Arney, Kopparapu, Lincowski, Mandell, de~Wit, Pidhorodetska, {et~al.}}]{fauchez2019impact}
Fauchez, T.~J., Turbet, M., Villanueva, G.~L., {et~al.} 2019, The Astrophysical Journal, 887, 194

\bibitem[{Feng {et~al.}(2020)Feng, Line, \& Fortney}]{feng20202d}
Feng, Y.~K., Line, M.~R., \& Fortney, J.~J. 2020, The Astronomical Journal, 160, 137

\bibitem[{{Feng} {et~al.}(2016){Feng}, {Line}, {Fortney}, {Stevenson}, {Bean}, {Kreidberg}, \& {Parmentier}}]{Feng2016nonuniform}
{Feng}, Y.~K., {Line}, M.~R., {Fortney}, J.~J., {et~al.} 2016, \apj, 829, 52, \dodoi{10.3847/0004-637X/829/1/52}

\bibitem[{Feng {et~al.}(2018)Feng, Robinson, Fortney, Lupu, Marley, Lewis, Macintosh, \& Line}]{feng2018characterizing}
Feng, Y.~K., Robinson, T.~D., Fortney, J.~J., {et~al.} 2018, The Astronomical Journal, 155, 200

\bibitem[{Feroz {et~al.}(2009)Feroz, Hobson, \& Bridges}]{feroz2009multinest}
Feroz, F., Hobson, M., \& Bridges, M. 2009, Monthly Notices of the Royal Astronomical Society, 398, 1601

\bibitem[{Feroz \& Hobson(2008)}]{feroz2008multimodal}
Feroz, F., \& Hobson, M.~P. 2008, Monthly Notices of the Royal Astronomical Society, 384, 449

\bibitem[{Feroz {et~al.}(2013)Feroz, Hobson, Cameron, \& Pettitt}]{feroz2013importance}
Feroz, F., Hobson, M.~P., Cameron, E., \& Pettitt, A.~N. 2013, arXiv preprint arXiv:1306.2144

\bibitem[{Fortney {et~al.}(2019)Fortney, Robinson, Domagal-Goldman, del Genio, Gordon, Gharib-Nezhad, Lewis, Sousa-Silva, Airapetian, Drouin, {et~al.}}]{fortney2019need}
Fortney, J., Robinson, T., Domagal-Goldman, S., {et~al.} 2019, Astro2020: Decadal Survey on Astronomy and Astrophysics, 2020, 146

\bibitem[{Fortney(2005)}]{fortney2005effect}
Fortney, J.~J. 2005, Monthly Notices of the Royal Astronomical Society, 364, 649

\bibitem[{Fujii {et~al.}(2017)Fujii, Lustig-Yaeger, \& Cowan}]{fujii2017rotational}
Fujii, Y., Lustig-Yaeger, J., \& Cowan, N.~B. 2017, The Astronomical Journal, 154, 189

\bibitem[{Gandhi \& Madhusudhan(2018)}]{gandhi2018retrieval}
Gandhi, S., \& Madhusudhan, N. 2018, Monthly Notices of the Royal Astronomical Society, 474, 271

\bibitem[{Gao {et~al.}(2015)Gao, Hu, Robinson, Li, \& Yung}]{gao2015stability}
Gao, P., Hu, R., Robinson, T.~D., Li, C., \& Yung, Y.~L. 2015, The Astrophysical Journal, 806, 249

\bibitem[{Gaudi {et~al.}(2020)Gaudi, Seager, Mennesson, Kiessling, Warfield, Cahoy, Clarke, Domagal-Goldman, Feinberg, Guyon, {et~al.}}]{gaudi2020habitable}
Gaudi, B.~S., Seager, S., Mennesson, B., {et~al.} 2020, arXiv preprint arXiv:2001.06683

\bibitem[{Gillon {et~al.}(2016)Gillon, Jehin, Lederer, Delrez, de~Wit, Burdanov, Van~Grootel, Burgasser, Triaud, Opitom, {et~al.}}]{gillon2016temperate}
Gillon, M., Jehin, E., Lederer, S.~M., {et~al.} 2016, Nature, 533, 221

\bibitem[{Gillon {et~al.}(2017)Gillon, Triaud, Demory, Jehin, Agol, Deck, Lederer, de~Wit, Burdanov, Ingalls, {et~al.}}]{gillon2017seven}
Gillon, M., Triaud, A.~H., Demory, B.-O., {et~al.} 2017, Nature, 542, 456

\bibitem[{Goldblatt {et~al.}(2013)Goldblatt, Robinson, Zahnle, \& Crisp}]{goldblatt2013low}
Goldblatt, C., Robinson, T.~D., Zahnle, K.~J., \& Crisp, D. 2013, Nature Geoscience, 6, 661

\bibitem[{Guimond \& Cowan(2018)}]{guimond2018direct}
Guimond, C.~M., \& Cowan, N.~B. 2018, The Astronomical Journal, 155, 230

\bibitem[{Guyon {et~al.}(2012)Guyon, Martinache, Cady, Belikov, Balasubramanian, Wilson, Clergeon, \& Mateen}]{guyon2012elts}
Guyon, O., Martinache, F., Cady, E.~J., {et~al.} 2012, in Adaptive Optics Systems III, Vol. 8447, International Society for Optics and Photonics, 84471X

\bibitem[{Guzm{\'a}n-Marmolejo {et~al.}(2013)Guzm{\'a}n-Marmolejo, Segura, \& Escobar-Briones}]{guzman2013abiotic}
Guzm{\'a}n-Marmolejo, A., Segura, A., \& Escobar-Briones, E. 2013, Astrobiology, 13, 550

\bibitem[{Harman {et~al.}(2015)Harman, Schwieterman, Schottelkotte, \& Kasting}]{harman2015abiotic}
Harman, C., Schwieterman, E., Schottelkotte, J.~C., \& Kasting, J. 2015, The Astrophysical Journal, 812, 137

\bibitem[{Harris {et~al.}(2020)Harris, Millman, van~der Walt, Gommers, Virtanen, Cournapeau, Wieser, Taylor, Berg, Smith, Kern, Picus, Hoyer, van Kerkwijk, Brett, Haldane, Fernández~del Río, Wiebe, Peterson, Gérard-Marchant, Sheppard, Reddy, Weckesser, Abbasi, Gohlke, \& Oliphant}]{NumPy2020}
Harris, C.~R., Millman, K.~J., van~der Walt, S.~J., {et~al.} 2020, Nature, 585, 357–362, \dodoi{10.1038/s41586-020-2649-2}

\bibitem[{Harrison {et~al.}(2021)Harrison, Kennicutt, Dalcanton, de~Zeeuw, Driesman, Fortney, Gonzalez, Goodman, Kamionkowski, Macintosh, {et~al.}}]{national2021pathways}
Harrison, F., Kennicutt, R., Dalcanton, J., {et~al.} 2021, Pathways to discovery in astronomy and astrophysics for the 2020s

\bibitem[{Hitchcock \& Lovelock(1967)}]{hitchcock1967life}
Hitchcock, D.~R., \& Lovelock, J.~E. 1967, Icarus, 7, 149

\bibitem[{Hu {et~al.}(2012)Hu, Seager, \& Bains}]{hu2012photochemistry}
Hu, R., Seager, S., \& Bains, W. 2012, The Astrophysical Journal, 761, 166

\bibitem[{Hunter(2007)}]{Hunter2007}
Hunter, J.~D. 2007, Computing In Science \& Engineering, 9, 90, \dodoi{10.1109/MCSE.2007.55}

\bibitem[{{Irwin} {et~al.}(2008){Irwin}, {Teanby}, {de Kok}, {Fletcher}, {Howett}, {Tsang}, {Wilson}, {Calcutt}, {Nixon}, \& {Parrish}}]{Irwin2008}
{Irwin}, P.~G.~J., {Teanby}, N.~A., {de Kok}, R., {et~al.} 2008, \jqsrt, 109, 1136, \dodoi{10.1016/j.jqsrt.2007.11.006}

\bibitem[{Jeffreys(1998)}]{jeffreys1998theory}
Jeffreys, H. 1998, The theory of probability (OUP Oxford)

\bibitem[{{Kaltenegger} {et~al.}(2020){Kaltenegger}, {MacDonald}, {Kozakis}, {Lewis}, {Mamajek}, {McDowell}, \& {Vanderburg}}]{kalteneggerWDs2020}
{Kaltenegger}, L., {MacDonald}, R.~J., {Kozakis}, T., {et~al.} 2020, \apjl, 901, L1, \dodoi{10.3847/2041-8213/aba9d3}

\bibitem[{{Kasting} \& {Donahue}(1980)}]{kastingozone}
{Kasting}, J.~F., \& {Donahue}, T.~M. 1980, \jgr, 85, 3255, \dodoi{10.1029/JC085iC06p03255}

\bibitem[{Kim {et~al.}(2011)Kim, Jung, Sim, Courtin, Bellucci, Sicardy, Song, \& Minh}]{kim2011retrieval}
Kim, S.~J., Jung, A., Sim, C., {et~al.} 2011, Planetary and Space Science, 59, 699

\bibitem[{{Kleinb{\"o}hl} {et~al.}(2009){Kleinb{\"o}hl}, {Schofield}, {Kass}, {Abdou}, {Backus}, {Sen}, {Shirley}, {Lawson}, {Richardson}, {Taylor}, {Teanby}, \& {McCleese}}]{Kleinb2009}
{Kleinb{\"o}hl}, A., {Schofield}, J.~T., {Kass}, D.~M., {et~al.} 2009, Journal of Geophysical Research (Planets), 114, E10006, \dodoi{10.1029/2009JE003358}

\bibitem[{Komacek {et~al.}(2020)Komacek, Fauchez, Wolf, \& Abbot}]{komacek2020clouds}
Komacek, T.~D., Fauchez, T.~J., Wolf, E.~T., \& Abbot, D.~S. 2020, The Astrophysical Journal Letters, 888, L20

\bibitem[{Kopparapu {et~al.}(2021)Kopparapu, Arney, Haqq-Misra, Lustig-Yaeger, \& Villanueva}]{kopparapu2021nitrogen}
Kopparapu, R., Arney, G., Haqq-Misra, J., Lustig-Yaeger, J., \& Villanueva, G. 2021, The Astrophysical Journal, 908, 164

\bibitem[{Kozakis {et~al.}(2022)Kozakis, Mendon{\c{c}}a, \& Buchhave}]{kozakis2022ozone}
Kozakis, T., Mendon{\c{c}}a, J.~M., \& Buchhave, L.~A. 2022, Bulletin of the American Astronomical Society, 54, 102

\bibitem[{Kreidberg {et~al.}(2014)Kreidberg, Bean, D{\'e}sert, Benneke, Deming, Stevenson, Seager, Berta-Thompson, Seifahrt, \& Homeier}]{kreidberg2014clouds}
Kreidberg, L., Bean, J.~L., D{\'e}sert, J.-M., {et~al.} 2014, Nature, 505, 69

\bibitem[{Kreidberg {et~al.}(2015)Kreidberg, Line, Bean, Stevenson, D{\'e}sert, Madhusudhan, Fortney, Barstow, Henry, Williamson, {et~al.}}]{kreidberg2015detection}
Kreidberg, L., Line, M.~R., Bean, J.~L., {et~al.} 2015, The Astrophysical Journal, 814, 66

\bibitem[{Krissansen-Totton {et~al.}(2016)Krissansen-Totton, Bergsman, \& Catling}]{krissansen2016detecting}
Krissansen-Totton, J., Bergsman, D.~S., \& Catling, D.~C. 2016, Astrobiology, 16, 39

\bibitem[{Krissansen-Totton {et~al.}(2018{\natexlab{a}})Krissansen-Totton, Garland, Irwin, \& Catling}]{krissansen2018detectability}
Krissansen-Totton, J., Garland, R., Irwin, P., \& Catling, D.~C. 2018{\natexlab{a}}, The Astronomical Journal, 156, 114

\bibitem[{Krissansen-Totton {et~al.}(2018{\natexlab{b}})Krissansen-Totton, Olson, \& Catling}]{krissansen2018disequilibrium}
Krissansen-Totton, J., Olson, S., \& Catling, D.~C. 2018{\natexlab{b}}, Science advances, 4, eaao5747

\bibitem[{Lavie {et~al.}(2017)Lavie, Mendon{\c{c}}a, Mordasini, Malik, Bonnefoy, Demory, Oreshenko, Grimm, Ehrenreich, \& Heng}]{lavie2017helios}
Lavie, B., Mendon{\c{c}}a, J.~M., Mordasini, C., {et~al.} 2017, The Astronomical Journal, 154, 91

\bibitem[{Lim {et~al.}(2023)Lim, Benneke, Doyon, MacDonald, Piaulet, Artigau, Coulombe, Radica, L’Heureux, Albert, {et~al.}}]{lim2023atmospheric}
Lim, O., Benneke, B., Doyon, R., {et~al.} 2023, The Astrophysical Journal Letters, 955, L22

\bibitem[{{Lin} {et~al.}(2021){Lin}, {MacDonald}, {Kaltenegger}, \& {Wilson}}]{zifanabiotic2021}
{Lin}, Z., {MacDonald}, R.~J., {Kaltenegger}, L., \& {Wilson}, D.~J. 2021, \mnras, 505, 3562, \dodoi{10.1093/mnras/stab1486}

\bibitem[{Lin {et~al.}(2021)Lin, MacDonald, Kaltenegger, \& Wilson}]{lin2021differentiating}
Lin, Z., MacDonald, R.~J., Kaltenegger, L., \& Wilson, D.~J. 2021, Monthly Notices of the Royal Astronomical Society, 505, 3562

\bibitem[{Lincowski {et~al.}(2019)Lincowski, Lustig-Yaeger, \& Meadows}]{lincowski2019observing}
Lincowski, A.~P., Lustig-Yaeger, J., \& Meadows, V.~S. 2019, The Astronomical Journal, 158, 26

\bibitem[{Lincowski {et~al.}(2018)Lincowski, Meadows, Crisp, Robinson, Luger, Lustig-Yaeger, \& Arney}]{lincowski2018evolved}
Lincowski, A.~P., Meadows, V.~S., Crisp, D., {et~al.} 2018, The Astrophysical Journal, 867, 76

\bibitem[{{Line} \& {Parmentier}(2016)}]{Line2016}
{Line}, M.~R., \& {Parmentier}, V. 2016, \apj, 820, 78, \dodoi{10.3847/0004-637X/820/1/78}

\bibitem[{Line {et~al.}(2021)Line, Brogi, Bean, Gandhi, Zalesky, Parmentier, Smith, Mace, Mansfield, Kempton, {et~al.}}]{line2021solar}
Line, M.~R., Brogi, M., Bean, J.~L., {et~al.} 2021, Nature, 598, 580

\bibitem[{Luger \& Barnes(2015)}]{luger2015extreme}
Luger, R., \& Barnes, R. 2015, Astrobiology, 15, 119

\bibitem[{Lustig-Yaeger(2020)}]{lustig2020detection}
Lustig-Yaeger, J. 2020, The Detection, Characterization, and Retrieval of Terrestrial Exoplanet Atmospheres (University of Washington)

\bibitem[{Lustig-Yaeger {et~al.}(2023)Lustig-Yaeger, Meadows, Crisp, Line, \& Robinson}]{lustigyaeger2023earth}
Lustig-Yaeger, J., Meadows, V.~S., Crisp, D., Line, M.~R., \& Robinson, T.~D. 2023, The Planetary Science Journal, 4, 170

\bibitem[{Lustig-Yaeger {et~al.}(2019{\natexlab{a}})Lustig-Yaeger, Meadows, \& Lincowski}]{lustig2019detectability}
Lustig-Yaeger, J., Meadows, V.~S., \& Lincowski, A.~P. 2019{\natexlab{a}}, The Astronomical Journal, 158, 27

\bibitem[{Lustig-Yaeger {et~al.}(2019{\natexlab{b}})Lustig-Yaeger, Meadows, \& Lincowski}]{lustig2019mirage}
---. 2019{\natexlab{b}}, The Astrophysical Journal Letters, 887, L11

\bibitem[{Lustig-Yaeger {et~al.}(2018)Lustig-Yaeger, Meadows, Mendoza, Schwieterman, Fujii, Luger, \& Robinson}]{lustig2018detecting}
Lustig-Yaeger, J., Meadows, V.~S., Mendoza, G.~T., {et~al.} 2018, The Astronomical Journal, 156, 301

\bibitem[{Lustig-Yaeger {et~al.}(2019{\natexlab{c}})Lustig-Yaeger, Robinson, \& Arney}]{lustig2019coronagraph}
Lustig-Yaeger, J., Robinson, T.~D., \& Arney, G. 2019{\natexlab{c}}, Journal of Open Source Software, 4, 1387

\bibitem[{{Lustig-Yaeger} {et~al.}(2022){Lustig-Yaeger}, {Sotzen}, {Stevenson}, {Luger}, {May}, {Mayorga}, {Mandt}, \& {Izenberg}}]{lustig-yaegerHBAR2021}
{Lustig-Yaeger}, J., {Sotzen}, K.~S., {Stevenson}, K.~B., {et~al.} 2022, \aj, 163, 140, \dodoi{10.3847/1538-3881/ac5034}

\bibitem[{{LUVOIR Mission Concept Study Team}(2019)}]{luvoir2019luvoir}
{LUVOIR Mission Concept Study Team}. 2019, Technical Report, NASA

\bibitem[{MacDonald {et~al.}(2020)MacDonald, Goyal, \& Lewis}]{macdonald2020so}
MacDonald, R.~J., Goyal, J.~M., \& Lewis, N.~K. 2020, The Astrophysical Journal Letters, 893, L43

\bibitem[{MacDonald \& Lewis(2021)}]{macdonald2021trident}
MacDonald, R.~J., \& Lewis, N.~K. 2021, arXiv preprint arXiv:2111.05862

\bibitem[{Madhusudhan \& Seager(2009)}]{madhusudhan2009temperature}
Madhusudhan, N., \& Seager, S. 2009, The Astrophysical Journal, 707, 24

\bibitem[{{Mahieux} {et~al.}(2010){Mahieux}, {Vandaele}, {Neefs}, {Robert}, {Wilquet}, {Drummond}, {Federova}, \& {Bertaux}}]{Mahieux2010}
{Mahieux}, A., {Vandaele}, A.~C., {Neefs}, E., {et~al.} 2010, Journal of Geophysical Research (Planets), 115, E12014, \dodoi{10.1029/2010JE003589}

\bibitem[{Mamajek \& Stapelfeldt(2023)}]{habworldstargets}
Mamajek, E., \& Stapelfeldt, K. 2023, {NASA} {ExEP} Mission Star List for the {Habitable Worlds Observatory}, Tech. rep., NASA Jet Propulsion Laboratory and California Institute of Technology

\bibitem[{{May} {et~al.}(2021){May}, {Taylor}, {Komacek}, {Line}, \& {Parmentier}}]{mayicecloud2021}
{May}, E.~M., {Taylor}, J., {Komacek}, T.~D., {Line}, M.~R., \& {Parmentier}, V. 2021, \apjl, 911, L30, \dodoi{10.3847/2041-8213/abeeff}

\bibitem[{{May and MacDonald} {et~al.}(2023){May and MacDonald}, Bennett, Moran, Wakeford, Peacock, Lustig-Yaeger, Highland, Stevenson, Sing, {et~al.}}]{may2023double}
{May and MacDonald}, Bennett, K.~A., Moran, S.~E., {et~al.} 2023, arXiv preprint arXiv:2310.10711

\bibitem[{Meadows {et~al.}(2017)Meadows, Arney, Schwieterman, Lustig-Yaeger, Lincowski, Robinson, Domagal-Goldman, Barnes, Fleming, Deitrick, {et~al.}}]{meadows2017proxima}
Meadows, V., Arney, G., Schwieterman, E., {et~al.} 2017, in American Astronomical Society Meeting Abstracts\# 229, Vol. 229, 120--03

\bibitem[{Meadows(2017)}]{meadows2017reflections}
Meadows, V.~S. 2017, Astrobiology, 17, 1022

\bibitem[{Meadows \& Crisp(1996)}]{meadows1996ground}
Meadows, V.~S., \& Crisp, D. 1996, Journal of Geophysical Research: Planets, 101, 4595

\bibitem[{Meadows {et~al.}(2023)Meadows, Lincowski, \& Lustig-Yaeger}]{Meadows2023}
Meadows, V.~S., Lincowski, A.~P., \& Lustig-Yaeger, J. 2023, The Planetary Science Journal, 4, 192

\bibitem[{Meadows {et~al.}(2018{\natexlab{a}})Meadows, Arney, Schwieterman, Lustig-Yaeger, Lincowski, Robinson, Domagal-Goldman, Deitrick, Barnes, Fleming, {et~al.}}]{meadows2018habitability}
Meadows, V.~S., Arney, G.~N., Schwieterman, E.~W., {et~al.} 2018{\natexlab{a}}, Astrobiology, 18, 133

\bibitem[{Meadows {et~al.}(2018{\natexlab{b}})Meadows, Reinhard, Arney, Parenteau, Schwieterman, Domagal-Goldman, Lincowski, Stapelfeldt, Rauer, DasSarma, {et~al.}}]{meadows2018exoplanet}
Meadows, V.~S., Reinhard, C.~T., Arney, G.~N., {et~al.} 2018{\natexlab{b}}, Astrobiology, 18, 630

\bibitem[{Meixner {et~al.}(2019)Meixner, Cooray, Leisawitz, Staguhn, Armus, Battersby, Bauer, Bergin, Bradford, Ennico-Smith, {et~al.}}]{meixner2019origins}
Meixner, M., Cooray, A., Leisawitz, D., {et~al.} 2019, arXiv preprint arXiv:1912.06213

\bibitem[{Mennesson {et~al.}(2016)Mennesson, Gaudi, Seager, Cahoy, Domagal-Goldman, Feinberg, Guyon, Kasdin, Marois, Mawet, {et~al.}}]{mennesson2016habitable}
Mennesson, B., Gaudi, S., Seager, S., {et~al.} 2016, in Space telescopes and instrumentation 2016: Optical, infrared, and millimeter wave, Vol. 9904, SPIE, 212--221

\bibitem[{Miguel \& Kaltenegger(2013)}]{miguel2013exploring}
Miguel, Y., \& Kaltenegger, L. 2013, The Astrophysical Journal, 780, 166

\bibitem[{Mikal-Evans(2022)}]{mikal2022detecting}
Mikal-Evans, T. 2022, Monthly Notices of the Royal Astronomical Society, 510, 980

\bibitem[{Misra {et~al.}(2014{\natexlab{a}})Misra, Meadows, Claire, \& Crisp}]{misra2014using}
Misra, A., Meadows, V., Claire, M., \& Crisp, D. 2014{\natexlab{a}}, Astrobiology, 14, 67

\bibitem[{Misra {et~al.}(2014{\natexlab{b}})Misra, Meadows, \& Crisp}]{misra2014effects}
Misra, A., Meadows, V., \& Crisp, D. 2014{\natexlab{b}}, The Astrophysical Journal, 792, 61

\bibitem[{{Molli{\`e}re} {et~al.}(2019){Molli{\`e}re}, {Wardenier}, {van Boekel}, {Henning}, {Molaverdikhani}, \& {Snellen}}]{mollierepetitRADTRANS2019}
{Molli{\`e}re}, P., {Wardenier}, J.~P., {van Boekel}, R., {et~al.} 2019, \aap, 627, A67, \dodoi{10.1051/0004-6361/201935470}

\bibitem[{Moran {et~al.}(2023)Moran, Stevenson, Sing, MacDonald, Kirk, Lustig-Yaeger, Peacock, Mayorga, Bennett, L{\'o}pez-Morales, {et~al.}}]{moran2023high}
Moran, S.~E., Stevenson, K.~B., Sing, D.~K., {et~al.} 2023, The Astrophysical Journal Letters, 948, L11

\bibitem[{Mu{\~n}oz {et~al.}(2012)Mu{\~n}oz, Osorio, Barrena, Monta{\~n}{\'e}s-Rodr{\'\i}guez, Mart{\'\i}n, \& Pall{\'e}}]{munoz2012glancing}
Mu{\~n}oz, A.~G., Osorio, M.~Z., Barrena, R., {et~al.} 2012, The Astrophysical Journal, 755, 103

\bibitem[{Nayak {et~al.}(2017)Nayak, Lupu, Marley, Fortney, Robinson, \& Lewis}]{nayak2017atmospheric}
Nayak, M., Lupu, R., Marley, M.~S., {et~al.} 2017, Publications of the Astronomical Society of the Pacific, 129, 034401

\bibitem[{{Niraula} {et~al.}(2022){Niraula}, {de Wit}, {Gordon}, {Hargreaves}, {Sousa-Silva}, \& {Kochanov}}]{Niraula2022}
{Niraula}, P., {de Wit}, J., {Gordon}, I.~E., {et~al.} 2022, Nature Astronomy, \dodoi{10.1038/s41550-022-01773-1}

\bibitem[{Nixon \& Madhusudhan(2022)}]{nixon2022aura}
Nixon, M.~C., \& Madhusudhan, N. 2022, arXiv preprint arXiv:2201.03532

\bibitem[{Olson {et~al.}(2016)Olson, Reinhard, \& Lyons}]{olson2016limited}
Olson, S.~L., Reinhard, C.~T., \& Lyons, T.~W. 2016, Proceedings of the National Academy of Sciences, 113, 11447

\bibitem[{Pall{\'e} {et~al.}(2009)Pall{\'e}, Osorio, Barrena, Monta{\~n}{\'e}s-Rodr{\'\i}guez, \& Mart{\'\i}n}]{palle2009earth}
Pall{\'e}, E., Osorio, M. R.~Z., Barrena, R., Monta{\~n}{\'e}s-Rodr{\'\i}guez, P., \& Mart{\'\i}n, E.~L. 2009, Nature, 459, 814

\bibitem[{Pfisterer {et~al.}(2018)Pfisterer, Harvey, \& Breckinridge}]{pfisterer2018role}
Pfisterer, R.~N., Harvey, J.~E., \& Breckinridge, J.~B. 2018, in Space Telescopes and Instrumentation 2018: Optical, Infrared, and Millimeter Wave, Vol. 10698, International Society for Optics and Photonics, 106985F

\bibitem[{Pluriel {et~al.}(2020)Pluriel, Whiteford, Edwards, Changeat, Yip, Baeyens, Al-Refaie, Bieger, Blain, Gressier, {et~al.}}]{pluriel2020ares}
Pluriel, W., Whiteford, N., Edwards, B., {et~al.} 2020, The Astronomical Journal, 160, 112

\bibitem[{{Price-Whelan} {et~al.}(2018){Price-Whelan}, {Sip{\H{o}}cz}, {G{\"u}nther}, {Lim}, {Crawford}, {Conseil}, {Shupe}, {Craig}, {Dencheva}, {Ginsburg}, {VanderPlas}, {Bradley}, {P{\'e}rez-Su{\'a}rez}, {de Val-Borro}, {Paper Contributors}, {Aldcroft}, {Cruz}, {Robitaille}, {Tollerud}, {Coordination Committee}, {Ardelean}, {Babej}, {Bach}, {Bachetti}, {Bakanov}, {Bamford}, {Barentsen}, {Barmby}, {Baumbach}, {Berry}, {Biscani}, {Boquien}, {Bostroem}, {Bouma}, {Brammer}, {Bray}, {Breytenbach}, {Buddelmeijer}, {Burke}, {Calderone}, {Cano Rodr{\'\i}guez}, {Cara}, {Cardoso}, {Cheedella}, {Copin}, {Corrales}, {Crichton}, {D{\textquoteright}Avella}, {Deil}, {Depagne}, {Dietrich}, {Donath}, {Droettboom}, {Earl}, {Erben}, {Fabbro}, {Ferreira}, {Finethy}, {Fox}, {Garrison}, {Gibbons}, {Goldstein}, {Gommers}, {Greco}, {Greenfield}, {Groener}, {Grollier}, {Hagen}, {Hirst}, {Homeier}, {Horton}, {Hosseinzadeh}, {Hu}, {Hunkeler}, {Ivezi{\'c}}, {Jain}, {Jenness}, {Kanarek}, {Kendrew}, {Kern}, {Kerzendorf}, {Khvalko},
  {King}, {Kirkby}, {Kulkarni}, {Kumar}, {Lee}, {Lenz}, {Littlefair}, {Ma}, {Macleod}, {Mastropietro}, {McCully}, {Montagnac}, {Morris}, {Mueller}, {Mumford}, {Muna}, {Murphy}, {Nelson}, {Nguyen}, {Ninan}, {N{\"o}the}, {Ogaz}, {Oh}, {Parejko}, {Parley}, {Pascual}, {Patil}, {Patil}, {Plunkett}, {Prochaska}, {Rastogi}, {Reddy Janga}, {Sabater}, {Sakurikar}, {Seifert}, {Sherbert}, {Sherwood-Taylor}, {Shih}, {Sick}, {Silbiger}, {Singanamalla}, {Singer}, {Sladen}, {Sooley}, {Sornarajah}, {Streicher}, {Teuben}, {Thomas}, {Tremblay}, {Turner}, {Terr{\'o}n}, {van Kerkwijk}, {de la Vega}, {Watkins}, {Weaver}, {Whitmore}, {Woillez}, {Zabalza}, \& {Contributors}}]{Astropy2018}
{Price-Whelan}, A.~M., {Sip{\H{o}}cz}, B.~M., {G{\"u}nther}, H.~M., {et~al.} 2018, \aj, 156, 123, \dodoi{10.3847/1538-3881/aabc4f}

\bibitem[{Price-Whelan {et~al.}(2022)Price-Whelan, Lim, Earl, Starkman, Bradley, Shupe, Patil, Corrales, Brasseur, N{\"o}the, {et~al.}}]{price2022astropy}
Price-Whelan, A.~M., Lim, P.~L., Earl, N., {et~al.} 2022, The Astrophysical Journal, 935, 167

\bibitem[{Quanz {et~al.}(2021)Quanz, Ottiger, Fontanet, Kammerer, Menti, Dannert, Gheorghe, Absil, Airapetian, Alei, {et~al.}}]{quanz2021large}
Quanz, S.~P., Ottiger, M., Fontanet, E., {et~al.} 2021, arXiv preprint arXiv:2101.07500

\bibitem[{Rackham {et~al.}(2018)Rackham, Apai, \& Giampapa}]{rackham2018transit}
Rackham, B.~V., Apai, D., \& Giampapa, M.~S. 2018, The Astrophysical Journal, 853, 122

\bibitem[{Ranjan {et~al.}(2023)Ranjan, Schwieterman, Leung, Harman, \& Hu}]{ranjan2023importance}
Ranjan, S., Schwieterman, E.~W., Leung, M., Harman, C.~E., \& Hu, R. 2023, The Astrophysical Journal Letters, 958, L15

\bibitem[{Ratner \& Walker(1972)}]{ratner1972atmospheric}
Ratner, M.~I., \& Walker, J.~C. 1972, Journal of Atmospheric Sciences, 29, 803

\bibitem[{Robinson(2017{\natexlab{a}})}]{robinson2017characterizing}
Robinson, T.~D. 2017{\natexlab{a}}, arXiv preprint arXiv:1701.05205

\bibitem[{Robinson(2017{\natexlab{b}})}]{robinson2017theory}
---. 2017{\natexlab{b}}, The Astrophysical Journal, 836, 236

\bibitem[{Robinson(2018)}]{robinson2018characterizing}
---. 2018, in Handbook of Exoplanets (Springer International Publishing), 3137--3157

\bibitem[{Robinson \& Catling(2014)}]{robinson2014common}
Robinson, T.~D., \& Catling, D.~C. 2014, Nature Geoscience, 7, 12

\bibitem[{Robinson {et~al.}(2010)Robinson, Meadows, \& Crisp}]{robinson2010detecting}
Robinson, T.~D., Meadows, V.~S., \& Crisp, D. 2010, The Astrophysical Journal Letters, 721, L67

\bibitem[{Robinson {et~al.}(2016)Robinson, Stapelfeldt, \& Marley}]{robinson2016characterizing}
Robinson, T.~D., Stapelfeldt, K.~R., \& Marley, M.~S. 2016, Publications of the Astronomical Society of the Pacific, 128, 025003

\bibitem[{Robinson {et~al.}(2011)Robinson, Meadows, Crisp, Deming, A'hearn, Charbonneau, Livengood, Seager, Barry, Hearty, {et~al.}}]{robinson2011earth}
Robinson, T.~D., Meadows, V.~S., Crisp, D., {et~al.} 2011, Astrobiology, 11, 393

\bibitem[{Rodler \& L{\'o}pez-Morales(2014)}]{rodler2014feasibility}
Rodler, F., \& L{\'o}pez-Morales, M. 2014, The Astrophysical Journal, 781, 54

\bibitem[{Rothman {et~al.}(2010)Rothman, Gordon, Barber, Dothe, Gamache, Goldman, Perevalov, Tashkun, \& Tennyson}]{rothman2010hitemp}
Rothman, L.~S., Gordon, I., Barber, R., {et~al.} 2010, Journal of Quantitative Spectroscopy and Radiative Transfer, 111, 2139

\bibitem[{Rothman {et~al.}(2013)Rothman, Gordon, Babikov, Barbe, Benner, Bernath, Birk, Bizzocchi, Boudon, Brown, {et~al.}}]{rothman2013hitran2012}
Rothman, L.~S., Gordon, I.~E., Babikov, Y., {et~al.} 2013, Journal of Quantitative Spectroscopy and Radiative Transfer, 130, 4

\bibitem[{Rotman {et~al.}(2022)Rotman, Komacek, Villanueva, Fauchez, \& May}]{rotman2022general}
Rotman, Y., Komacek, T.~D., Villanueva, G.~L., Fauchez, T.~J., \& May, E.~M. 2022, The Astrophysical Journal Letters, 942, L4

\bibitem[{Rugheimer {et~al.}(2015)Rugheimer, Kaltenegger, Segura, Linsky, \& Mohanty}]{rugheimer2015effect}
Rugheimer, S., Kaltenegger, L., Segura, A., Linsky, J., \& Mohanty, S. 2015, The Astrophysical Journal, 809, 57

\bibitem[{Rugheimer {et~al.}(2013)Rugheimer, Kaltenegger, Zsom, Segura, \& Sasselov}]{rugheimer2013spectral}
Rugheimer, S., Kaltenegger, L., Zsom, A., Segura, A., \& Sasselov, D. 2013, Astrobiology, 13, 251

\bibitem[{Schaefer {et~al.}(2016)Schaefer, Wordsworth, Berta-Thompson, \& Sasselov}]{schaefer2016predictions}
Schaefer, L., Wordsworth, R.~D., Berta-Thompson, Z., \& Sasselov, D. 2016, The Astrophysical Journal, 829, 63

\bibitem[{Schwieterman {et~al.}(2019)Schwieterman, Reinhard, Olson, Ozaki, Harman, Hong, \& Lyons}]{schwieterman2019rethinking}
Schwieterman, E.~W., Reinhard, C.~T., Olson, S.~L., {et~al.} 2019, The Astrophysical Journal, 874, 9

\bibitem[{Schwieterman {et~al.}(2015)Schwieterman, Robinson, Meadows, Misra, \& Domagal-Goldman}]{schwieterman2015detecting}
Schwieterman, E.~W., Robinson, T.~D., Meadows, V.~S., Misra, A., \& Domagal-Goldman, S. 2015, The Astrophysical Journal, 810, 57

\bibitem[{Schwieterman {et~al.}(2016)Schwieterman, Meadows, Domagal-Goldman, Deming, Arney, Luger, Harman, Misra, \& Barnes}]{schwieterman2016identifying}
Schwieterman, E.~W., Meadows, V.~S., Domagal-Goldman, S.~D., {et~al.} 2016, The Astrophysical Journal Letters, 819, L13

\bibitem[{Schwieterman {et~al.}(2018)Schwieterman, Kiang, Parenteau, Harman, DasSarma, Fisher, Arney, Hartnett, Reinhard, Olson, {et~al.}}]{schwieterman2018exoplanet}
Schwieterman, E.~W., Kiang, N.~Y., Parenteau, M.~N., {et~al.} 2018, Astrobiology, 18, 663

\bibitem[{Seager(2011)}]{seager2011exoplanets}
Seager, S. 2011, Exoplanets (University of Arizona Press)

\bibitem[{Segura {et~al.}(2005)Segura, Kasting, Meadows, Cohen, Scalo, Crisp, Butler, \& Tinetti}]{segura2005biosignatures}
Segura, A., Kasting, J.~F., Meadows, V., {et~al.} 2005, Astrobiology, 5, 706

\bibitem[{Segura {et~al.}(2003)Segura, Krelove, Kasting, Sommerlatt, Meadows, Crisp, Cohen, \& Mlawer}]{segura2003ozone}
Segura, A., Krelove, K., Kasting, J.~F., {et~al.} 2003, Astrobiology, 3, 689

\bibitem[{Segura {et~al.}(2007)Segura, Meadows, Kasting, Crisp, \& Cohen}]{segura2007abiotic}
Segura, A., Meadows, V., Kasting, J., Crisp, D., \& Cohen, M. 2007, Astronomy \& Astrophysics, 472, 665

\bibitem[{Selsis {et~al.}(2002)Selsis, Despois, \& Parisot}]{selsis2002signature}
Selsis, F., Despois, D., \& Parisot, J.-P. 2002, Astronomy \& Astrophysics, 388, 985

\bibitem[{Skilling(2004)}]{skilling2004nested}
Skilling, J. 2004, in AIP Conference Proceedings, Vol. 735, American Institute of Physics, 395--405

\bibitem[{Skilling(2006)}]{skilling2006nested}
Skilling, J. 2006, Bayesian analysis, 1, 833

\bibitem[{Snellen {et~al.}(2013)Snellen, De~Kok, Le~Poole, Brogi, \& Birkby}]{snellen2013finding}
Snellen, I., De~Kok, R., Le~Poole, R., Brogi, M., \& Birkby, J. 2013, The Astrophysical Journal, 764, 182

\bibitem[{{Spurr} {et~al.}(2001){Spurr}, {Kurosu}, \& {Chance}}]{Spurr2001}
{Spurr}, R.~J.~D., {Kurosu}, T.~P., \& {Chance}, K.~V. 2001, \jqsrt, 68, 689, \dodoi{10.1016/S0022-4073(00)00055-8}

\bibitem[{Stamnes {et~al.}(1988)Stamnes, Tsay, Wiscombe, \& Jayaweera}]{stamnes1988numerically}
Stamnes, K., Tsay, S.-C., Wiscombe, W., \& Jayaweera, K. 1988, Applied optics, 27, 2502

\bibitem[{Stark {et~al.}(2019)Stark, Belikov, Bolcar, Cady, Crill, Ertel, Groff, Hildebrandt, Krist, Lisman, {et~al.}}]{stark2019exoearth}
Stark, C.~C., Belikov, R., Bolcar, M.~R., {et~al.} 2019, Journal of Astronomical Telescopes, Instruments, and Systems, 5, 024009

\bibitem[{Suissa {et~al.}(2020)Suissa, Mandell, Wolf, Villanueva, Fauchez, \& kumar Kopparapu}]{suissa2020dim}
Suissa, G., Mandell, A.~M., Wolf, E.~T., {et~al.} 2020, The Astrophysical Journal, 891, 58

\bibitem[{{Taylor} {et~al.}(2020){Taylor}, {Parmentier}, {Irwin}, {Aigrain}, {Lee}, \& {Krissansen-Totton}}]{Taylor2020biases}
{Taylor}, J., {Parmentier}, V., {Irwin}, P. G.~J., {et~al.} 2020, \mnras, 493, 4342, \dodoi{10.1093/mnras/staa552}

\bibitem[{{Tremblay} {et~al.}(2020){Tremblay}, {Line}, {Stevenson}, {Kataria}, {Zellem}, {Fortney}, \& {Morley}}]{tremblaydetectability2020}
{Tremblay}, L., {Line}, M.~R., {Stevenson}, K., {et~al.} 2020, \aj, 159, 117, \dodoi{10.3847/1538-3881/ab64dd}

\bibitem[{Tsai {et~al.}(2023)Tsai, Lee, Powell, Gao, Zhang, Moses, H{\'e}brard, Venot, Parmentier, Jordan, {et~al.}}]{tsai2023photochemically}
Tsai, S.-M., Lee, E.~K., Powell, D., {et~al.} 2023, Nature, 617, 483

\bibitem[{Tsamalis {et~al.}(2013)Tsamalis, Ch{\'e}din, Pelon, \& Capelle}]{tsamalis2013seasonal}
Tsamalis, C., Ch{\'e}din, A., Pelon, J., \& Capelle, V. 2013, Atmospheric Chemistry and Physics, 13, 11235

\bibitem[{Tsiaras {et~al.}(2016)Tsiaras, Rocchetto, Waldmann, Venot, Varley, Morello, Damiano, Tinetti, Barton, Yurchenko, {et~al.}}]{tsiaras2016detection}
Tsiaras, A., Rocchetto, M., Waldmann, I., {et~al.} 2016, The Astrophysical Journal, 820, 99

\bibitem[{Turbet {et~al.}(2020)Turbet, Bolmont, Bourrier, Demory, Leconte, Owen, \& Wolf}]{turbet2020review}
Turbet, M., Bolmont, E., Bourrier, V., {et~al.} 2020, Space science reviews, 216, 1

\bibitem[{{van der Walt} {et~al.}(2011){van der Walt}, {Colbert}, \& {Varoquaux}}]{Walt2011}
{van der Walt}, S., {Colbert}, S.~C., \& {Varoquaux}, G. 2011, Computing in Science and Engineering, 13, 22, \dodoi{10.1109/MCSE.2011.37}

\bibitem[{Vinatier {et~al.}(2007)Vinatier, B{\'e}zard, Fouchet, Teanby, de~Kok, Irwin, Conrath, Nixon, Romani, Flasar, {et~al.}}]{vinatier2007vertical}
Vinatier, S., B{\'e}zard, B., Fouchet, T., {et~al.} 2007, Icarus, 188, 120

\bibitem[{{Virtanen} {et~al.}(2019){Virtanen}, {Gommers}, {Oliphant}, {Haberland}, {Reddy}, {Cournapeau}, {Burovski}, {Peterson}, {Weckesser}, {Bright}, {van der Walt}, {Brett}, {Wilson}, {Jarrod Millman}, {Mayorov}, {Nelson}, {Jones}, {Kern}, {Larson}, {Carey}, {Polat}, {Feng}, {Moore}, {Vand erPlas}, {Laxalde}, {Perktold}, {Cimrman}, {Henriksen}, {Quintero}, {Harris}, {Archibald}, {Ribeiro}, {Pedregosa}, {van Mulbregt}, \& {Contributors}}]{Virtanen2019scipy}
{Virtanen}, P., {Gommers}, R., {Oliphant}, T.~E., {et~al.} 2019, arXiv e-prints, arXiv:1907.10121.
\newblock \doarXiv{1907.10121}

\bibitem[{Virtanen {et~al.}(2020)Virtanen, Gommers, Oliphant, Haberland, Reddy, Cournapeau, Burovski, Peterson, Weckesser, Bright, {van der Walt}, Brett, Wilson, Millman, Mayorov, Nelson, Jones, Kern, Larson, Carey, Polat, Feng, Moore, {VanderPlas}, Laxalde, Perktold, Cimrman, Henriksen, Quintero, Harris, Archibald, Ribeiro, Pedregosa, {van Mulbregt}, \& {SciPy 1.0 Contributors}}]{SciPy2020}
Virtanen, P., Gommers, R., Oliphant, T.~E., {et~al.} 2020, Nature Methods, 17, 261, \dodoi{10.1038/s41592-019-0686-2}

\bibitem[{Wakeford \& Sing(2015)}]{wakeford2015transmission}
Wakeford, H.~R., \& Sing, D.~K. 2015, Astronomy \& Astrophysics, 573, A122

\bibitem[{Wakeford {et~al.}(2017)Wakeford, Sing, Deming, Lewis, Goyal, Wilson, Barstow, Kataria, Drummond, Evans, {et~al.}}]{wakeford2017complete}
Wakeford, H.~R., Sing, D.~K., Deming, D., {et~al.} 2017, The Astronomical Journal, 155, 29

\bibitem[{Waldmann {et~al.}(2015)Waldmann, Tinetti, Rocchetto, Barton, Yurchenko, \& Tennyson}]{waldmann2015tau}
Waldmann, I.~P., Tinetti, G., Rocchetto, M., {et~al.} 2015, The Astrophysical Journal, 802, 107

\bibitem[{Walker(1977)}]{walker1977evolution}
Walker, J. 1977, Evolution of the Atmosphere (Macmillan Co., New York)

\bibitem[{Wang {et~al.}(2018)Wang, Forget, Bertrand, Spiga, Millour, \& Navarro}]{wang2018parameterization}
Wang, C., Forget, F., Bertrand, T., {et~al.} 2018, Journal of Geophysical Research: Planets, 123, 982

\bibitem[{Williams \& Gaidos(2008)}]{williams2008detecting}
Williams, D.~M., \& Gaidos, E. 2008, Icarus, 195, 927

\bibitem[{Wogan {et~al.}(2020)Wogan, Krissansen-Totton, \& Catling}]{wogan2020abundant}
Wogan, N., Krissansen-Totton, J., \& Catling, D.~C. 2020, The Planetary Science Journal, 1, 58

\bibitem[{Wordsworth \& Pierrehumbert(2014)}]{wordsworth2014abiotic}
Wordsworth, R., \& Pierrehumbert, R. 2014, The Astrophysical Journal Letters, 785, L20

\bibitem[{Wunderlich {et~al.}(2019)Wunderlich, Godolt, Grenfell, St{\"a}dt, Smith, Gebauer, Schreier, Hedelt, \& Rauer}]{wunderlich2019detectability}
Wunderlich, F., Godolt, M., Grenfell, J.~L., {et~al.} 2019, Astronomy \& Astrophysics, 624, A49

\bibitem[{Young {et~al.}(2024)Young, Crouse, Arney, Domagal-Goldman, Robinson, \& Bastelberger}]{young2024retrievals}
Young, A.~V., Crouse, J., Arney, G., {et~al.} 2024, The Planetary Science Journal, 5, 7

\bibitem[{Young {et~al.}(2023)Young, Robinson, Krissansen-Totton, Schwieterman, Wogan, Way, Sohl, Arney, Reinhard, Line, {et~al.}}]{young2023inferring}
Young, A.~V., Robinson, T.~D., Krissansen-Totton, J., {et~al.} 2023, arXiv preprint arXiv:2311.06083

\bibitem[{Zerkle {et~al.}(2012)Zerkle, Claire, Domagal-Goldman, Farquhar, \& Poulton}]{zerkle2012bistable}
Zerkle, A.~L., Claire, M.~W., Domagal-Goldman, S.~D., Farquhar, J., \& Poulton, S.~W. 2012, Nature Geoscience, 5, 359

\end{thebibliography}
\bibliographystyle{aasjournal}

\appendix{}
\vspace{-10pt}
\section{Corner Plots of Posterior Distributions}
\renewcommand{\figurename}{Figure Set}

\label{sec:ap:corners}

We display the complete set of posteriors for the retrievals presented in this study in four figure sets. The complete figure sets (16 images) are available in the online journal.  Figure sets 1 and 2 include the posteriors for the two transmission cases, and figure sets 3 and 4 includes the posteriors for the two direct imaging cases. Figure set 1 includes the complete set of posterior distributions for the transmission retrievals of the Earth-like TRAPPIST-1 e with the evenly-mixed ($\mathcal{M}_{iso}$), three fixed points ($\mathcal{M}_{3P}$), five fixed points ($\mathcal{M}_{5P}$), and the three free points ($\mathcal{M}_{3F}$) forward models. This pattern repeats for the remaining planetary cases investigated here, with the transmission retrievals of the abiotic, false-positive O$_3$ TRAPPIST-1 e (figure set 2), and the direct imaging retrievals of the clear-sky (figure set 3) and cloudy Earth (figure set 4).

\setcounter{figure}{0}


\figsetstart
\figsetnum{1}
\figsettitle{\textit{Origins}-Like Transmission Retrieval of Earth-Like TRAPPIST-1 e}
\figsetgrpstart
\figsetgrpnum{figurenumber.1}
\figsetgrptitle{Evenly-Mixed Forward Model ($\mathcal{M}_{iso})$}
\figsetplot{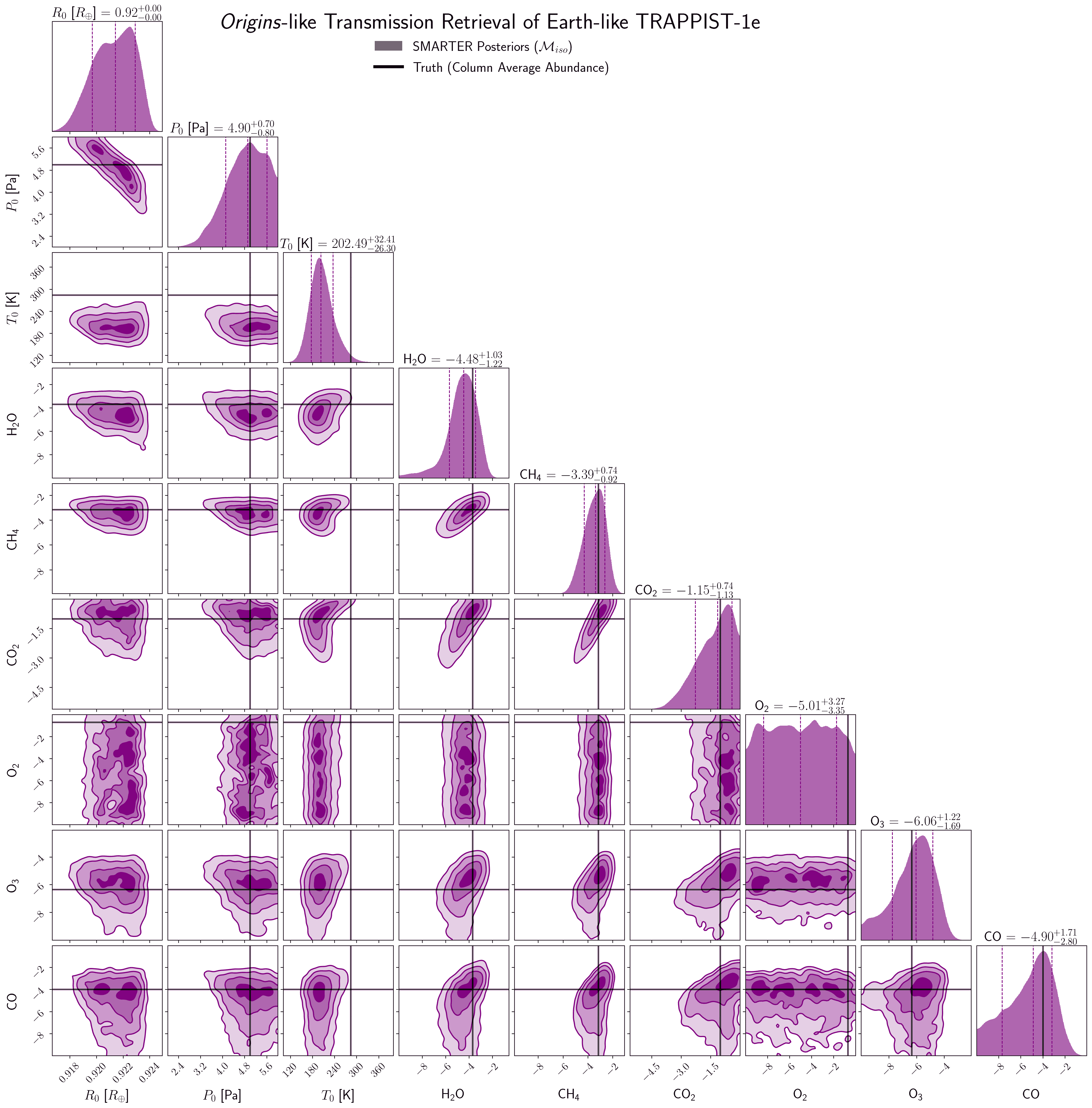}
\figsetgrpnote{The results of the retrieval on the Earth-like TRAPPIST-1 e transmission spectrum with the evenly-mixed forward model, $\mathcal{M}_{iso}$. The plots on the diagonal are the posterior distributions for each free parameter in the retrieval, while the off-diagonal plots indicate the covariances between the retrieval parameters. Truth values are indicated by solid black lines. For the gas abundances, these truth values represent the column average abundance for each gas in the true spectrum.}
\figsetgrpend

\figsetgrpstart
\figsetgrpnum{figurenumber.2}
\figsetgrptitle{Three-Fixed Points Forward Model ($\mathcal{M}_{3P})$}
\figsetplot{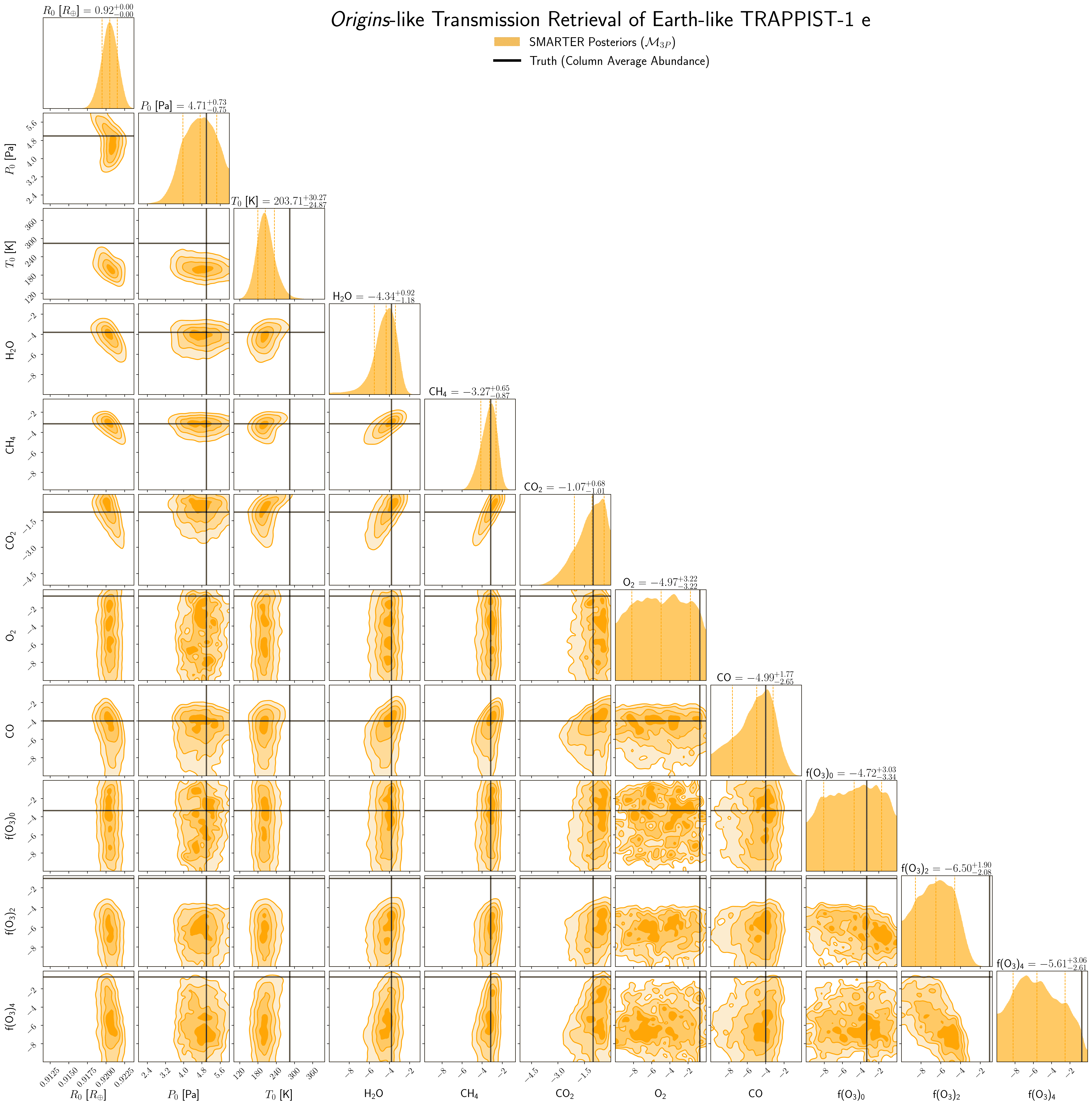}
\figsetgrpnote{The results of the retrieval on the Earth-like TRAPPIST-1 e transmission spectrum with the three-fixed points forward model, $\mathcal{M}_{3P}$. Truth values are indicated by solid black lines. For the non-O$_3$ gas abundances, these truth values represent the column average abundance for each gas in the true spectrum. For the O$_3$ gas abundances, the truth values represent the true O$_3$ value at the corresponding pressure level based on the true profile.}
\figsetgrpend

\figsetgrpstart
\figsetgrpnum{figurenumber.3}
\figsetgrptitle{Five-Fixed Points Forward Model ($\mathcal{M}_{5P})$}
\figsetplot{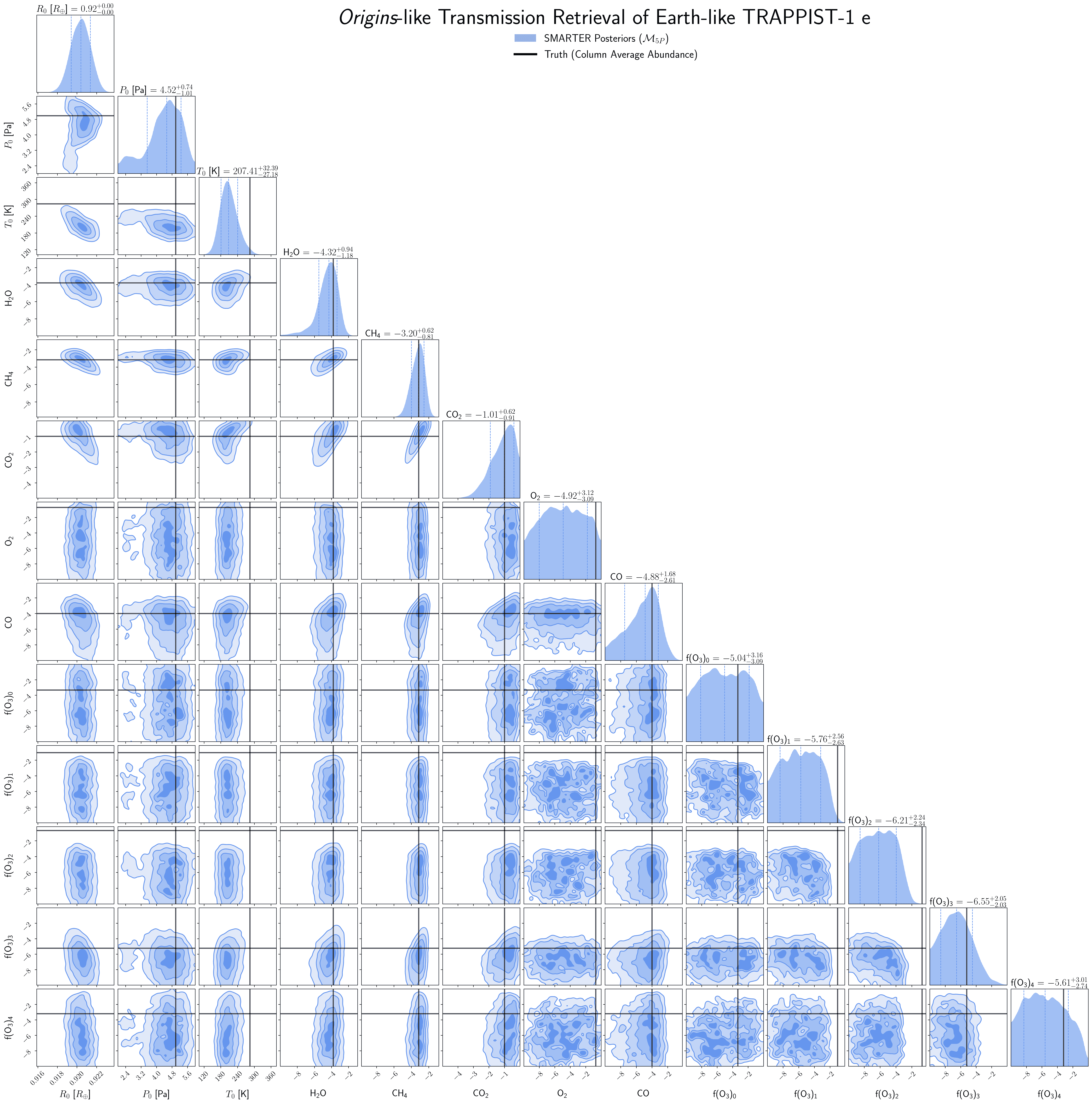}
\figsetgrpnote{The results of the retrieval on the Earth-like TRAPPIST-1 e transmission spectrum with the five-fixed points forward model, $\mathcal{M}_{5P}$.}
\figsetgrpend

\figsetgrpstart
\figsetgrpnum{figurenumber.4}
\figsetgrptitle{Three-Free Points Forward Model ($\mathcal{M}_{3F})$}
\figsetplot{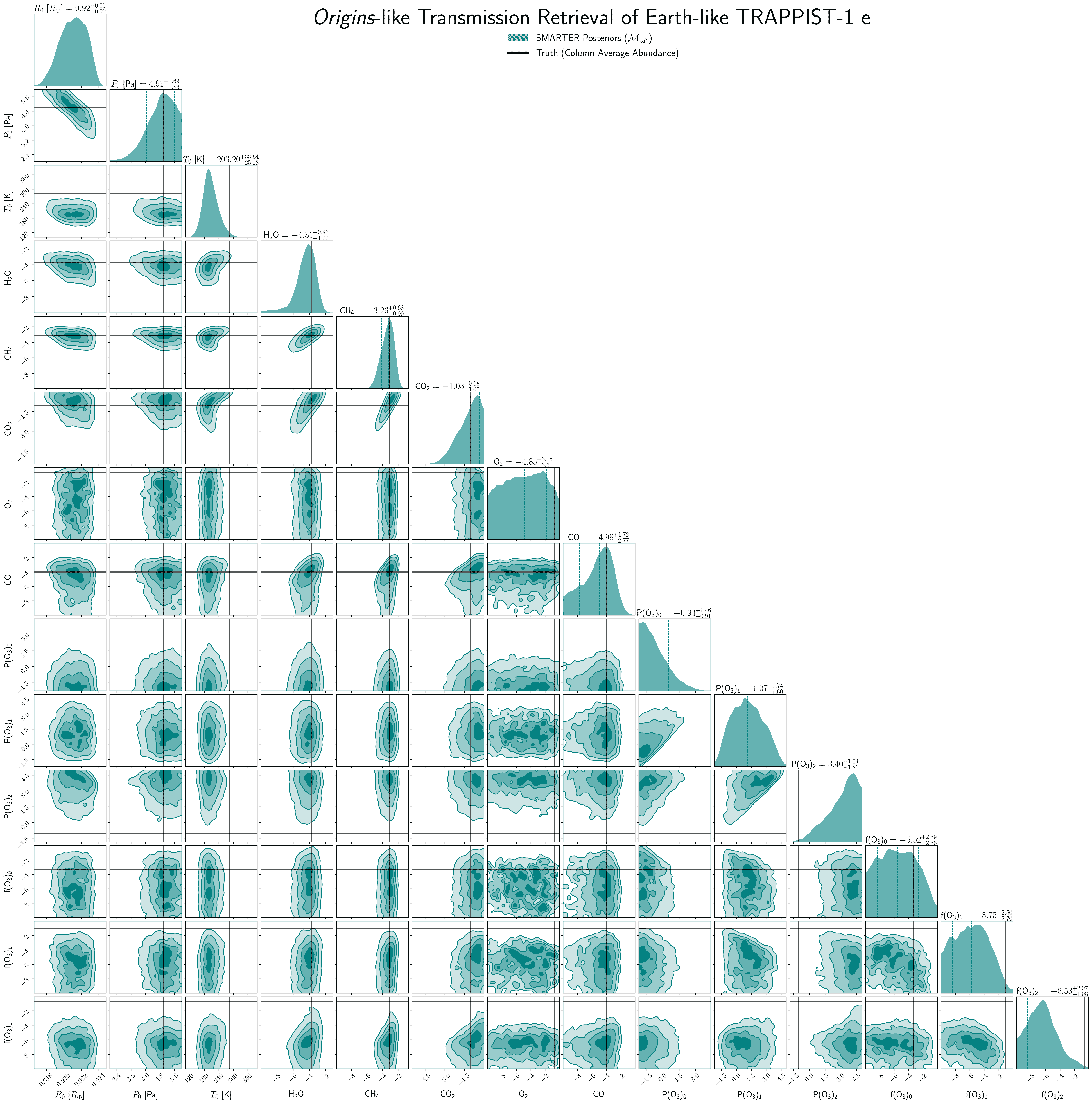}
\figsetgrpnote{The results of the retrieval on the Earth-like TRAPPIST-1 e transmission spectrum with the three-free points forward model, $\mathcal{M}_{3F}$.}
\figsetgrpend
\figsetend

\begin{figure}[htbp]
    \centering
    \includegraphics[width=0.8\textwidth]{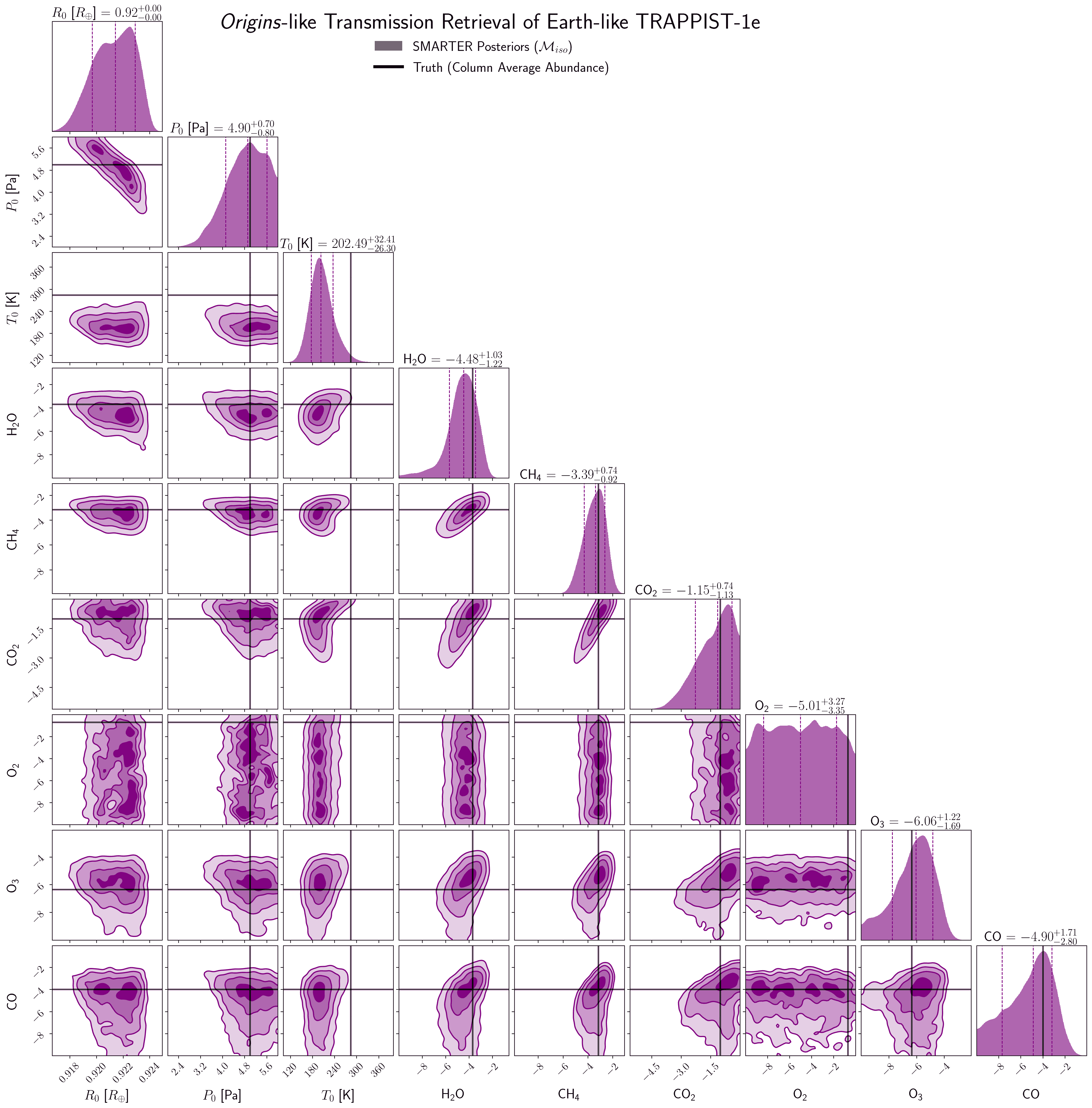}
    \caption{The results of the retrieval on the Earth-like TRAPPIST-1 e transmission spectrum with the evenly-mixed forward model, $\mathcal{M}_{iso}$. The plots on the diagonal are the posterior distributions for each free parameter in the retrieval, while the off-diagonal plots indicate the covariances between the retrieval parameters. Truth values are indicated by solid black lines. For the gas abundances, these truth values represent the column average abundance for each gas in the true spectrum.}
\end{figure}

\newpage


\figsetstart
\figsetnum{2}
\figsettitle{\textit{Origins}-Like Transmission Retrieval of False-Positive O$_3$ TRAPPIST-1 e}
\figsetgrpstart
\figsetgrpnum{figurenumber.1}
\figsetgrptitle{Evenly-Mixed Forward Model ($\mathcal{M}_{iso})$}
\figsetplot{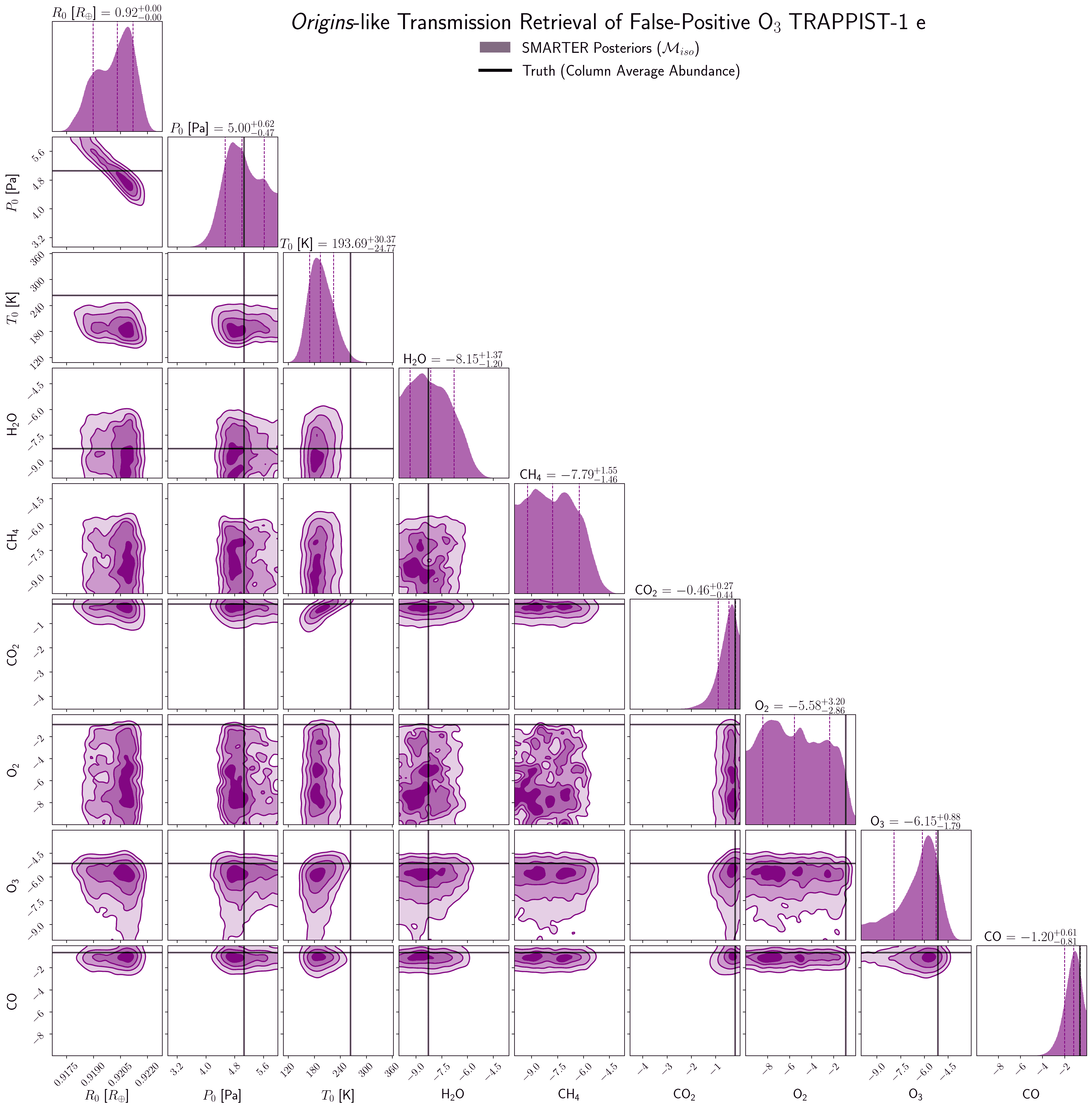}
\figsetgrpnote{The results of the retrieval on the abiotic, false-positive O$_3$ TRAPPIST-1 e transmission spectrum with the evenly-mixed forward model, $\mathcal{M}_{iso}$.}
\figsetgrpend

\figsetgrpstart
\figsetgrpnum{figurenumber.2}
\figsetgrptitle{Three-Fixed Points Forward Model ($\mathcal{M}_{3P})$}
\figsetplot{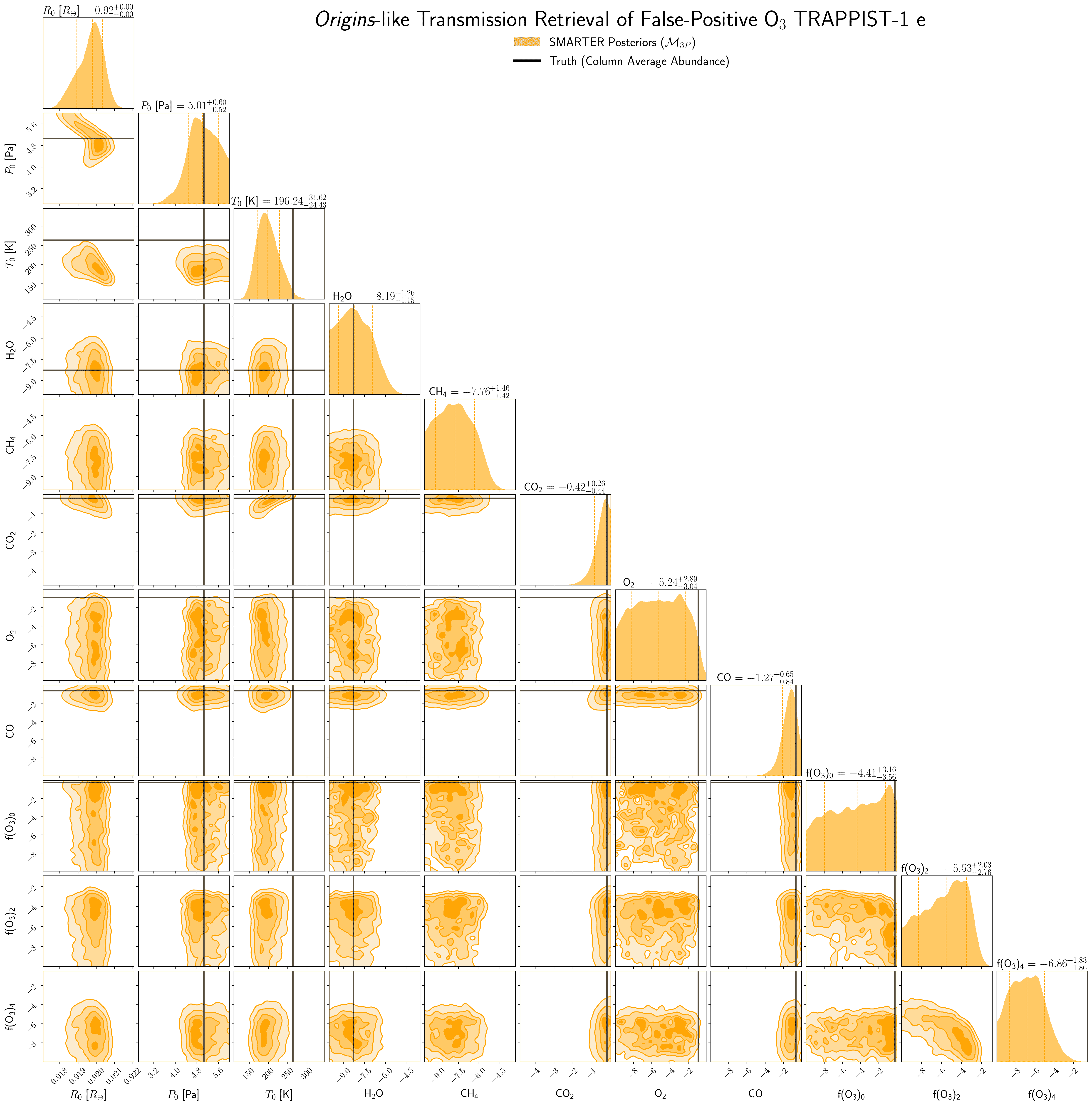}
\figsetgrpnote{The results of the retrieval on the abiotic, false-positive O$_3$ TRAPPIST-1 e transmission spectrum with the three-fixed points forward model, $\mathcal{M}_{3P}$.}
\figsetgrpend

\figsetgrpstart
\figsetgrpnum{figurenumber.3}
\figsetgrptitle{Five-Fixed Points Forward Model ($\mathcal{M}_{5P})$}
\figsetplot{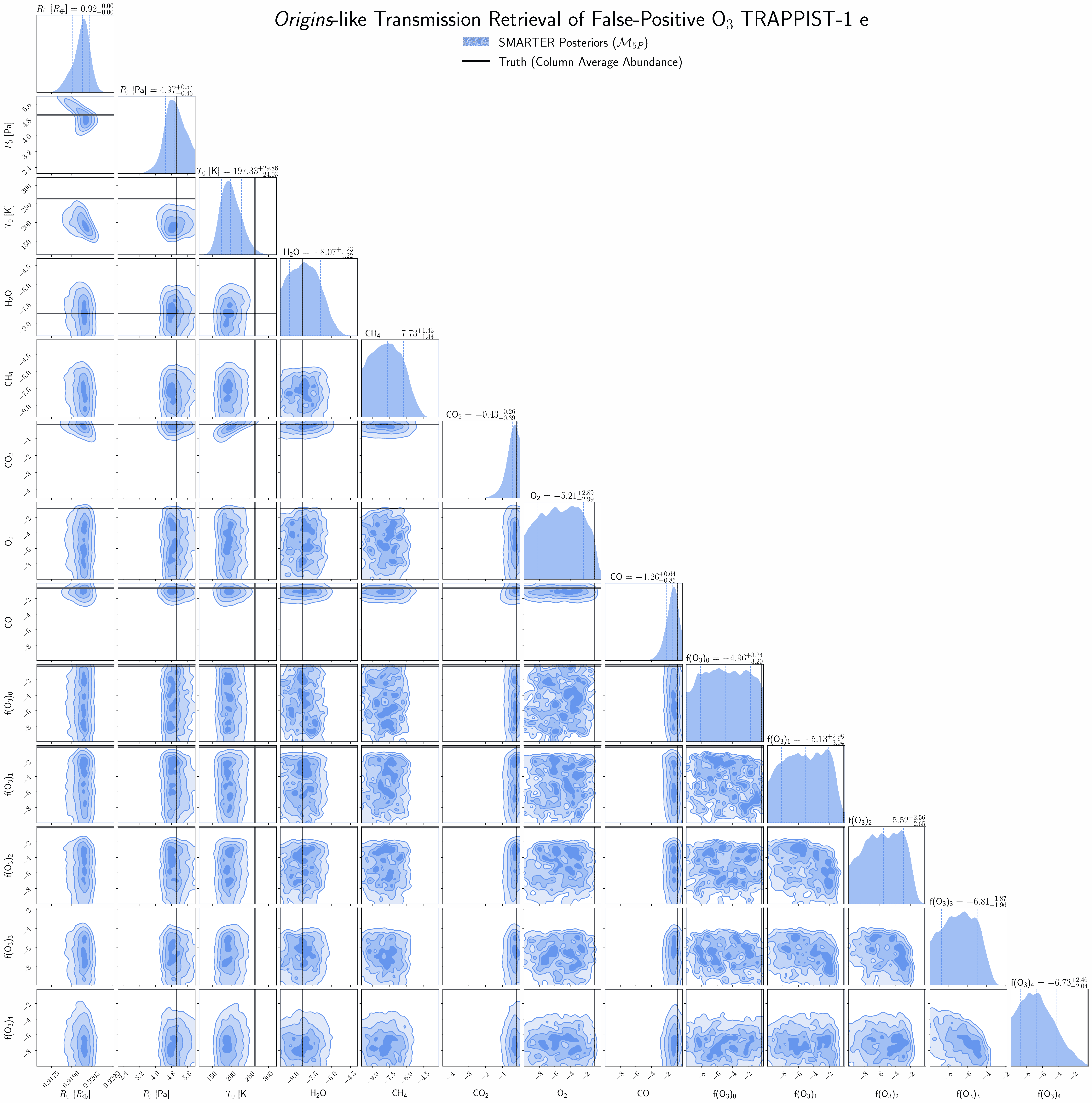}
\figsetgrpnote{The results of the retrieval on the on the abiotic, false-positive O$_3$ TRAPPIST-1 e transmission spectrum with the five-fixed points forward model, $\mathcal{M}_{5P}$.}
\figsetgrpend

\figsetgrpstart
\figsetgrpnum{figurenumber.4}
\figsetgrptitle{Three-Free Points Forward Model ($\mathcal{M}_{3F})$}
\figsetplot{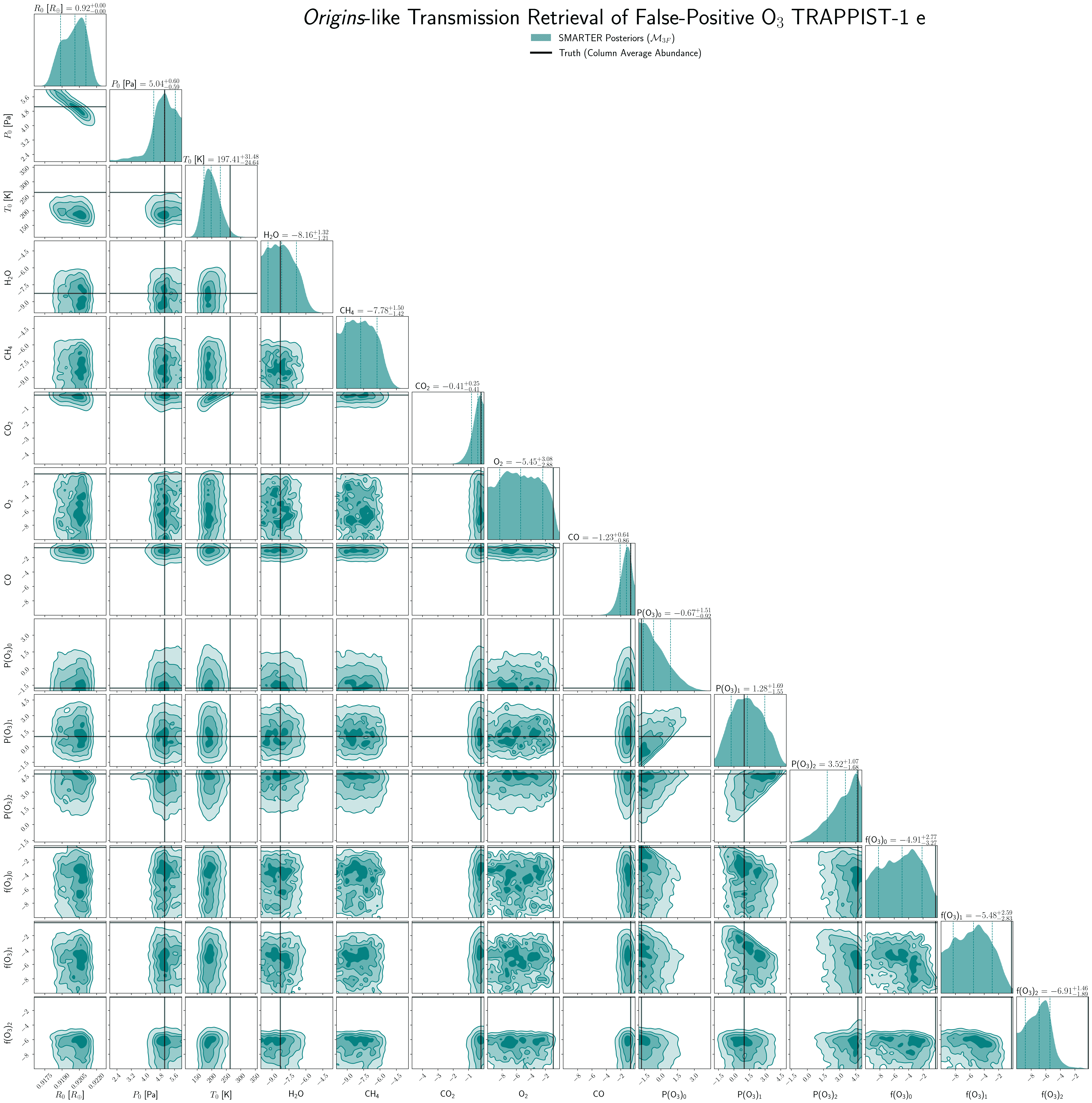}
\figsetgrpnote{The results of the retrieval on the on the abiotic, false-positive O$_3$ TRAPPIST-1 e transmission spectrum with the three-free points forward model, $\mathcal{M}_{3F}$.}
\figsetgrpend
\figsetend

\begin{figure}[htbp]
    \centering
    \includegraphics[width=\textwidth]{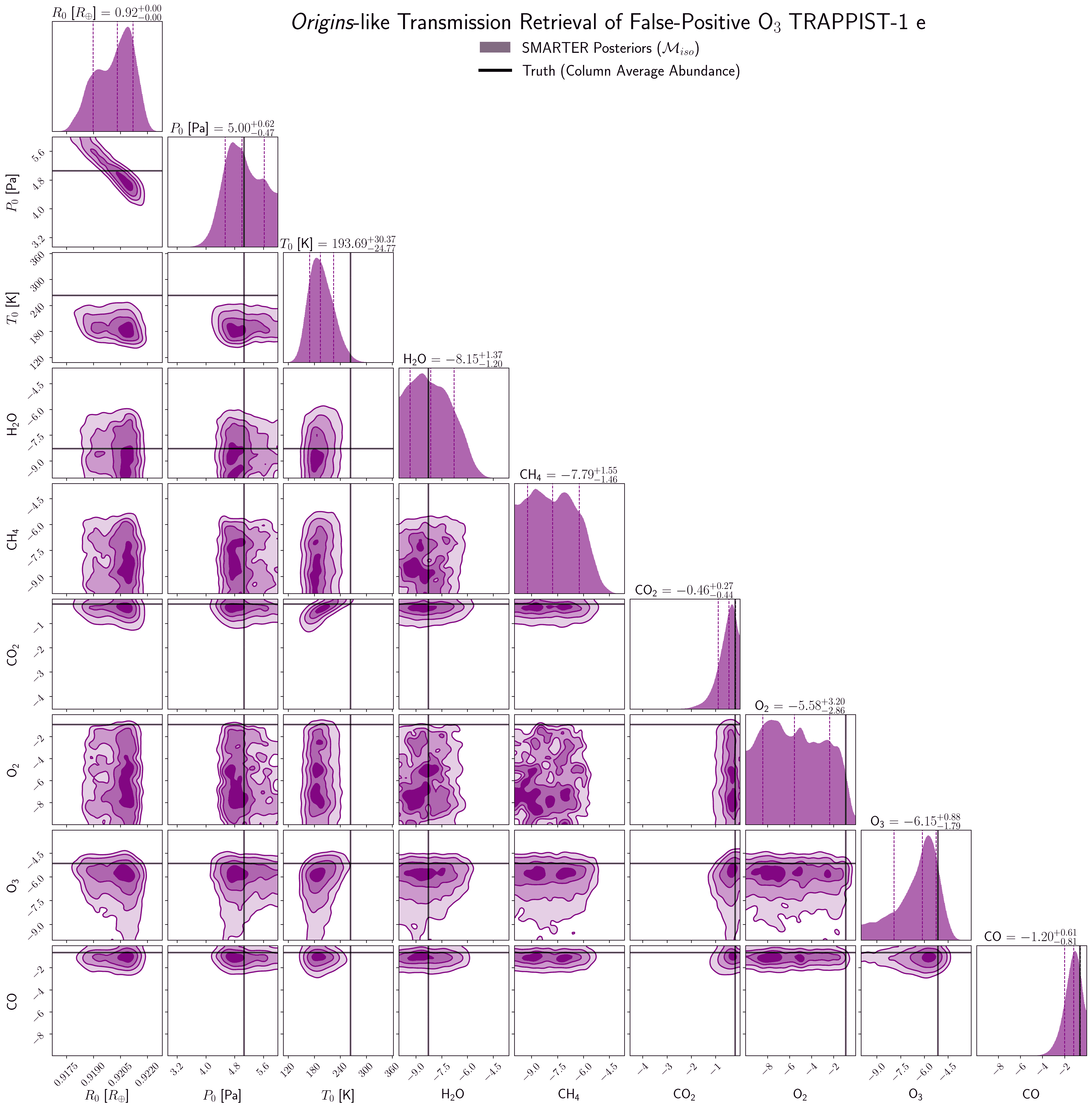}
    \caption{The results of the retrieval on the abiotic, false-positive O$_3$ TRAPPIST-1 e transmission spectrum with the evenly-mixed forward model, $\mathcal{M}_{iso}$.}
\end{figure}
\newpage


\figsetstart
\figsetnum{3}
\figsettitle{LUVOIR-Like Direct Imaging Retrieval of Clear-Sky Earth at 10 pc}
\figsetgrpstart
\figsetgrpnum{figurenumber.1}
\figsetgrptitle{Evenly-Mixed Forward Model ($\mathcal{M}_{iso})$}
\figsetplot{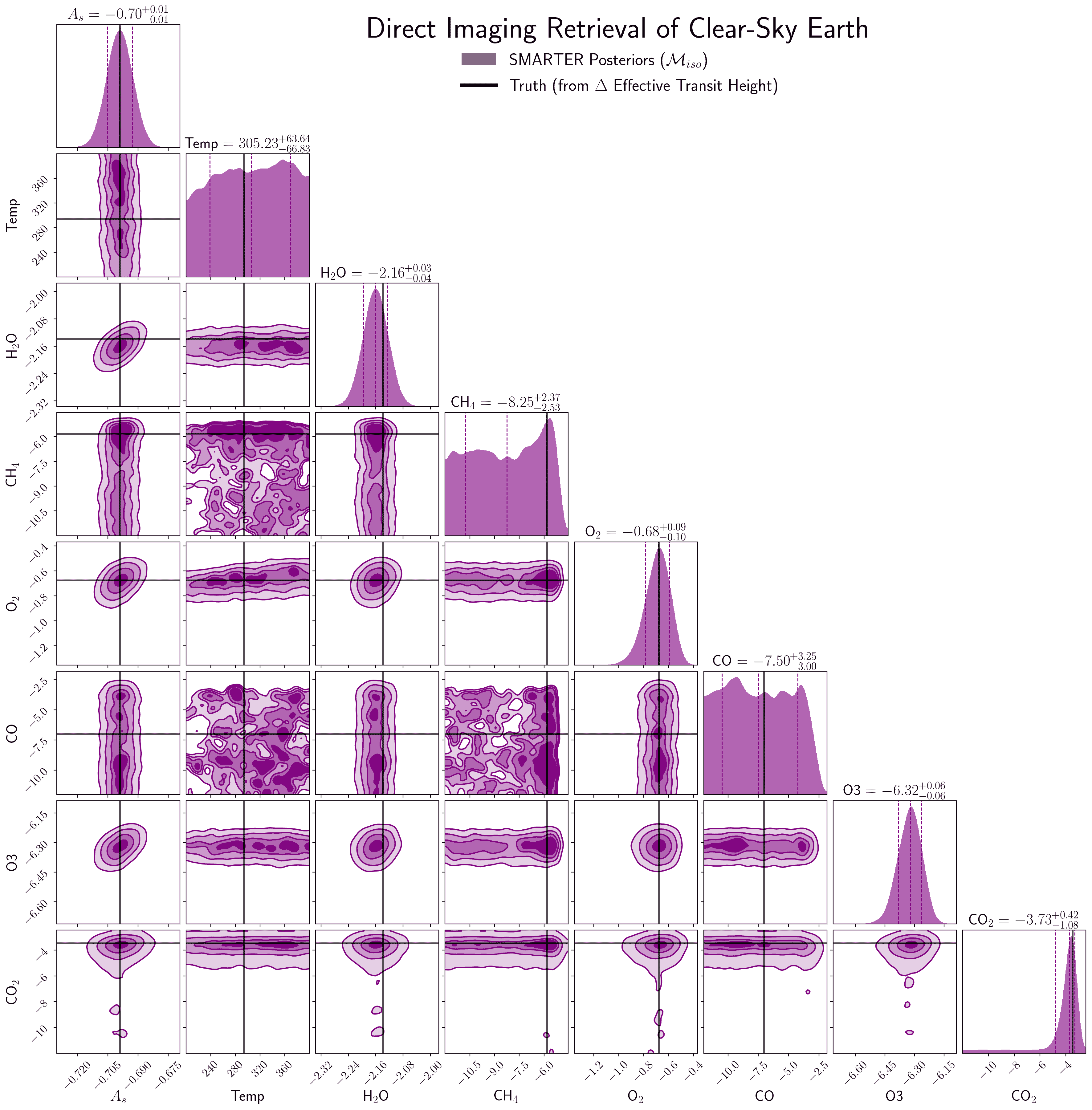}
\figsetgrpnote{The results of the retrieval on the clear-sky Earth direct imaging spectrum with the evenly-mixed forward model, $\mathcal{M}_{iso}$.}
\figsetgrpend

\figsetgrpnum{figurenumber.2}
\figsetgrptitle{Three-Fixed Points Forward Model ($\mathcal{M}_{3P})$}
\figsetplot{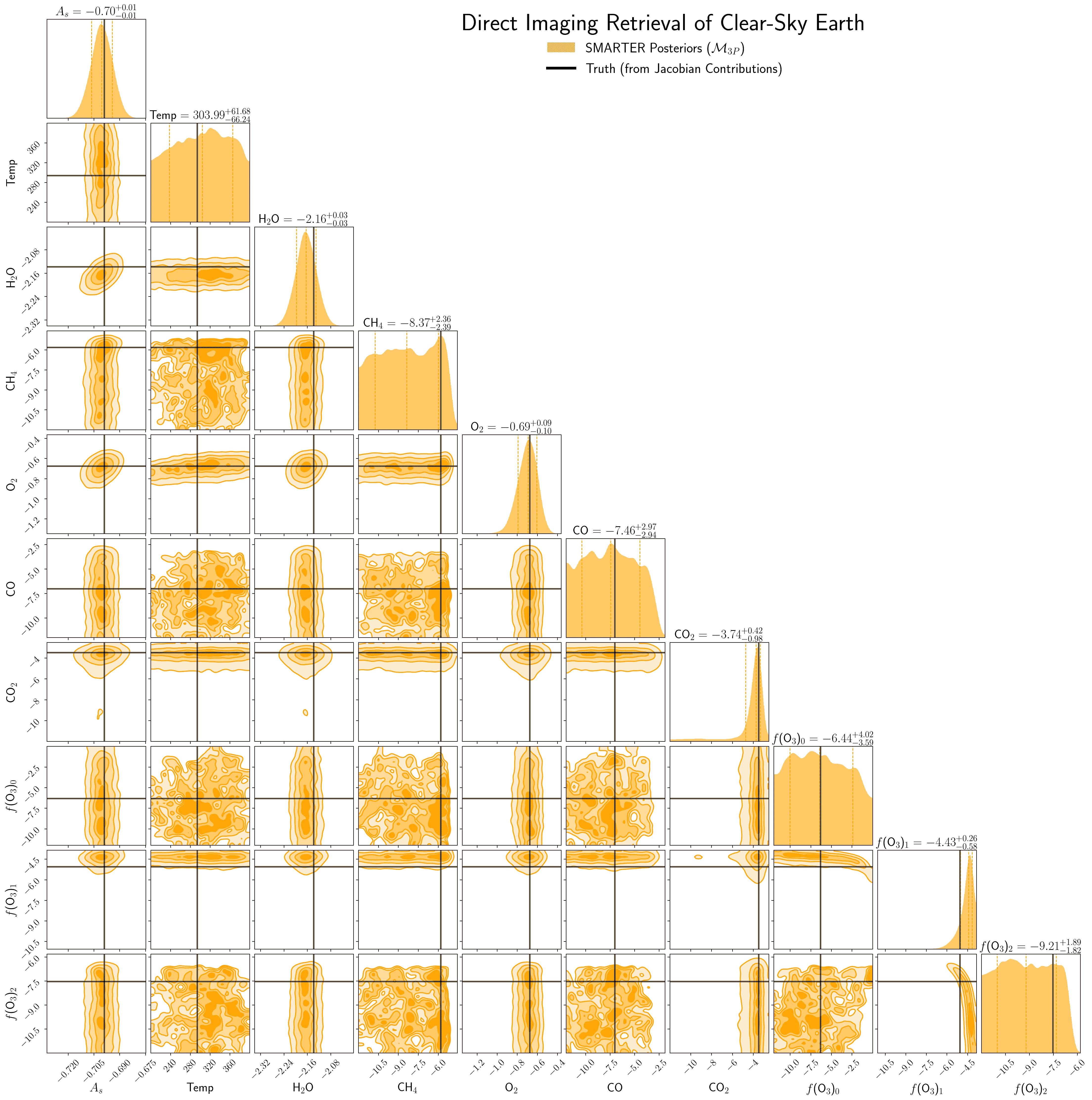}
\figsetgrpnote{The results of the retrieval on the clear-sky Earth direct imaging spectrum with the three-fixed points forward model, $\mathcal{M}_{3P}$.}
\figsetgrpend

\figsetgrpnum{figurenumber.3}
\figsetgrptitle{Five-Fixed Points Forward Model ($\mathcal{M}_{5P})$}
\figsetplot{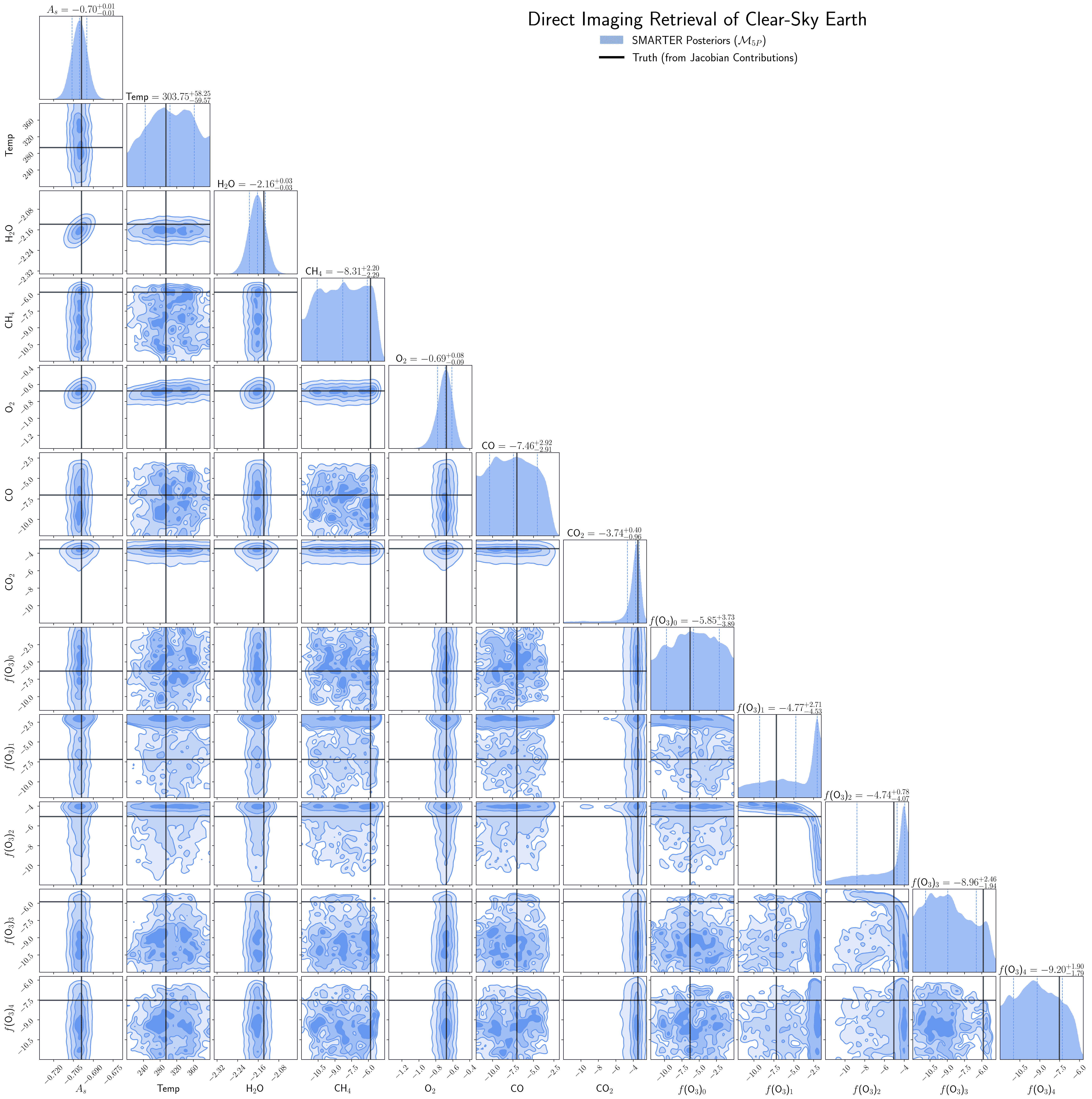}
\figsetgrpnote{The results of the retrieval on the clear-sky Earth direct imaging spectrum with the five-fixed points forward model, $\mathcal{M}_{5P}$.}
\figsetgrpend

\figsetgrpnum{figurenumber.4}
\figsetgrptitle{Three-Free Points Forward Model ($\mathcal{M}_{3F})$}
\figsetplot{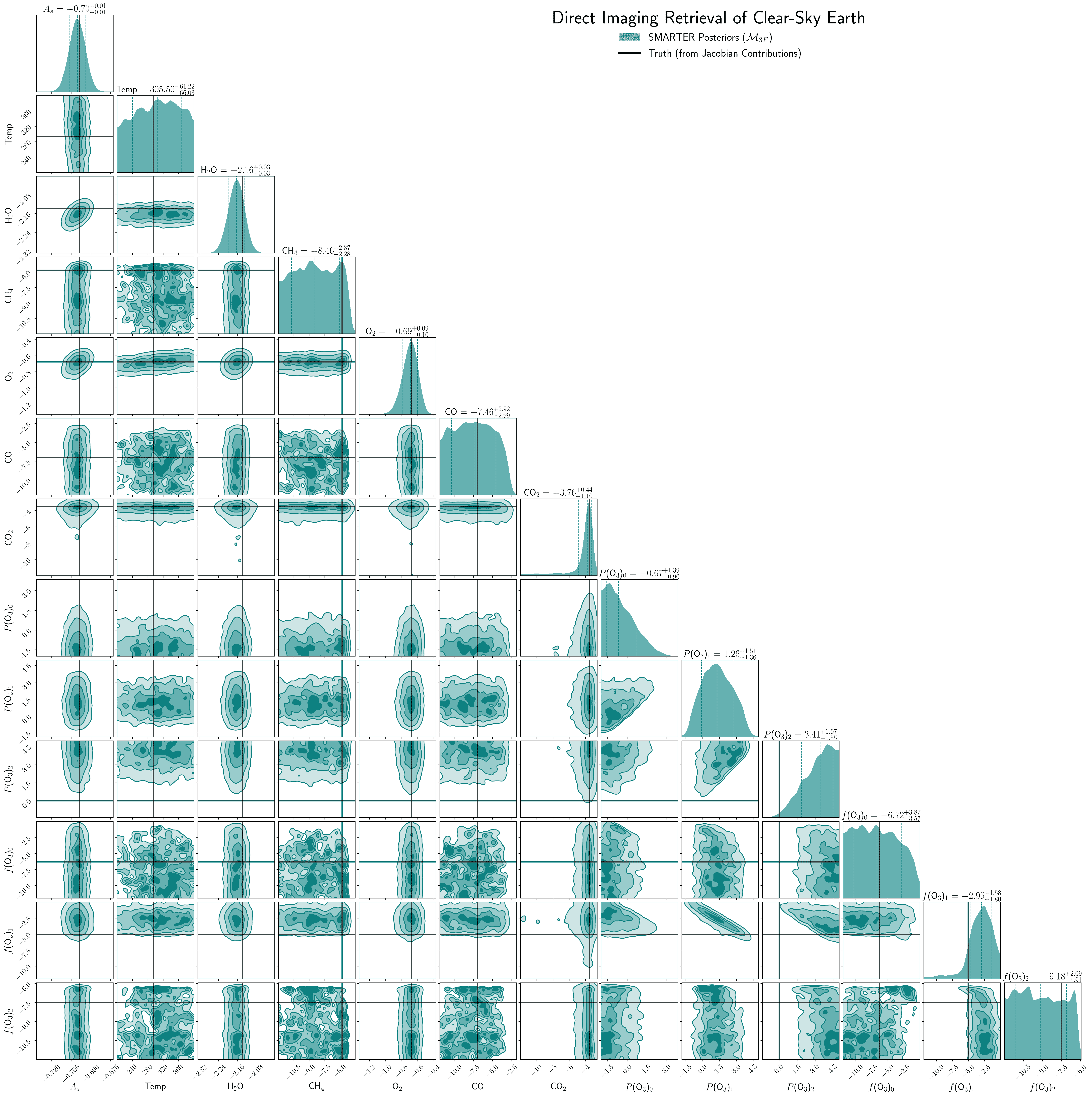}
\figsetgrpnote{The results of the retrieval on the clear-sky Earth direct imaging spectrum with the three-free points forward model, $\mathcal{M}_{3F}$.}
\figsetgrpend
\figsetend

\begin{figure}[htbp]
    \centering
    \includegraphics[width=\textwidth]{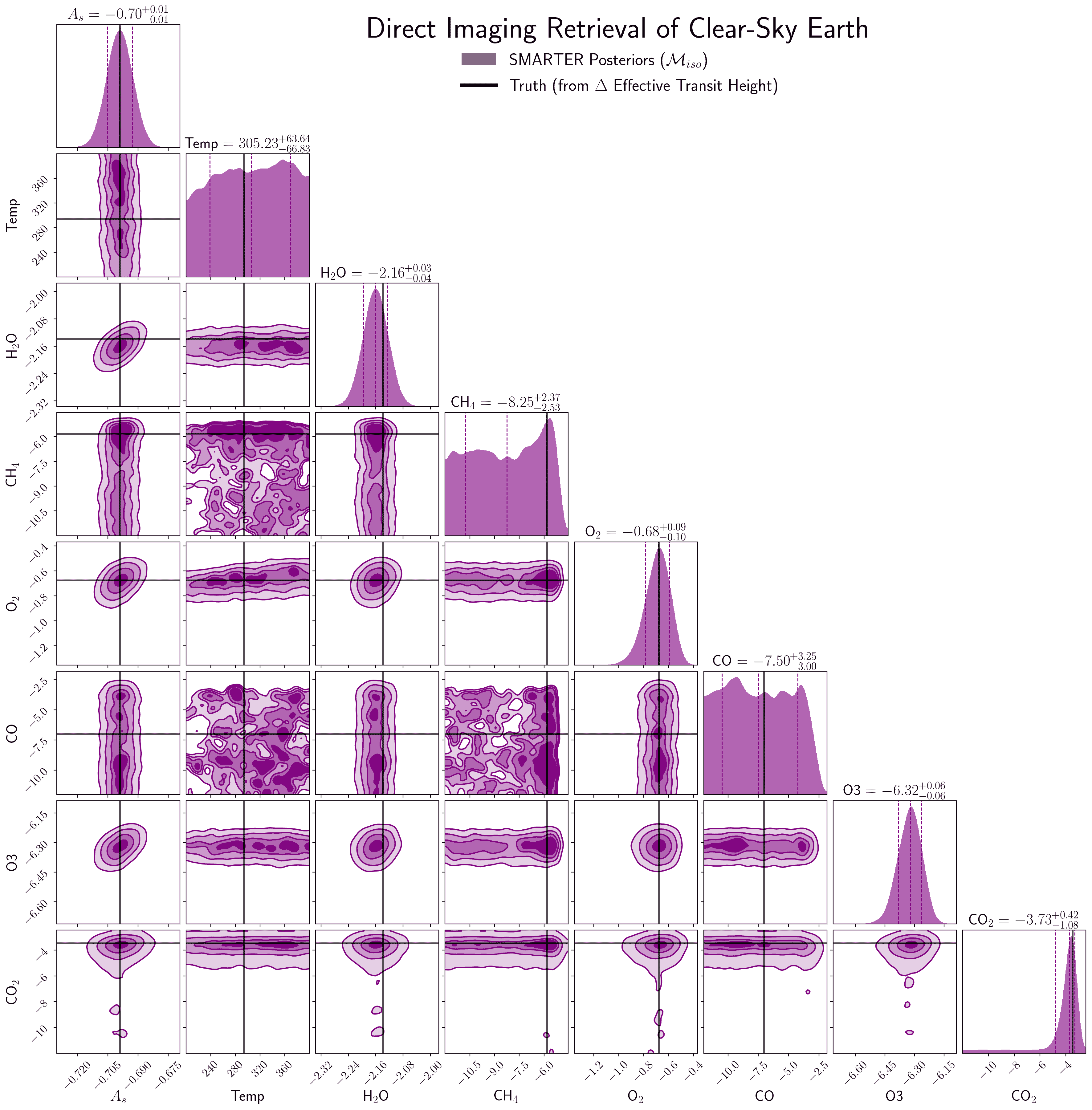}
    \caption{The results of the retrieval on the clear-sky Earth direct imaging spectrum with the evenly-mixed forward model, $\mathcal{M}_{iso}$.}
\end{figure}
\newpage


\figsetstart
\figsetnum{4}
\figsettitle{LUVOIR-Like Direct Imaging Retrieval of 50\% Cloudy Earth at 10 pc}
\figsetgrpstart
\figsetgrpnum{figurenumber.1}
\figsetgrptitle{Evenly-Mixed Forward Model ($\mathcal{M}_{iso})$}
\figsetplot{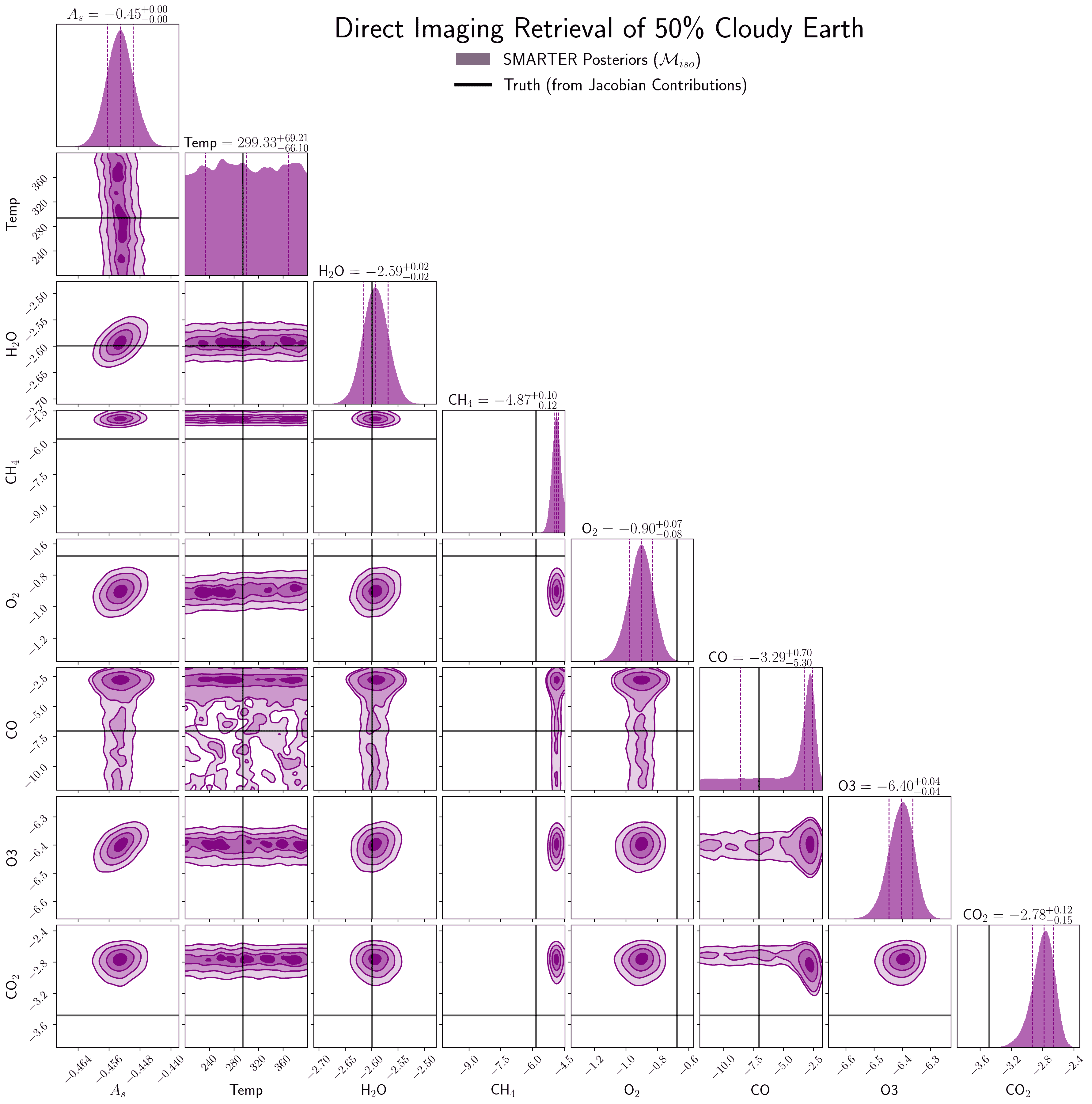}
\figsetgrpnote{The results of the retrieval on the 50\% cloudy (25\% stratocumulus, 25\% cirrus, 50\% clear) Earth direct imaging spectrum with the evenly-mixed forward model, $\mathcal{M}_{iso}$.}
\figsetgrpend

\figsetgrpstart
\figsetgrpnum{figurenumber.2}
\figsetgrptitle{Three-Fixed Points Forward Model ($\mathcal{M}_{3P})$}
\figsetplot{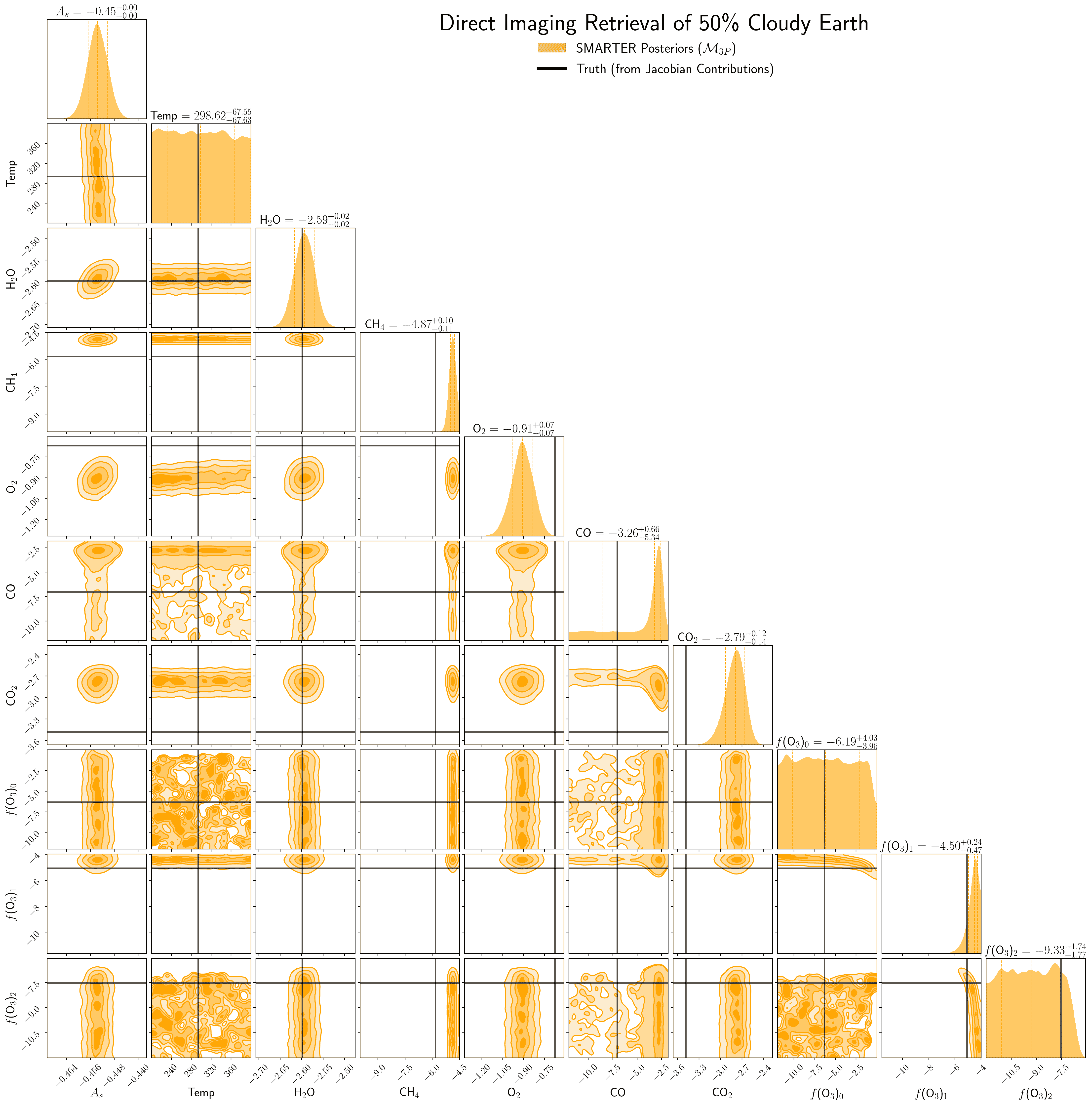}
\figsetgrpnote{The results of the retrieval on the 50\% cloudy Earth direct imaging spectrum with the three-fixed points forward model, $\mathcal{M}_{3P}$.}
\figsetgrpend

\figsetgrpstart
\figsetgrpnum{figurenumber.3}
\figsetgrptitle{Five-Fixed Points Forward Model ($\mathcal{M}_{5P})$}
\figsetplot{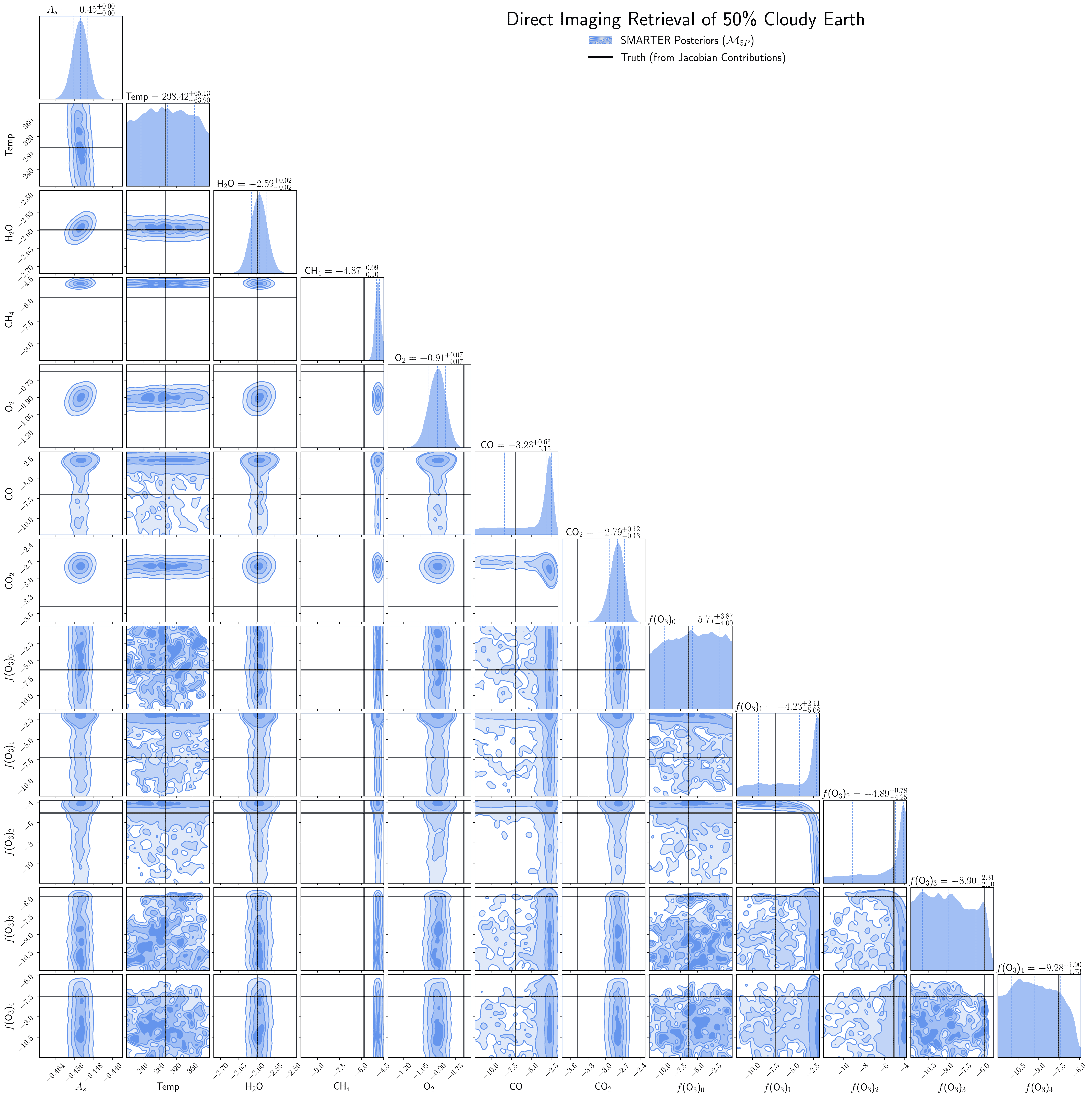}
\figsetgrpnote{The results of the retrieval on the 50\% cloudy Earth direct imaging spectrum with the five-fixed points forward model, $\mathcal{M}_{5P}$.}
\figsetgrpend

\figsetgrpstart
\figsetgrpnum{figurenumber.4}
\figsetgrptitle{Three-Free Points Forward Model ($\mathcal{M}_{3F})$}
\figsetplot{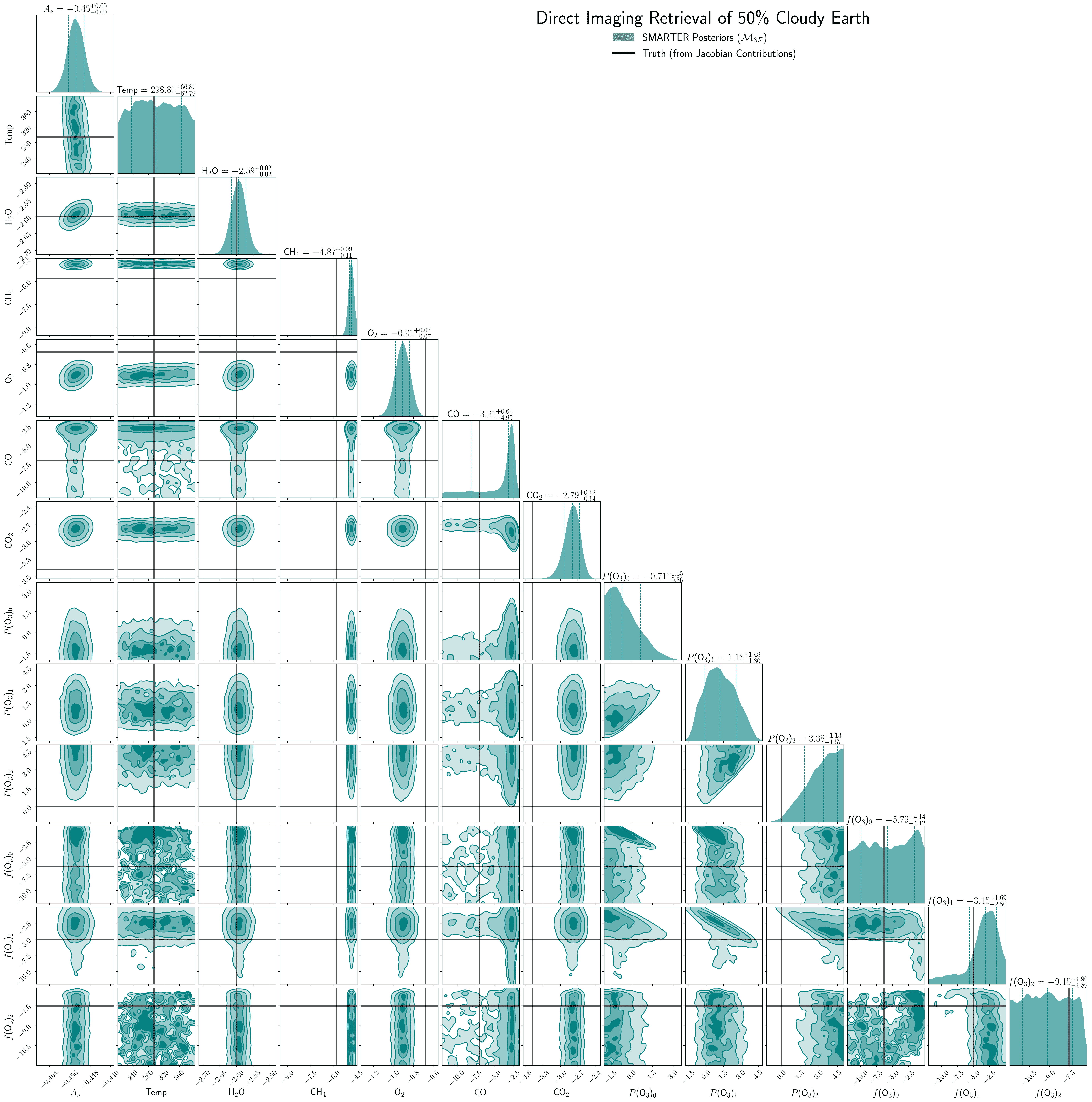}
\figsetgrpnote{The results of the retrieval on the 50\% cloudy Earth direct imaging spectrum with the three-free points forward model, $\mathcal{M}_{3F}$.}
\figsetgrpend
\figsetend

\begin{figure}[htbp]
    \centering
    \includegraphics[width=\textwidth]{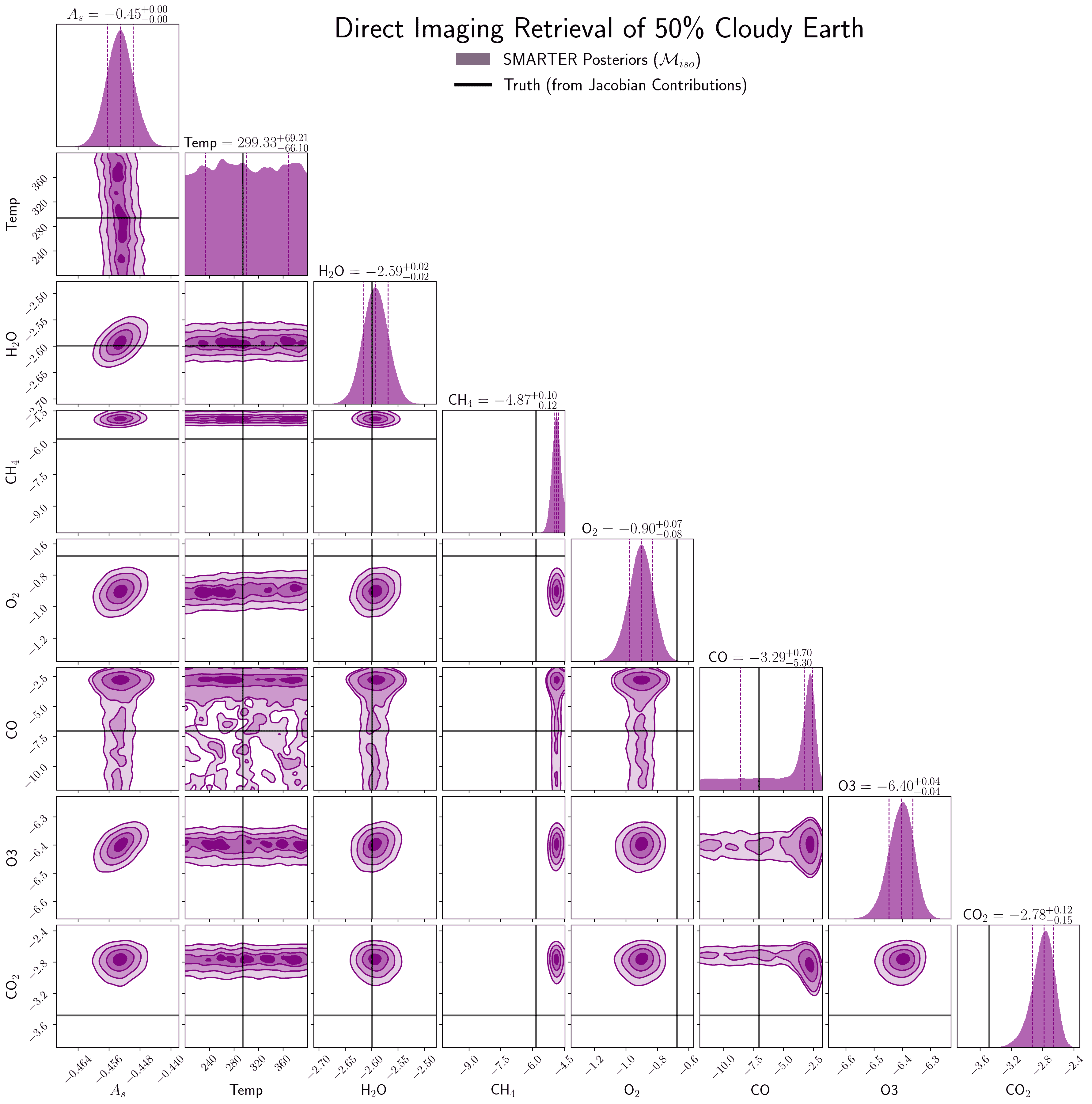}
    \caption{The results of the retrieval on the 50\% cloudy (25\% stratocumulus, 25\% cirrus, 50\% clear) Earth direct imaging spectrum with the evenly-mixed forward model, $\mathcal{M}_{iso}$.}
\end{figure}

\end{document}